\documentclass[twocolumn,superscriptaddress,longbibliography]{revtex4-1}

\usepackage{array}
\usepackage{graphicx}
\usepackage{verbatim}
\usepackage{color}
\usepackage{multirow}

\newcommand{\ket}[1]{\left\vert{#1}\right\rangle}

\begin{document}

\title{Layered architecture for quantum computing}
\author{N. Cody Jones}
\email{ncodyjones@gmail.com}
\affiliation{Edward L. Ginzton Laboratory,
         Stanford University,
         Stanford, California 94305-4088, USA}

\author{Rodney {Van Meter}}
\affiliation{Faculty of Environment and Information Studies, Keio University, Japan}
\author{Austin G. Fowler}
\affiliation{Centre for Quantum Computation and Communication Technology, University of Melbourne, Victoria, Australia}
\author{Peter L. McMahon}
\affiliation{Edward L. Ginzton Laboratory,
         Stanford University,
         Stanford, California 94305-4088, USA}
\author{Jungsang Kim}
\affiliation{Fitzpatrick Institute for Photonics, Duke University, Durham, NC, USA}
\author{Thaddeus D. Ladd}
\affiliation{Edward L. Ginzton Laboratory,
         Stanford University,
         Stanford, California 94305-4088, USA}
\affiliation{National Institute of Informatics,
         2-1-2 Hitotsubashi, Chiyoda-ku, Tokyo 101-8430, Japan}
\author{Yoshihisa Yamamoto}
\affiliation{Edward L. Ginzton Laboratory,
         Stanford University,
         Stanford, California 94305-4088, USA}
\affiliation{National Institute of Informatics,
         2-1-2 Hitotsubashi, Chiyoda-ku, Tokyo 101-8430, Japan}

\begin{abstract}
We develop a layered quantum computer architecture, which is a systematic framework for tackling the individual challenges of developing a quantum computer while constructing a cohesive device design.  We discuss many of the prominent techniques for implementing circuit-model quantum computing and introduce several new methods, with an emphasis on employing surface code quantum error correction.  In doing so, we propose a new quantum computer architecture based on optical control of quantum dots.  The timescales of physical hardware operations and logical, error-corrected quantum gates differ by several orders of magnitude.  By dividing functionality into layers, we can design and analyze subsystems independently, demonstrating the value of our layered architectural approach.  Using this concrete hardware platform, we provide resource analysis for executing fault-tolerant quantum algorithms for integer factoring and quantum simulation, finding that the quantum dot architecture we study could solve such problems on the timescale of days.
\end{abstract}


\maketitle

\section{Introduction}
\label{Intro}
Quantum computing as an engineering discipline is still in its infancy.  Although the physics is well understood, developing devices which compute with quantum mechanics is technologically daunting.  While experiments to date manipulate only a handful of quantum bits~\cite{Ladd10}, we consider what effort is required to build a large-scale quantum computer.  This objective demands more than a cursory estimate of the number of qubits and gates required for a given algorithm.  One must consider the faulty quantum hardware, with errors caused by both the environment and deliberate control operations; when error correction is invoked, classical processing is required; constructing arbitrary gate sequences from a limited fault-tolerant set requires special treatment, and so on.  This paper provides a framework to address the complete challenge of designing a quantum computer.

Many researchers have presented and examined components of large-scale quantum computing.  We study here how these components may be combined in an efficient design, and we introduce new methods which improve the quantum computer we propose.  This engineering pursuit is quantum computer architecture, which we develop here in layers.  An \emph{architecture} decomposes complex system behaviors into a manageable set of operations.  A \emph{layered architecture} does this through layers of abstraction where each embodies a critical set of related functions.  For our purposes, each ascending layer brings the system closer to an ideal quantum computing environment.

The paper is organized as follows.  The remainder of Section~\ref{Intro} provides a global view of a layered quantum computer architecture, indicating how each of the topics we examine are connected.  Section~\ref{Physical_layer} enumerates the essential components of a quantum computer by examining a new hardware platform based on the optical control of quantum dots.  Section~\ref{virtualization} discusses control techniques for suppressing hardware errors prior to using active error correction.  Section~\ref{QEC} demonstrates how to implement and account for the resources of quantum error correction, with particular emphasis on the surface code~\cite{Fowler09}.  Section~\ref{Logical_layer} analyzes the necessary techniques for constructing universal quantum gates from the limited set of operations provided by error correction.  Section~\ref{application_layer} calculates the computer resources necessary to implement two prominent quantum algorithms: integer factoring and quantum simulation.  Section~\ref{Timing} discusses timing issues which affect how the layers in the architecture interact with each other.  Finally, Section~\ref{discussion} discusses how our findings are applicable to future work in quantum computing.

\subsection{Prior work on quantum computer architecture}
Many different quantum computing technologies are under experimental investigation~\cite{Ladd10}, but for each a scalable system architecture remains an open research problem.  Since DiVincenzo introduced his fundamental criteria for a viable quantum computing technology~\cite{DiVincenzo00} and Steane emphasized the difficulty of designing systems capable of running quantum error correction (QEC) adequately~\cite{Steane02,Steane2007}, several groups of researchers have outlined various additional taxonomies addressing the architectural needs of large-scale systems~\cite{Spiller05,VanMeter06}.  As an example, small-scale interconnects have been proposed for many technologies, but the problems of organizing subsystems using these techniques into a complete architecture for a large-scale system have been addressed by only a few researchers.  In particular, the issue of heterogeneity in system architecture has received relatively little attention.

The most important subroutine in fault-tolerant quantum computers considered thus far is the preparation of ancilla states for fault-tolerant circuits, because these circuits often require very many ancillas.  Taylor \emph{et al.} proposed a design with alternating ``ancilla blocks'' and ``data blocks'' in the device layout~\cite{Taylor2005}.  Steane introduced the idea of ``factories'' for creating ancillas~\cite{Steane1998}, which we examine for the case of the surface code in this work.  Isailovic \emph{et al.}~\cite{Isailovic08} studied this problem for ion trap architectures and found that, for typical quantum circuits, approximately 90\% of the quantum computer must be devoted to such factories in order to calculate ``at the speed of data,'' or where ancilla-production is not the rate-limiting process.  The findings we present here are in close agreement with this estimate.  Metodi \emph{et al.} also considered production of ancillas in ion trap designs, focusing instead on a 3-qubit ancilla state used for the Toffoli gate~\cite{Metodi2005}, which is an alternative pathway to a universal fault-tolerant set of gates.

Some researchers have studied the difficulty of moving data in a quantum processor.  Kielpinski \emph{et al.}~proposed a scalable ion trap technology utilizing separate memory and computing areas~\cite{Kielpinski02}.  Because quantum error correction requires rapid cycling across all physical qubits in the system, this approach is best used as a unit cell replicated across a larger system.   Other researchers have proposed homogeneous systems built around this basic concept.  One common structure is a recursive H tree, which works well with a small number of layers of a Calderbank-Shor-Steane (CSS) code, targeted explicitly at ion trap systems~\cite{Copsey03,Svore06}.   Oskin \emph{et al.}~\cite{Oskin03}, building on the Kane solid-state NMR technology~\cite{Kane98}, proposed a loose lattice of sites, explicitly considering the issues of classical control and movement of quantum data in scalable systems, but without a specific plan for QEC.  In the case of quantum computing with superconducting circuits, the quantum von Neumann architecture specifically considers dedicated hardware for quantum memories, zeroing registers, and a quantum bus~\cite{Mariantoni2011}.

Long-range coupling and communication is a significant challenge for quantum computers.  Cirac~\emph{et al.} proposed the use of photonic qubits to distribute entanglement between distant atoms~\cite{Cirac1997}, and other researchers have investigated the prospects for optically-mediated nonlocal gates~\cite{vanEnk1999,Steane2000,Duan01,VanMeter2007,DuanMonroe}.  Such photonic channels could be utilized to realize a modular, scalable distributed quantum computer~\cite{KimKim}.  Conversely, Metodi \emph{et al.} consider how to use local gates and quantum teleportation to move logical qubits throughout their ion-trap QLA architecture~\cite{Metodi2005}.  Fowler \emph{et al.}~\cite{Fowler07} investigated a Josephson junction flux qubit architecture considering the extreme difficulties of routing both the quantum couplers and large numbers of classical control lines, producing a structure with support for CSS codes and logical qubits organized in a line.  Whitney \emph{et al.}~\cite{Whitney07,Whitney09} have investigated automated layout and optimization of circuit designs specifically for ion trap architectures, and Isailovic \emph{et al.}~\cite{Isailovic06,Isailovic08} have studied interconnection and data throughput issues in similar ion trap systems, with an emphasis on preparing ancillas for teleportation gates~\cite{Gottesman1999}.

Other work has studied quantum computer architectures with only nearest-neighbor coupling between qubits in an array~\cite{Levy2001,Fowler2004,Aliferis2007b,Levy2009,Levy2011}, which is appealing from a hardware design perspective.  With the recent advances in the operation of the topological codes and their desirable characteristics such as having a high practical threshold and requiring only nearest-neighbor interactions, research effort has shifted toward architectures capable of building and maintaining large two- and three-dimensional cluster states~\cite{Weinstein2005,Stock09,Devitt10,Devitt2011}. These systems rely on topological error correction models~\cite{Kitaev2002}, whose higher tolerance to error often comes at the cost of a larger physical system, relative to, for example, implementations based on the Steane code~\cite{Oskin2002}.  The surface code~\cite{Fowler09}, which we examine in this work for its impact on architecture, belongs to the topological family of codes.

Recent attention has been directed at distributed models of quantum computing.  Devitt~\emph{et~al.} studied how to distribute a photonic cluster-state quantum computing network over different geographic regions~\cite{Devitt2011b}.  The abstract framework of a quantum multicomputer recognizes that large-scale systems demand heterogeneous interconnects~\cite{VanMeter06b}; in most quantum computing technologies, it may not be possible to build monolithic systems that contain, couple, and control billions of physical qubits.  Van Meter \emph{et al.}~\cite{VanMeter09} extended this architectural framework with a design based on nanophotonic coupling of electron spin quantum dots that explicitly uses multiple levels of interconnect with varying coupling fidelities (resulting in varying purification requirements), as well as the ability to operate with a very low yield of functional devices.  Although that proposed system has many attractive features, concerns about the difficulty of fabricating adequately high quality optical components and the desire to reduce the surface code lattice cycle time led to the system design proposed in this paper.

\subsection{Layered framework}
A good architecture must have a simple structure while also efficiently managing the complex array of resources in a quantum computer.  Layered architectures are a conventional approach to solving such engineering problems in many fields of information technology, and Ref.~\cite{Svore06} presents a layered architecture for quantum computer design software.  Our architecture, which describes the physical design of the quantum computer, consists of five layers, where each layer has a prescribed set of duties to accomplish.  The interface between two layers is defined by the services a lower layer provides to the one above it.  To execute an operation, a layer must issue commands to the layer below and process the results.  Designing a system this way ensures that related operations are grouped together and that the system organization is hierarchical.  Such an approach allows quantum engineers to focus on individual challenges, while also seeing how a process fits into the overall design.  By organizing the architecture in layers, we deliberately create a \emph{modular} design for the quantum computer.

\begin{figure}
  \centering
  \includegraphics[width=8.5cm]{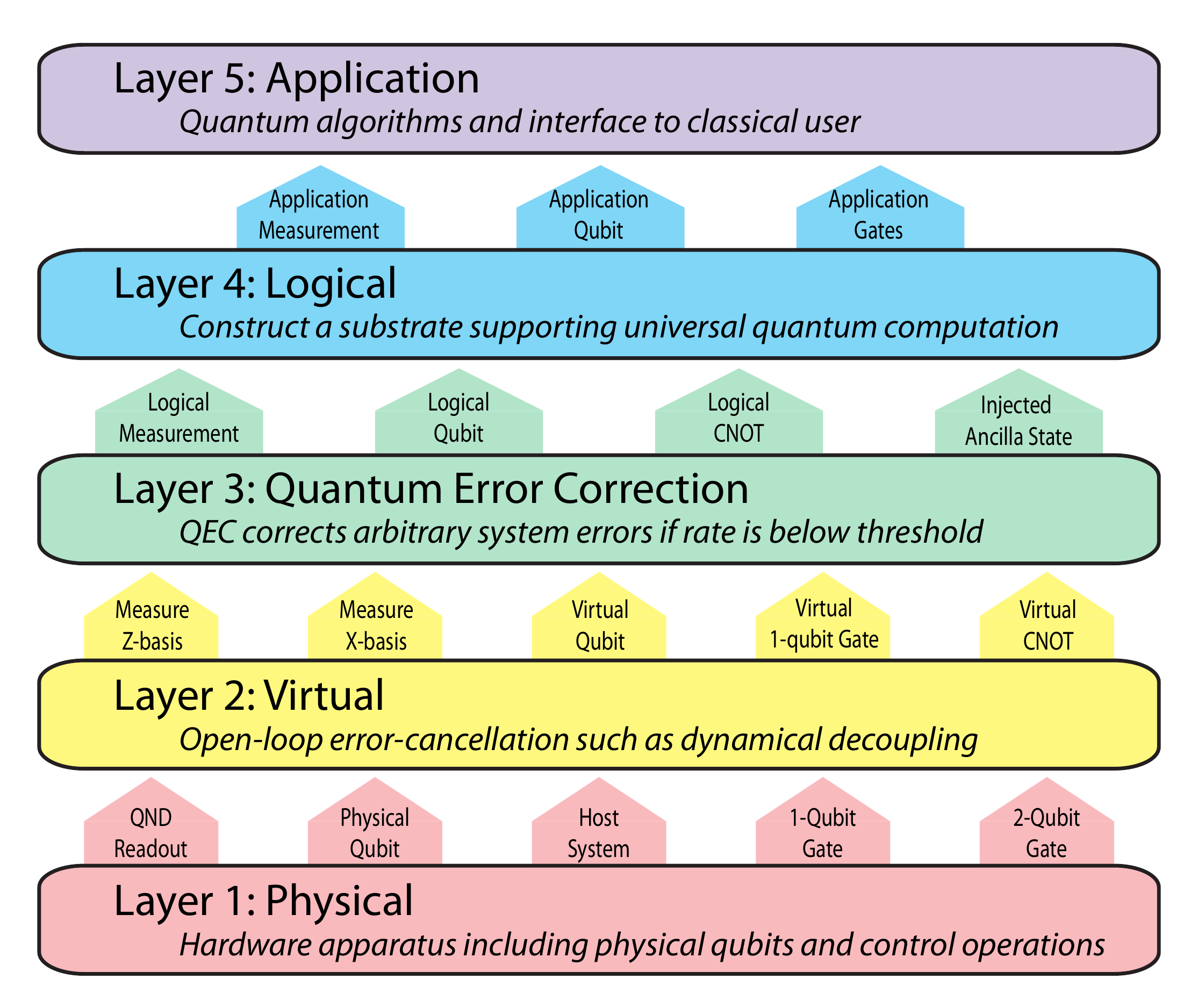}\\
  \caption{Color.  Layered control stack which forms the framework of a quantum computer architecture.  Vertical arrows indicate services provided to a higher layer.}
  \label{LayeredControlStack}
\end{figure}

The layered framework can be understood by a control stack composed of the five layers in the architecture.  Fig.~\ref{LayeredControlStack} shows an example of the control stack for the quantum dot architecture we propose here, but the particular interfaces between layers will vary according to the physical hardware, quantum error correction scheme, \emph{etc.} that one chooses to implement.  At the top of the control stack is the Application layer, where a quantum algorithm is implemented and results are provided to the user. The bottom Physical layer hosts the raw physical processes supporting the quantum computer. The layers between (Virtual, Quantum Error Correction, and Logical) are essential for shaping the faulty quantum processes in the Physical layer into a system of high-accuracy \emph{fault-tolerant}~\cite{Preskill97} qubits and quantum gates at the Application layer.

\subsection{Interaction between layers}
\begin{figure*}
  \centering
  \includegraphics[width=15cm]{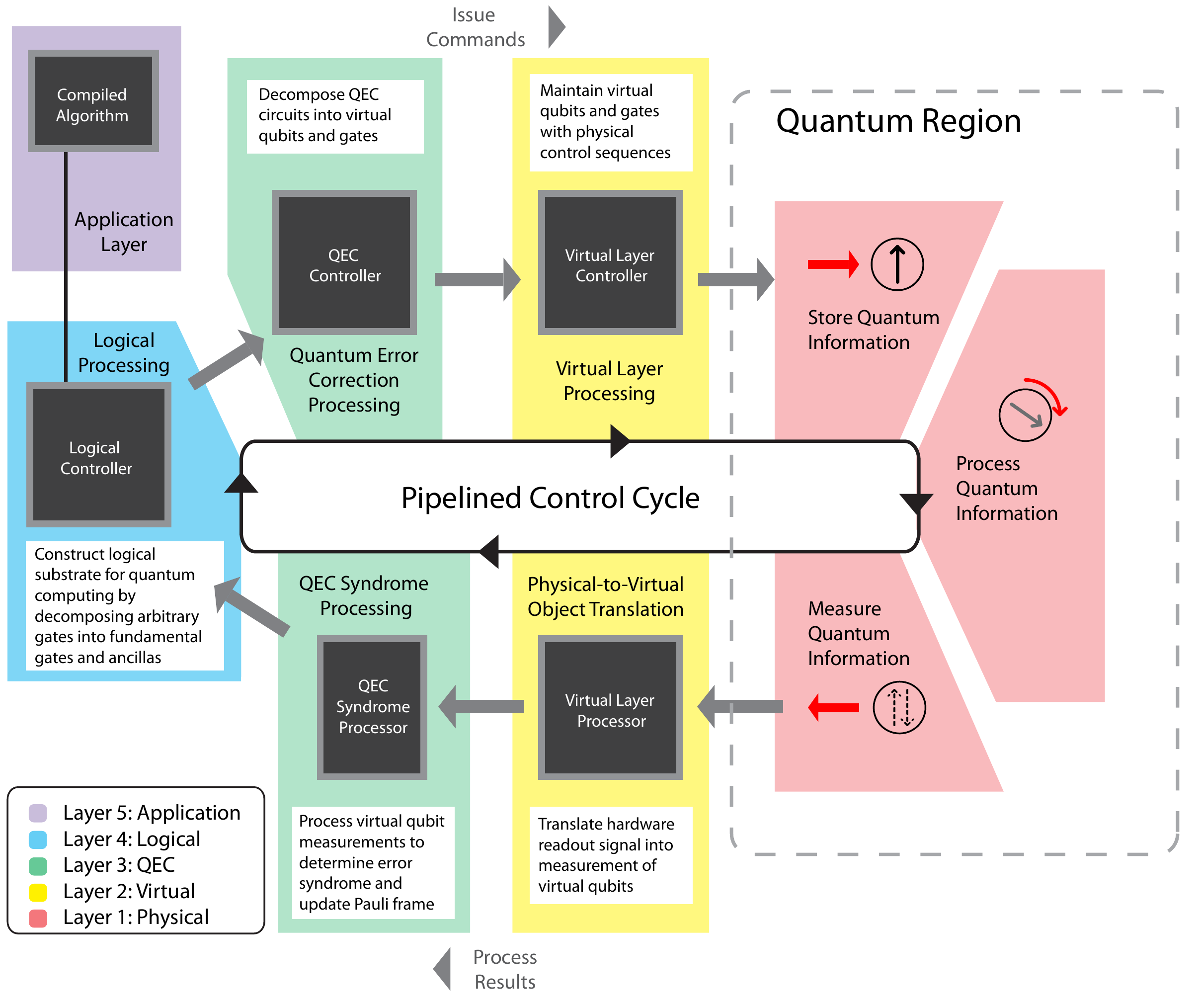}\\
  \caption{Color.  Primary control cycle of the layered architecture quantum computer.  Whereas the control stack in Fig.~\ref{LayeredControlStack} dictates the interfaces between layers, the control cycle determines the timing and sequencing of operations.  The dashed box encircling the Physical layer indicates that all quantum processes happen exclusively here, and the layers above process and organize the operations of the Physical layer.  The Application layer is external to the loop since it functions without any dependence on the specific quantum computer design.}
  \label{LayeredControlCycle_Abstract}
\end{figure*}

Two layers meet at an interface, which defines how they exchange instructions or the results of those instructions.  Many different commands are being executed and processed simultaneously, so we must also consider how the layers interact dynamically. For the quantum computer to function efficiently, each layer must issue instructions to layers below in a tightly defined sequence. However, a robust system must also be able to handle errors caused by faulty devices. To satisfy both criteria, a control loop must handle operations at all layers simultaneously while also processing syndrome measurement to correct errors that occur.  A prototype for this control loop is shown in Fig.~\ref{LayeredControlCycle_Abstract}.

The primary control cycle defines the dynamic behavior of the quantum computer in this architecture since all operations must interact with this loop. The principal purpose of the control cycle is to successfully implement quantum error correction.  The quantum computer must operate fast enough to correct errors; still, some control operations necessarily incur delays, so this cycle does not simply issue a single command and wait for the result before proceeding --- pipelining is essential~\cite{Isailovic08,Shen05}.  A related issue is that operations in different layers occur on drastically different timescales, as discussed later in Section~\ref{Timing}.  Fig.~\ref{LayeredControlCycle_Abstract} also describes the control structure needed for the quantum computer. Processors at each layer track the current operation and issue commands to lower layers.  Layers~1 to 4 interact in the loop, whereas the Application layer interfaces only with the Logical layer since it is agnostic about the underlying design of the quantum computer, which is explained in Section~\ref{application_layer}.

\subsection{The QuDOS hardware platform}
\begin{figure*}
  \centering
  \includegraphics[width=\textwidth]{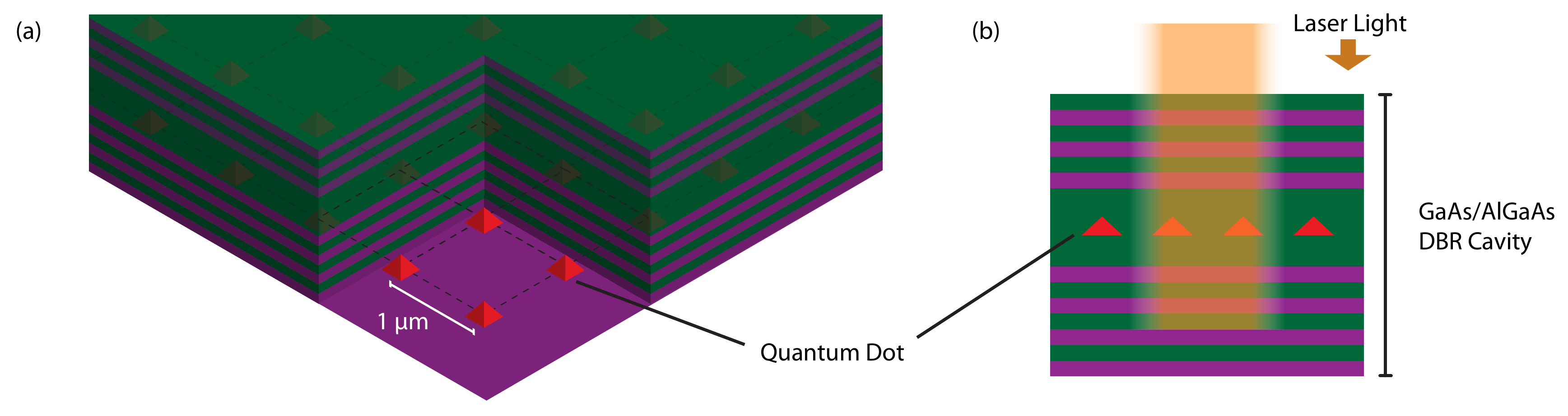}\\
  \caption{Color.  Quantum dots in a planar optical microcavity form the basis of the \mbox{QuDOS} hardware platform.  (a)~The quantum dots are arranged 1~$\mu$m apart in a two-dimensional square array.  The quantum dots trap single electrons, whose spins will be used for quantum information processing.  (b)~Side view. The electron spins are manipulated with laser pulses sent into the optical cavity from above, and two neighboring quantum dots can be coupled by a laser optical field which overlaps them.  The purple and green layers are AlGaAs and GaAs, grown by molecular beam epitaxy.  The alternating layers form a distributed Bragg reflector (DBR) optical cavity which is planar, confining light in the vertical direction and extending across the entire system in horizontal directions.}
  \label{Cavity_Perspective}
\end{figure*}

The layered framework for quantum computing was developed in tandem with a specific hardware platform, known as \mbox{QuDOS} (\textbf{qu}antum \textbf{d}ots with \textbf{o}ptically-controlled \textbf{s}pins).  The \mbox{QuDOS} platform uses electron spins within quantum dots for qubits. The quantum dots are arranged in a two-dimensional array; Fig.~\ref{Cavity_Perspective} shows a cut-away rendering of the quantum dot array inside an optical microcavity, which facilitates control of the electron spins with laser pulses.  We demonstrate that the \mbox{QuDOS} design is a promising candidate for large-scale quantum computing, beginning with an analysis of the hardware in the Physical layer.

\section{Layer 1: Physical}
\label{Physical_layer}
The essential requirements for the Physical layer are embodied by the DiVincenzo criteria~\cite{DiVincenzo00}, but we are also interested in performance of the quantum hardware.  The timescale of operations and the degrees of errors, both systematic and random, are critical parameters which determine the size and speed of the computer.  This section discusses the essential hardware components of a quantum computer, accompanied by the \mbox{QuDOS} platform we introduce as an example.  We conclude by analyzing the performance of the \mbox{QuDOS} hardware.  We caution that many of the required hardware elements are still under experimental development, but we choose those discussed below as examples to establish timescales which will impact higher layers of the architecture.

\subsection{Physical qubit}
\label{Quantum_memory}
A quantum computer must have the ability to store information between processing steps; the object fulfilling this role is conventionally known as the physical qubit.  A physical qubit may be more complex than a two-level system, and this issue is addressed by Layer 2 in the architecture, where control operations are used to form a true quantum bit as an information unit (see Section~\ref{virtual_qubit}).  Examples of physical qubits include trapped ions, photon polarization modes, electron spins, and quantum states in superconducting circuits~\cite{Ladd10}.  The remainder of the Physical layer is devoted to controlling and measuring the physical qubit.

The layered architecture design is flexible in the sense that the Physical layer can be tailored to a specific hardware, such as superconducting circuit qubits, with minimal change to higher layers such as error correction.  The physical qubit we consider in \mbox{QuDOS} is the spin of an electron bound within an InGaAs self-assembled quantum dot (QD) surrounded by GaAs substrate~\cite{Bjork94,Imamoglu99,Bonadeo00,Guest02,Hours05,Yamamoto09}. These QDs can be optically excited to trion states (a bound electron and exciton), which emit light of wavelength $\sim 900$ nm when they decay.  A transverse magnetic field splits the spin levels into two metastable ground states~\cite{Bayer02}, which will later form a two-level system for a virtual qubit in Layer 2.  The energy separation of the spin states is important for two reasons related to controlling the electron spin.  First, the energy splitting facilitates control with optical pulses as explained in Section~\ref{1qubit_gates}.  Second, there is continuous phase rotation between spin states $\ket{\uparrow}$ and $\ket{\downarrow}$ around the $\hat{Z}$-axis on the qubit Bloch sphere, which in conjunction with timed optical pulses provides complete unitary control of the electron spin vector.

\subsection{Host system}
For our purposes, the \emph{host system} is the engineered environment of the physical qubit which supports computing.  Examples include the trapping fields in ion-trap designs, the waveguides in optical quantum computing, and the diamond crystal surrounding nitrogen-vacancy centers~\cite{Ladd10}.  The host system will define the immediate environment of the physical qubit, which will be important for characterizing noise affecting quantum operations.

We noted above that, in \mbox{QuDOS}, the electron spin is bound within a quantum dot.  These quantum dots are embedded in an optical microcavity, which will facilitate quantum gate operations via laser pulses.  To accommodate the two-dimensional array of the surface code detailed in Layer 3, this microcavity must be planar in design, so the cavity is constructed from two distributed Bragg reflector (DBR) mirrors stacked vertically with a $\lambda/2$ cavity layer in between, as shown in Fig.~\ref{Cavity_Perspective}.  This cavity is grown by molecular beam epitaxy (MBE). The QDs are embedded at the center of this cavity to maximize interaction with antinodes of the cavity field modes.  Using MBE, high-quality (\(Q > 10^{5}\)) microcavities can be grown with alternating layers of GaAs/AlAs~\cite{Reitz07}.  The nuclei in the quantum dot and surrounding substrate have nonzero spin, which is an important source of noise (see Section~\ref{Phys_noise}).

\subsection{1-qubit gate mechanism}
\label{1qubit_gates}
The 1-qubit gate manipulates the state of a single physical qubit.  This 1-qubit gate mechanism is still a physical process, and only later will these physical control operations be combined into a ``virtual gate'' (see Section~\ref{virtual_gate}).  Still, it is important that the Physical layer delivers sufficient control of the physical qubit.  Full unitary control of a qubit requires at least two adjustable degrees of freedom, such as rotation around two axes on the Bloch sphere, and three freely adjustable parameters~\cite{Nielsen00}.

\begin{figure}
  \centering
  \includegraphics[width=8cm]{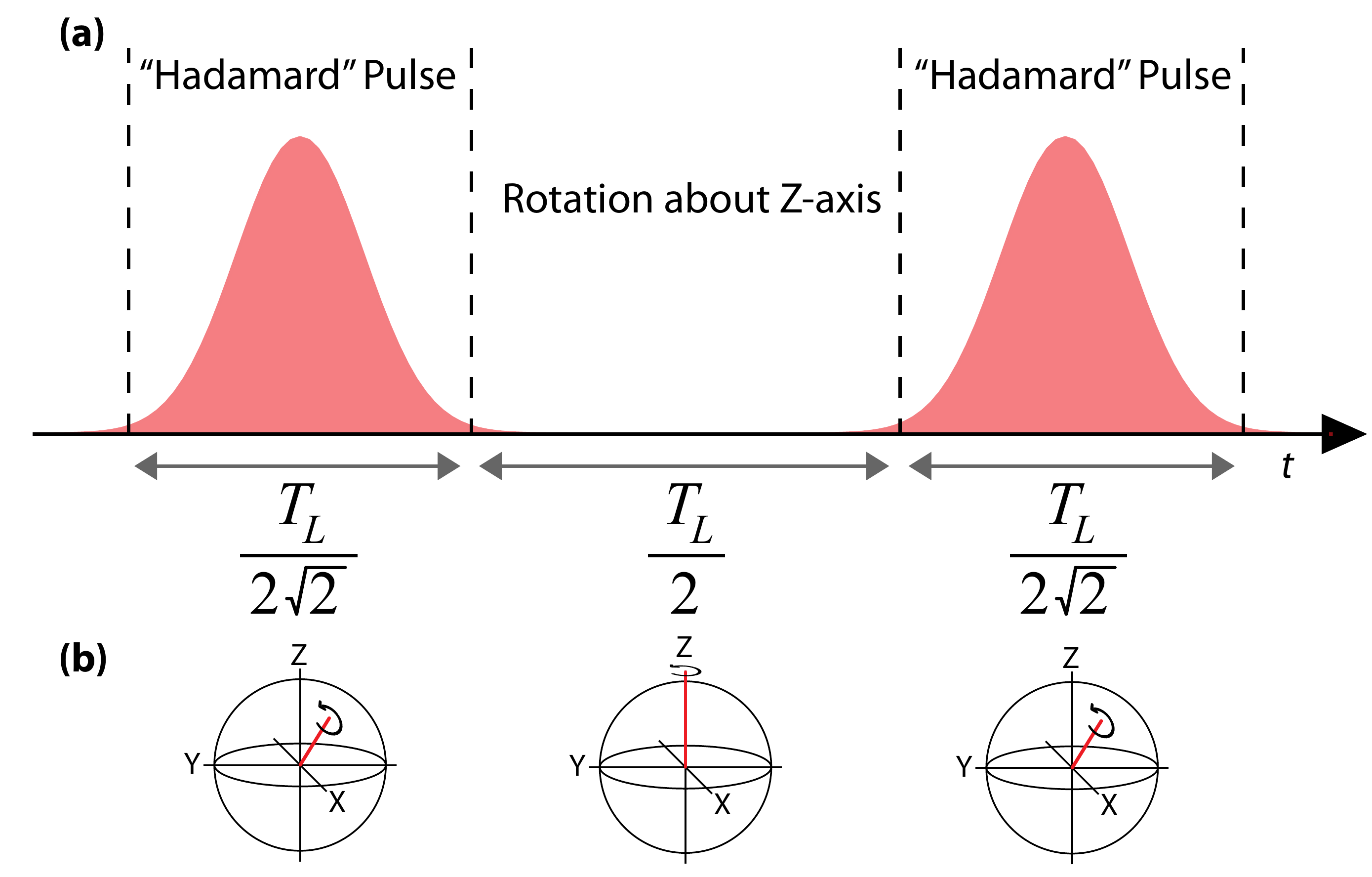}\\
  \caption{Color.  Hadamard pulses in \mbox{QuDOS}.  (a)~A short pulse sequence generates an $\hat{X}$-axis rotation on the spin Bloch sphere with two Hadamard pulses and $\hat{Z}$-axis precession from the magnetic field.  The duration of the pulses and the delay between them is proportional to the Larmor period, $T_L$.  (b)~Bloch sphere diagrams showing the axis of rotation at each time during the sequence.}
  \label{Hadamard}
\end{figure}

The 1-qubit operations in \mbox{QuDOS} are developed using a transverse magnetic field and ultrafast laser pulses~\cite{Press08,Yamamoto09}.  The magnetic field provides a constant-angular-frequency rotation around the $\hat{Z}$ axis, while laser pulses enact a power-dependent rotation around an orthogonal axis, which we label $\hat{X}$.  The first non-ideal behavior we consider is that the laser pulse has some finite duration, so that $\hat{X}$ and $\hat{Z}$ precession happen concurrently, which impairs manipulation of the spin Bloch vector.  To remedy this, we introduce ``Hadamard pulses''~\cite{DeGreve2012}---one tunes the laser pulse power and duration to make the pulse-driven $\hat{X}$-axis rotation equal in angular frequency to the $\hat{Z}$-axis precession from the magnetic field, so that the axis of rotation becomes $H = \frac{1}{\sqrt{2}}\left(\hat{X} + \hat{Z}\right)$.  A ``$\pi$-pulse'' around this axis is a Hadamard gate, and by using two Hadamard pulses and rotation $R_{\hat{Z}}(\theta)$ via free precession in the magnetic field, we can construct any $\hat{X}$-axis rotation by $R_{\hat{X}}(\theta) = H \cdot R_{\hat{Z}}(\theta) \cdot H$, as shown in Fig.~\ref{Hadamard}.  By implementing Hadamard pulses, we can obtain high-fidelity operations with pulses of finite duration.  A challenging problem for \mbox{QuDOS} is how the system executes millions of control operations in parallel.  We envision an optical imaging system consisting of an array of MEMS mirrors to individually steer laser control beams toward or away from quantum dots, along with electro-optic modulators to precisely control laser pulse timing; we discuss this approach in Appendix~\ref{App_Physical}, but rigorously engineering such a system is beyond our scope.

\subsection{2-qubit gate mechanism}
\label{2qubit}
The 2-qubit operation couples two physical qubits, which can generate entanglement.  This mechanism is crucial for quantum computing, yet it is often difficult to implement experimentally.  For example, entangling gates like \texttt{CNOT} are used frequently in quantum error correction, so developing fast, high-fidelity 2-qubit gate mechanisms is imperative for large-scale quantum information processing.  In many cases, the 2-qubit gate is the process which defines the speed and accuracy of a quantum computer.

The construction of a practical, scalable 2-qubit gate in \mbox{QuDOS} remains the most challenging element of the hardware, and various methods are currently under development.  A fast, all-optically controlled 2-qubit gate would certainly be attractive, and early proposals~\cite{Imamoglu99} identified the importance of employing the nonlinearities of cavity QED. Ref.~\cite{Imamoglu99} suggests the application of two lasers for both single-qubit and 2-qubit control; more recent developments have indicated that both single-qubit gates~\cite{Economou2006,Clark07,Press08} and 2-qubit gates~\cite{Ladd2011} can be accomplished using only a single optical pulse.

We consider a 2-qubit gate via the dispersive interaction proposed in Ref.~\cite{Ladd2011}. The critical figure of merit for the cavity QED system is the cooperativity factor $C$, which is proportional to the cavity quality factor $Q$ divided by the cavity volume $V$.  For \mbox{QuDOS}, we envision transverse cavity confinement entirely due to the extended microplanar microcavity arrangement, in which cooperativity factors are enhanced by the angle-dependence of the cavity response, an effect which is enlarged by high index of refraction contrast in the alternating mirrors of the DBR stack~\cite{Bjork94}.  While existing cooperativity factors achieved this way are not estimated to be high enough to produce quantum gates with error rates sufficiently low for fault-tolerant quantum computing, advanced control techniques and multi-spin encodings (such as for ``virtual qubits''; see Section~\ref{virtualization}) may enable this technology to function with acceptable error rates.  Ref.~\cite{Ladd2011} estimates that this gate will require 10--100 ns to execute, and for the present analysis we assume the value 32 ns, which coincides with a virtual gate in Section~\ref{virtual_gate}.  Further enhancements to speed and/or gate fidelity may be available by introducing exchange interactions using microcavity polaritons~\cite{polariton_mediate}; studying this possibility is the subject of future work.

\subsection{Measurement readout}
Measurement is another essential component of quantum computing.  At a bare minimum, one must be able to read the final result of a calculation, but typically measurement is used extensively in fault-tolerant quantum error correction.  For this reason, quantum computers may require measurement which is comparable in speed and accuracy to the control operations.  Moreover, many situations call for \emph{quantum non-demolition} (QND) measurement, where the physical qubit is projected into an eigenstate of the measurement operator.  To illustrate a counter-example, consider a qubit defined by the ground and first optically excited states of a quantum dot.  A possible measurement scheme is to detect a photon emission, which would indicate the qubit was in the excited state.  However, the final state of the qubit is the ground state for either measurement outcome, which is destructive measurement, and this procedure cannot be repeated.  Conversely, QND measurement is highly desirable because it can be repeated, so that classical readout noise can be reduced by time-averaging.

\begin{figure}
  \centering
  \includegraphics[width=8cm]{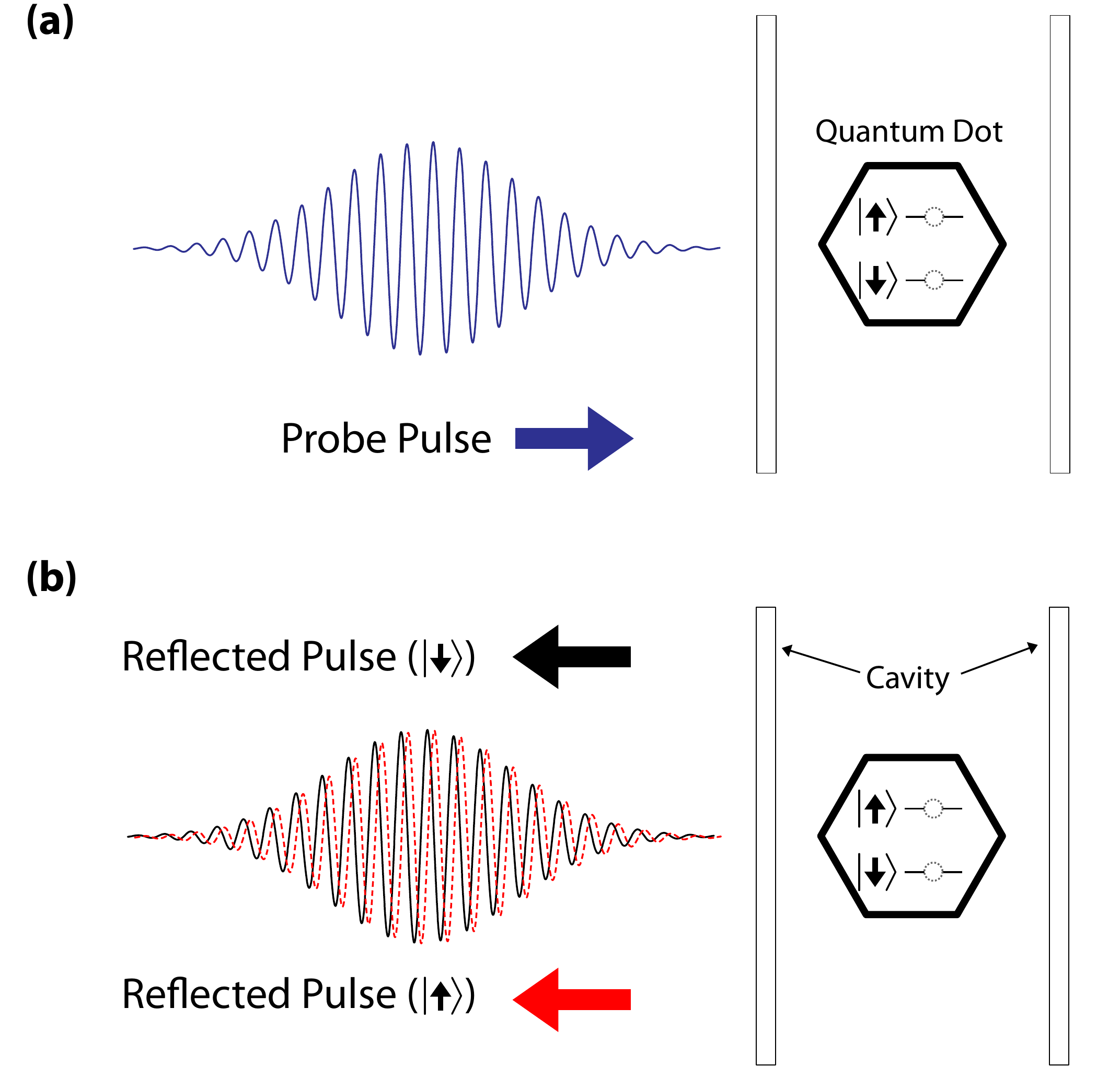}\\
  \caption{Color.  A dispersive quantum non-demolition (QND) readout scheme for \mbox{QuDOS}.  (a)~A probe pulse is sent into a microcavity containing a charged quantum dot.  (b)~The cavity-enhanced dispersive interaction between the pulse and the electron spin creates a state-dependent phase shift in the light which leaves the cavity.  Measurement of the phase shift can perform projective measurement on the electron spin.}
  \label{QND}
\end{figure}

\mbox{QuDOS} will require a QND measurement scheme which is still under experimental development.  The proposed mechanism (shown in Fig.~\ref{QND}) is based on Faraday/Kerr rotation.  The underlying physical principle is as follows: an off-resonant probe pulse impinges on a quantum dot, and it receives a different phase shift depending on whether the quantum dot electron is in the spin-up or spin-down state (these are separated in energy by the external magnetic field).  Sensitive photodetectors combined with homodyne detection measure the phase shift to enact a projective QND measurement on the electron spin. Several results in recent years have demonstrated the promise of this mechanism for measurement: multi-shot experiments by Berezovsky \emph{et al.}~\cite{Berez06} and Atat\"{u}re \emph{et al.}~\cite{Atature07} have measured spin-dependent phase shifts in charged quantum dots, and Fushman \emph{et al.}~\cite{Fushman08} observed a large phase shift induced by a neutral quantum dot in a photonic crystal cavity.  Most recently, Young \emph{et al.} observed a significantly enhanced phase shift from a quantum dot embedded in a micropillar cavity~\cite{Young2011}.

\subsection{Noise sources and errors}
\label{Phys_noise}
Noise and decoherence are the biggest obstacles to scalable quantum computing.  In general, the noise sources which corrupt the physical qubit or degrade the fidelity of control operations should be characterized as well as possible.  For the present analysis, we consider the noisy environment for an electron spin in \mbox{QuDOS}.  The primary noise in this system is dephasing, likely caused by the inhomogeneous distribution of nuclear spins in the quantum dot.  The ensemble dephasing is characterized by $T_2^* \approx$~2~ns, while the intrinsic dephasing is characterized by $T_2 \approx$~3~$\mu$s~\cite{Press10}.  When the noise experienced by a qubit is dominated by dephasing, one can counteract decoherence with control sequences tailored to this noise source~\cite{Aliferis2008}.  Section~\ref{virtual_qubit} introduces a decoupling scheme designed specifically for \mbox{QuDOS}.

\subsection{Hardware performance summary}
\label{Layer1_summary}
We summarize the execution times for the essential Layer 1 operations in \mbox{QuDOS} in Table~\ref{L1_Parameters}.  These are the quantum processes which are the building blocks of quantum information operations in Layers 2 and above.  For a complete quantum processor, however, one would also have to consider the classical control hardware and the engineering concerns, such as delays, which may occur in a large system.  For example, Ref.~\cite{Levy2011} considers the implications of classical control wires, such as routing concerns, signal timing, and the generation of heat in low-temperature devices.  Although engineering of classical control hardware is an important problem, it lies outside the scope of our present analysis, and we reserve it for future work.


\begin{table*}
  \centering
      \begin{tabular}{| m{3.5cm} | m{4.5cm} | >{\centering\arraybackslash}m{1.6cm} | m{4.5cm} |}
      \hline
      \textbf{Operation} & \textbf{Mechanism} & \textbf{Duration} & \textbf{Notes}
      \\ \hline
      \raggedright{Spin phase precession ($\hat{Z}$-axis)\rule{0Ex}{2.5Ex}} & \raggedright{Magnetic field splitting of spin energy levels} & 40 ps & \footnotesize{Inhomogeneous nuclear environment causes spectral broadening in Larmor frequency, which is the source of ${T_{2}}^{*}$ processes.} \\ \hline
      \raggedright{Spin state rotation pulse} & \raggedright{Stimulated Raman transition with broadband optical pulse} & 14 ps & \footnotesize{Red-detuned from spin ground state-trion transitions.} \\ \hline
      \raggedright{Entangling operation} & \raggedright{Nonlinear phase shift of spin states via coupling to a common cavity mode} & 32 ns	& \footnotesize{CW laser signal modulated by an electro-optic modulator (EOM).} \\ \hline
      \raggedright{QND measurement} & \raggedright{Dispersive phase-shift of light reflected from planar cavity} & 1 ns & \footnotesize{CW laser signal modulated by an EOM.} \\ \hline

      \end{tabular}
  \caption{Parameters for Layer 1 quantum operations.  Spin phase precession is determined by the spin-state energy splitting due to an external magnetic field.  To implement a Hadamard gate, the broadband pulse time is $1/\sqrt{8}$ of the Larmor period ($T_{\text{Larmor}}$).  Times for entangling operation and QND measurement are estimated from simulation.}
  \label{L1_Parameters}
\end{table*}


\section{Layer 2: Virtual}
\label{virtualization}
\begin{figure*}
  \centering
  \includegraphics[width=15cm]{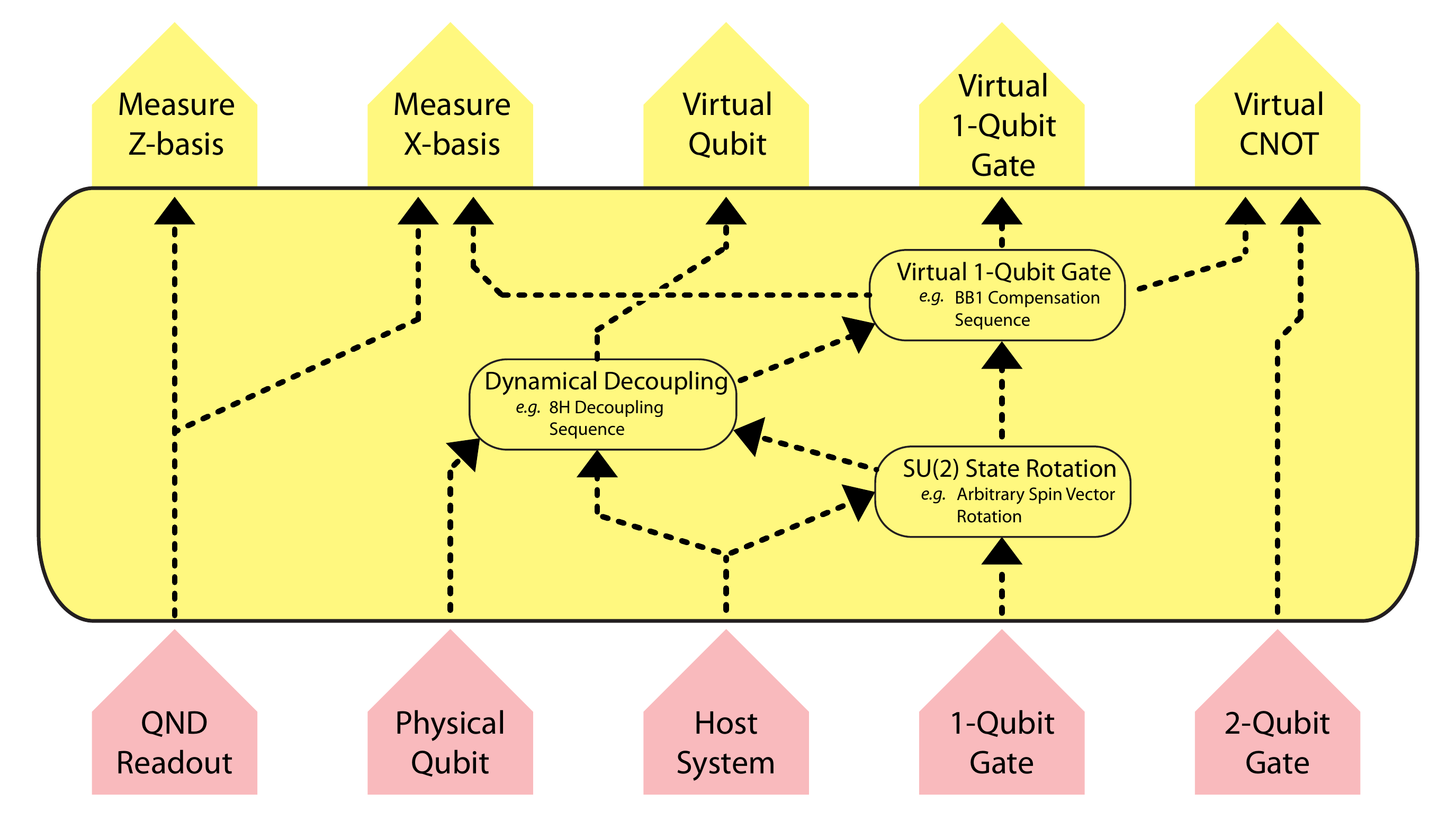}\\
  \caption{Color.  The mechanics of the Virtual layer. The outputs of Layer 1 are combined in controlled sequences to produce virtual qubits and gates.  Arrows indicate how the output of one process is used by another process.}
  \label{L2_ProcessTranslation}
\end{figure*}

The Virtual layer is where quantum effects in the Physical layer are first cast into information primitives --- virtual qubits and quantum gates.  We use ``virtual'' as it is defined in the field of computer science, where a virtual object obeys a pre-determined set of behaviors, without specifying the structure of this object.  As an example, a virtual qubit may be defined by a decoherence-free subspace~\cite{Lidar1998,Lidar03,Grace2006} constructed from three electron spins; when considered as a whole, three spins have many more degrees of freedom than a single qubit.  Similar behavior is seen in the quantum gates in \mbox{QuDOS}, which actually consist of a sequence of laser pulses.  This transcription process of converting many physical elements into a virtual information unit is the task of Layer~2, and we clarify the functions of this layer below.  Fig.~\ref{L2_ProcessTranslation} gives an overview of the Virtual layer processes in \mbox{QuDOS}.


In a general sense, the Virtual layer makes the Physical layer robust to systematic errors.  This effect will be seen in both virtual qubits and gates, where we enforce symmetries in the system (by careful design of control operations) which cause correlated errors to cancel by interference.  The simplest example of this behavior is the Hahn spin-echo sequence~\cite{Hahn50}, and in fact decoupling techniques will play a prominent role in how we construct a virtual qubit.

\subsection{Virtual qubit}
\label{virtual_qubit}
The virtual qubit shapes the underlying physical qubit into a two-level system which approximates an ideal qubit. However, the virtual qubit is modeled as having some finite amount of decoherence, such as the depolarizing channel~\cite{Nielsen00}.  Where applicable, dynamical decoupling~\cite{Viola1999,Khodjasteh2005,Biercuk2009} and/or decoherence-free subspaces~\cite{Lidar1998,Lidar03,Grace2006} are used to create long-lived virtual qubits, and the residual decoherence characterizes the lifetime of the virtual qubit.  In what follows, we consider how to construct a virtual qubit with a charged quantum dot, including the mitigation of several non-ideal effects in this system.

In \mbox{QuDOS}, the virtual qubit is created from the two metastable spin states of an electron confined to a QD.  As discussed in Section~\ref{Phys_noise}, the raw physical system has dephasing time ${T_{2}}^{*} \approx 2$~ns~\cite{Press10} caused by an inhomogeneous distribution of nuclear spins in the environment of the electron. This dephasing time is insufficient for quantum error correction in Layer~3, so this system must be augmented with dynamical decoupling techniques~\cite{Viola03,Ng09}, which extend the dephasing time of the virtual qubit into the microsecond regime~\cite{Press10}.


\begin{figure*}
  \centering
  \includegraphics[width=\textwidth]{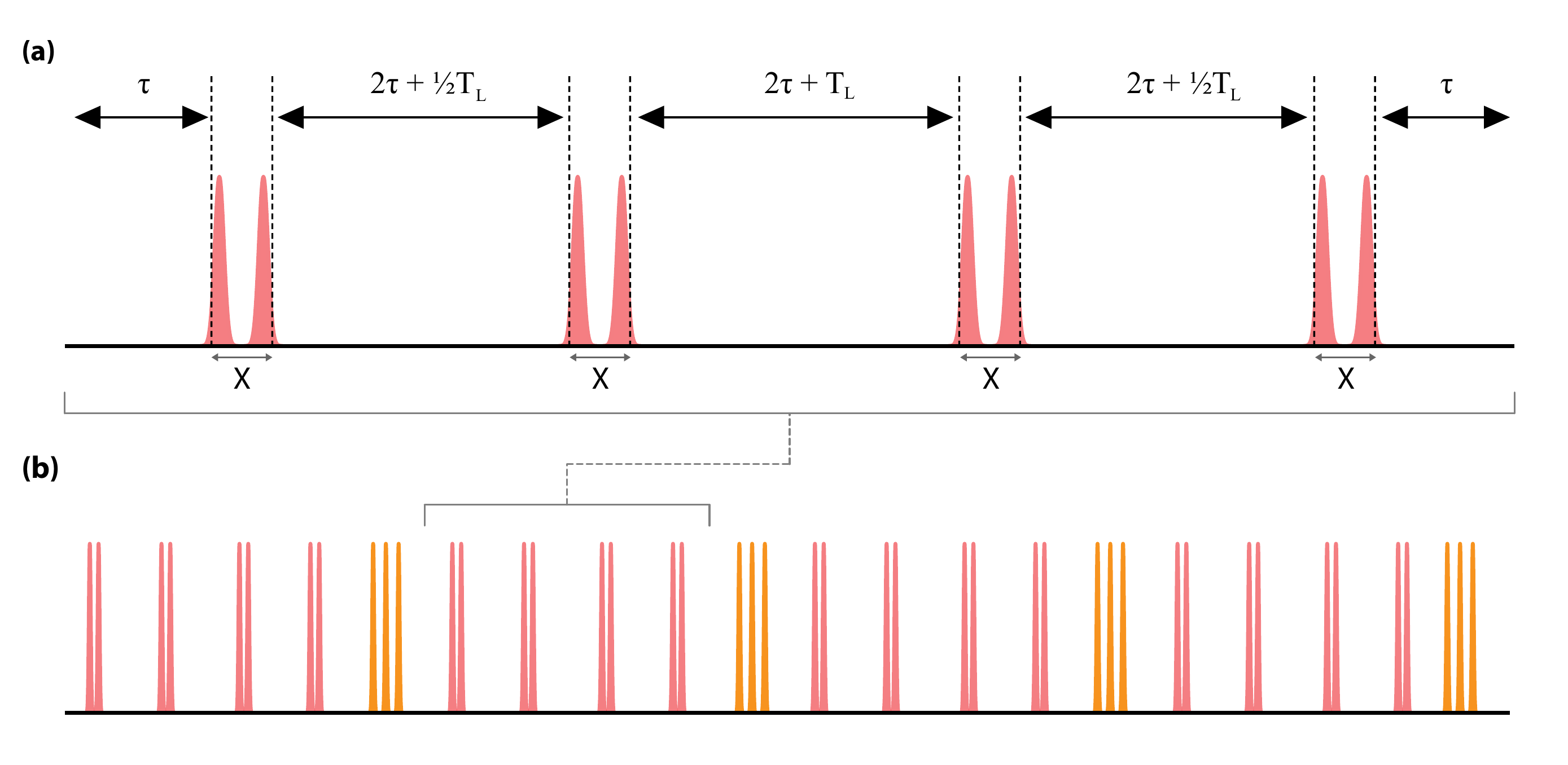}\\
  \caption{Color.  A special dynamical decoupling sequence for \mbox{QuDOS}, known as 8H since it requires eight Hadamard pulses.  $T_L$ is the Larmor period determined by the external magnetic field (see Table~\ref{L1_Parameters}).  (a)~Timing specification for the 8H sequence, where $\tau$ is an arbitrary time.  Each of the pulse pairs enacts a $\pi$-rotation around the $X$-axis of the virtual qubit Bloch sphere, as shown in Fig.~\ref{Hadamard}.  For 8H to work efficiently, $\tau \ll T_2$.  (b)~Four 8H sequences in a row interleaved with arbitrary gates formed from three Hadamard pulses (orange).  The overall sequence forms a virtual gate by way of a BB1 compensation sequence.}
  \label{8H_Sequence}
\end{figure*}


Constructing the virtual qubit in \mbox{QuDOS} requires Layer~2 to conceal the complexity of controlling the QD spin state. Because the physical qubit Bloch vector continuously rotates around the $\hat{Z}$-axis, control pulses must be accurately timed so that they perform the desired operation.  Furthermore, control of the QD spin is complicated by the inhomogeneous nuclear-spin environment which causes the $\hat{Z}$-axis rotation to proceed at a somewhat uncertain angular frequency. This problem is mitigated by a dynamical decoupling (DD) sequence, so that the system is decoupled from environmental noise and brought into a precisely controlled reference frame at a predictable time.  Fig.~\ref{8H_Sequence}a illustrates the ``8H'' decoupling sequence (so named because it uses eight Hadamard pulses), which is appropriate for use in \mbox{QuDOS}.  This control sequence is designed both to decouple a qubit from dephasing noise and to compensate for systematic pulse errors in the presence of a strong but slowly-fluctuating drift term in the qubit Hamiltonian, which is the case for optically-controlled quantum dots in a strong magnetic field.  Although longer sequences consisting of more pulses may in theory decouple to higher fidelity, we have chosen a sequence of just eight Hadamard pulses to minimize execution time.  Instead of using a more common sequence like Carr-Purcell (CP)~\cite{Carr54,Haeberlen68} or Uhrig dynamical decoupling (UDD)~\cite{Uhrig07}, the sequence in Fig.~\ref{8H_Sequence} is custom-designed to eliminate to first-order the errors which occur in both the free evolution and control of the virtual qubit (CP and UDD cannot accomplish the latter).  We note, however, that the 8H sequence does have a structure similar to the CP sequence.


\begin{figure}
  \centering
  \includegraphics[width=8cm]{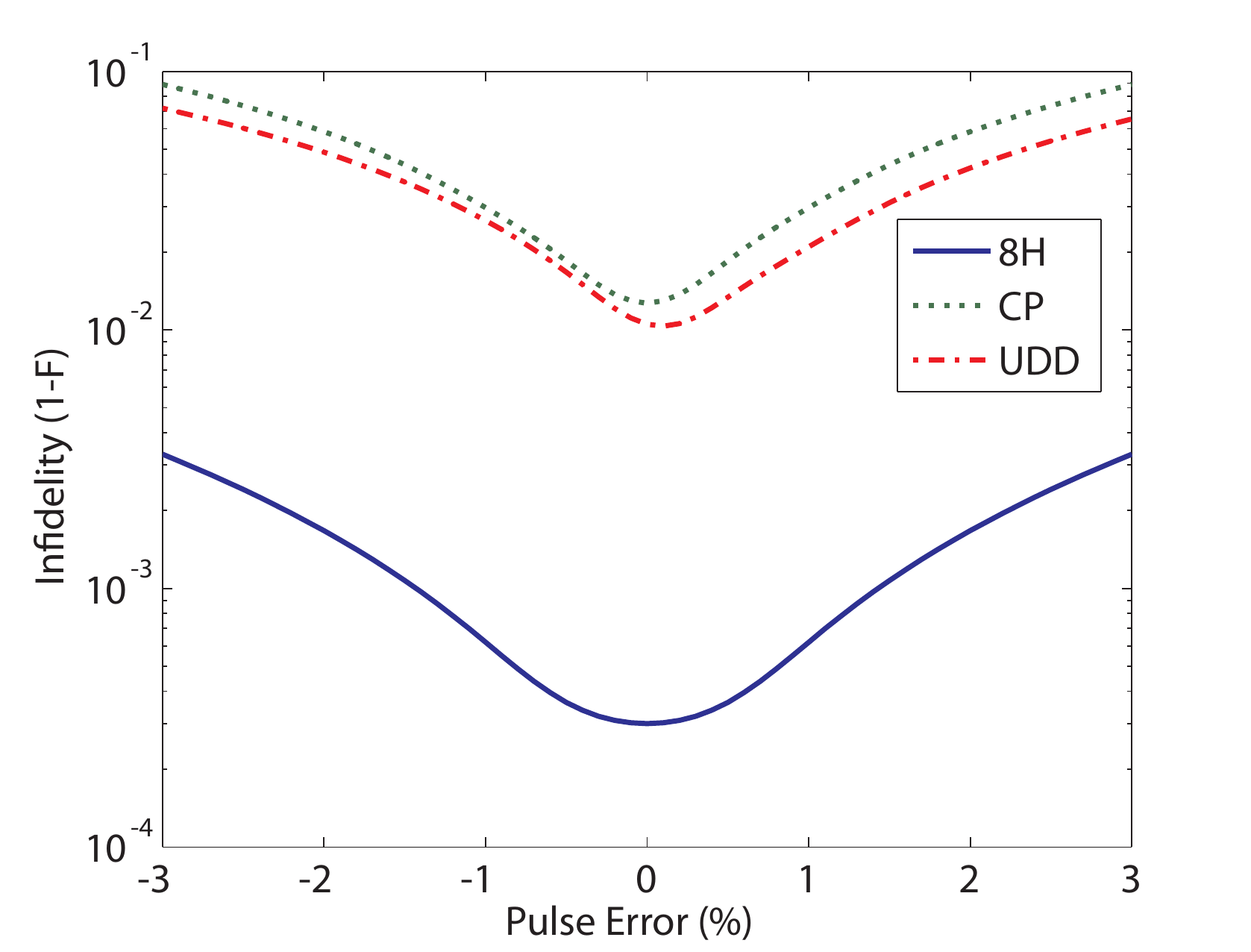}\\
  \caption{Color.  Simulation of the decoupling effectiveness of the 8H sequence compared to CP and UDD (each using 4 \texttt{X} gates) in the presence of dephasing noise and control errors.  Here, ``pulse error'' is a systematic, relative deviation in the energy of every pulse.  In all cases, two Hadamard pulses are combined to produce an approximate \texttt{X} gate, as in Fig.~\ref{Hadamard}.  The vertical axis is infidelity after evolution of the sequence in Fig.~\ref{8H_Sequence}a with $\tau$ = 1 ns; here infidelity is $1-F = 1-\chi_{II}$, where $\chi_{II}$ is the identity-to-identity matrix element in process tomography for the decoupling gate sequence with random noise.  Since we aim to execute virtual gates with $1-F < 10^{-3}$, laser pulse errors must be less than 1\% in order for the virtual qubit memory error rate to be adequately low.}
  \label{VQ_Sim}
\end{figure}


Fig.~\ref{VQ_Sim} shows the simulated effectiveness of 8H as compared to CP and UDD.  We have selected $\tau = 1$ ns (from Fig.~\ref{8H_Sequence}a), so one iteration of the sequence requires 8 ns.  Because this sequence is specifically designed to account for the errors particular to \mbox{QuDOS}, the performance exceeds that of the more common dynamical decoupling schemes.  Nonetheless, 8H may be very effective in other quantum information systems where the physical qubit states are separated in energy \emph{and} the control pulses have a duration which is comparable to the free precession (\emph{e.g.} Larmor) period of the qubit Bloch vector.

\subsection{Virtual gate}
\label{virtual_gate}
Virtual gates manipulate the state of the virtual qubit by combining physical control operations in Layer~1 in a manner which creates destructive interference of control errors. Quantum operations must be implemented by physical hardware which is ultimately faulty to some extent.  Many errors are \emph{systematic}, so that they are correlated in time, even if they are unknown to the quantum computer designer. Virtual gates suppress systematic errors as much as possible in order to satisfy the demands of the error correction system in Layer~3.

Efficient schemes exist for eliminating systematic errors. Compensation sequences can correct correlated errors in the gate operations in Layer~1~\cite{Brown04,Tomita10}. This situation arises often for errors due to imperfections in the control operations, such as laser intensity fluctuations or the coupling strength of a quantum dot electron to an optical field (caused by fabrication imperfections). If these errors are correlated on timescales longer than operations in this architecture, a compensation sequence is effective for generating a virtual gate with lower net error than each of the constituent gates in the sequence. Many compensation sequences are quite general, so that error reduction works without knowledge of the type or magnitude of error.  Dynamically corrected gates are an alternative scheme where one tunes the time-dependent Hamiltonian of the control operations~\cite{Khodjasteh2009}.  Beyond such open-loop control techniques, it is also desirable to characterize the accuracy of operations in the Virtual layer, especially multi-qubit gates and entangled states.  Systematically evaluating quantum operations is an important component of a research program to develop quantum computers and merits further investigation; however, it is beyond our present scope.

In \mbox{QuDOS}, the ultrafast pulses in Layer~1 would ideally induce a state rotation in the spin basis (two-level system), but inevitably the physical system will suffer from some loss of fidelity by both systematic and random processes.  We attempt to cause destructive interference of any systematic errors---both from the environment and control pulses---by embedding a BB1 compensation sequence within a train of 8H dynamical decoupling sequences, as shown in Fig.~\ref{8H_Sequence}b.  This approach is motivated by the properties of the physical qubit.  The electron spin has a strong but slowly fluctuating drift term in its Hamiltonian because of the magnetic field and the nuclear spin environment.  The 8H sequence brings the qubit ``into focus'' (analogous to a ``spin echo'') only at prescribed instants, which are when the BB1 pulses are applied.  This approach is more accurate than a BB1 sequence without refocusing because of the time required to implement rotations on the physical qubit Bloch sphere using Hadamard pulses, for the same reasons that 8H is more effective at decoupling than CP or UDD sequences, as in Figure~\ref{VQ_Sim}.  The BB1 compensation sequence requires four arbitrary gates~\cite{Brown04}; hence the virtual gate with error cancellation requires 32 ns.

\subsection{Measurement of virtual qubits}
Measurement is a crucial operation which must also be applied to the virtual qubit in a manner consistent with other control processes.  For example, dynamical decoupling prevents measurement by isolating a qubit from environment interactions, so DD may have to be suspended during readout.  Since measurement plays a crucial role in error correction, this mechanism should be made as fast and efficient as possible, and a slow measurement process may suffer loss of fidelity if the physical qubit decoheres quickly without DD.

Even if the measurement process is much faster than qubit decoherence, classical noise in the measurement readout signal could be a concern.  If the Physical layer provides QND measurement, then the Virtual layer can repeat the measurement of a virtual qubit multiple times and overcome noise in readout circuitry by a majority poll of discrete measurement outcomes.  This is a simple yet robust way to suppress measurement errors.  For example, if the optical measurement pulse in \mbox{QuDOS} requires 1 ns, then measurement could be repeated about 30 times in the same window of time as a virtual gate.  Another possibility is to couple a virtual qubit to one or more ancilla qubits which facilitate measurement~\cite{Capellaro2005}.  In such a scheme, the measurement process at the Physical layer could be destructive, but since only the ancilla is destroyed, the back-action on the original qubit is QND measurement, which can be repeated.

Measurement of the virtual qubit in \mbox{QuDOS} requires that the DD sequence be halted, because the 8H sequence interferes with readout. Since the measurement pulse is in the $\hat{Z}$-basis, rotations around the $\hat{Z}$-axis from the magnetic environment do not affect the outcome. Neglecting DD during measurement is acceptable because the longitudinal (\(T_{1}\)) relaxation time is very long compared with the measurement pulse duration~\cite{Elzerman04,Kroutvar04}.

\section{Layer 3: Quantum Error Correction}
\label{QEC}
Fault-tolerant quantum error correction (QEC) is essential for large-scale quantum computing.  In Section~\ref{Shor} we analyze an implementation of Shor's factoring algorithm requiring an error-per-gate of order $10^{-15}$, which is simply infeasible on faulty hardware, even using Layer~2 techniques like dynamical decoupling.  The action of error correction on a quantum information system is to pump entropy out in the form of an error syndrome; in the process, new resources---\emph{logical} qubits and gates---are created.  Whereas Layer~2 causes correlated errors to cancel, Layer~3 isolates and removes arbitrary errors, so long as the error rate is below a threshold~\cite{Aharonov1997}.  If this condition is met, QEC can in principle produce arbitrarily low-error logical qubits and gates.  Such complete error suppression is necessary because quantum algorithms in the Application layer assume logical qubits and gates are error-free.

\begin{figure*}
  \centering
  \includegraphics[width=15cm]{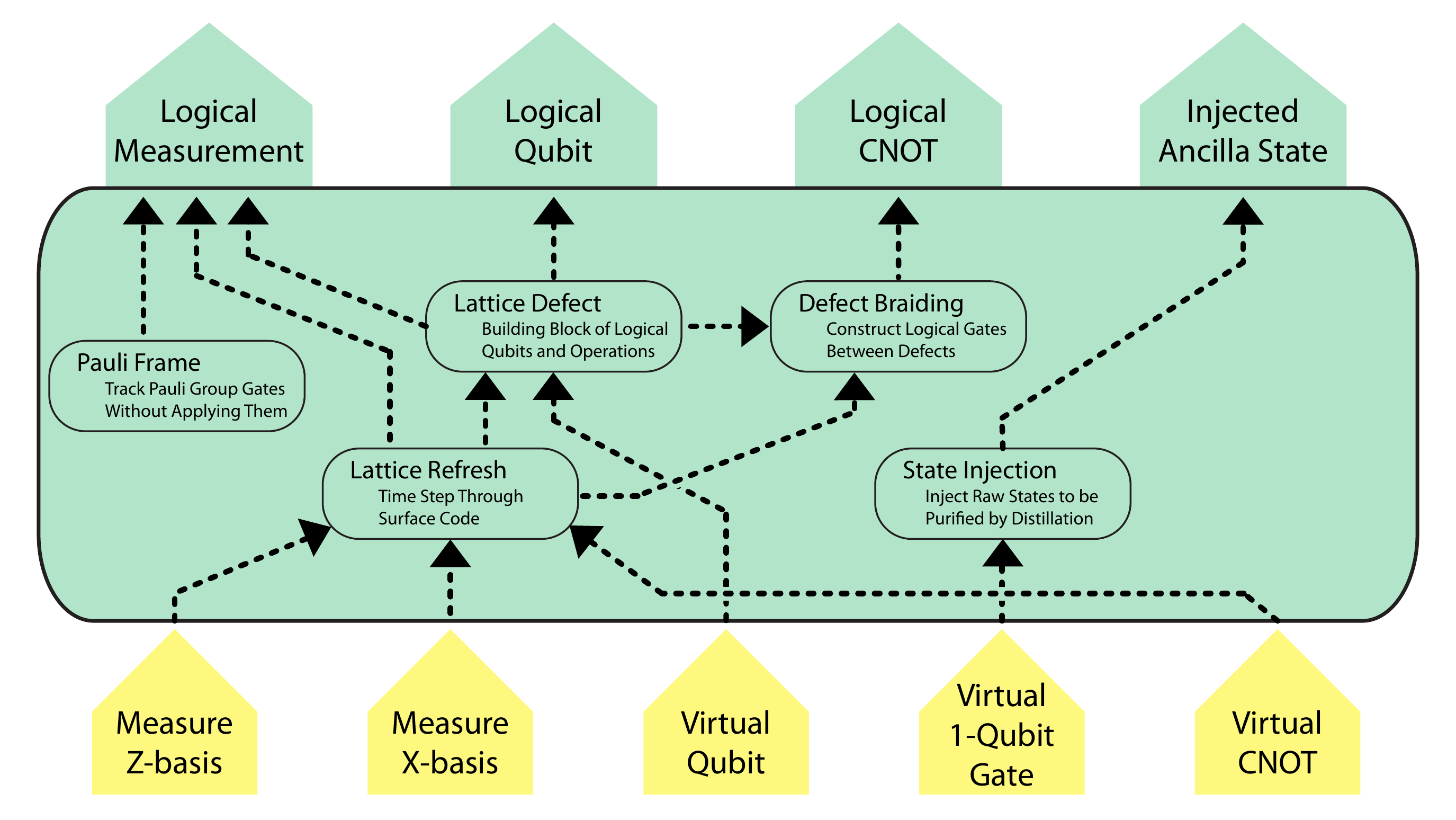}\\
  \caption{Color.  Process translation in Layer~3 in \mbox{QuDOS}.  A surface code is constructed with virtual qubits and gates, ultimately yielding logical qubits and operations.  The arrows in yellow along the bottom are outputs of Layer~2, whereas the green arrows at the top are the outputs of Layer~3.  Small dashed arrows indicate that the output of one process is used by another process.}
  \label{L3_ProcessTranslation}
\end{figure*}

The field of quantum error correction has become too broad to cover in its entirety~\cite{Preskill1998,Nielsen00,Devitt2009}.  Instead, we analyze the case of stabilizer codes~\cite{Gottesman97}, and we specifically consider the surface code~\cite{Fowler09,Rauss07b,Rauss07} for \mbox{QuDOS}.  We select the surface code for its high threshold and two-dimensional nearest-neighbor interaction geometry.  This section focuses on the aspects of quantum error correction that are relevant for a quantum computer architecture, such as determining the size of a code sufficient for a certain application, as well as how the errors are tracked by Pauli frames in classical hardware.  Fig.~\ref{L3_ProcessTranslation} shows the surface code operations in Layer~3 for \mbox{QuDOS}.

Other error-correction schemes besides the surface code could also be implemented in a layered architecture.  As examples, the $C4/C6$ code~\cite{Knill2005,Aliferis2008b} and Bacon-Shor codes~\cite{Bacon2006,Aliferis2007b,Levy2009,Levy2011} have also received significant attention as viable schemes for fault-tolerant quantum computation.  Because the layered architecture is modular, replacing the surface code with another QEC scheme is possible as long as the Virtual layer supports the necessary operations of the new code.  In general, the QEC code chosen is likely to impact many aspects of a quantum computing system, such as device geometry, connectivity, and sensitivity to defective components, so that the structure and behavior of the computer is defined in large part by the selected code.  For this reason, much attention should be devoted to optimizing Layer~3 in any quantum computer architecture.


One key message is worth emphasizing.  The \emph{threshold error rate} of an error-correcting code is defined as the error rate at which error correction begins to show a net gain in protecting information.  A functioning quantum error correction system must operate below threshold, and a \emph{practical} system must operate well below threshold.  We show in this section that the resources required for error correction become manageable when the hardware error rate is about an order of magnitude below the threshold of the chosen code.

\subsection{Estimating the strength of error correction needed}
\label{code_distance}
We consider how to estimate the degree of error correction required for a given application because this determines the necessary amount of resources in the computer.  Quantum error correction schemes generate protected codespaces within a larger Hilbert space formed from many qubits.  The tradeoff for reducing logical errors is that instead of requiring a single qubit, the quantum computer now requires many virtual qubits to produce a logical qubit.  The number of virtual qubits required for a single logical qubit is an important resource-usage quantity, and it depends on the performance aspects of the quantum computer:
\begin{itemize}
\item error per virtual gate ($\varepsilon_{\text{V}}$), which is an input to Layer~3 from Layer~2,
\item threshold error per virtual gate of the error-correcting code ($\varepsilon_{\text{thresh}}$),
\item distance ($d$) of the code,
\item maximum error per logical gate ($\varepsilon_{\text{L}}$), which is upper-bounded by the performance requirements of the quantum algorithm in Layer~5.
\end{itemize}
To determine $\varepsilon_{\text{L}}$, the simplest approach, $KQ$ product, assumes the worst case. If the quantum
algorithm has a circuit with logical depth $K$ acting on $Q$ logical qubits, then the maximum failure probability is given by
\begin{equation}
P_{\text{fail}} = 1 - (1 - \varepsilon_{\text{L}})^{KQ} \approx KQ\varepsilon_{\text{L}}
\end{equation}
for small $\varepsilon_{\text{L}}$.  Therefore, we demand that $\varepsilon_{\text{L}} \ll 1/KQ$.  Given these quantities, the average error per logical gate in a code operating well below threshold may be closely approximated~\cite{Aharonov1997,Preskill1998,Nielsen00,Fowler10,Wang10,Fowler2011b} by
\begin{equation}
\varepsilon_{\text{L}} \approx C_1\left(C_2\frac{\varepsilon_{\mathrm{V}}}{\varepsilon_{\text{thresh}}}\right)^{\lfloor\frac{d+1}{2}\rfloor},
\label{logical_error_eqn}
\end{equation}
where $C_1$ is a constant determined by the specific implementation of the code, $C_2 \sim 1$, and, by assumption, $\varepsilon_{\mathrm{V}} \ll \varepsilon_{\mathrm{thresh}}$.  The data in Ref.~\cite{Fowler2011b} suggests $C_1 \approx 0.13$ and $C_2 \approx 0.61$ for the surface code, which we now use as an example.  Given a known $\varepsilon_{\text{V}}$ and code-specific quantities $\{\varepsilon_{\text{thresh}},C_1,C_2\}$, one can determine the necessary distance $d$ such that the probability of failure of an entire quantum algorithm is sufficiently small.  For comparison, Aliferis presents similar analysis for concatenated codes such as the Bacon-Shor code~\cite{Aliferis2007}.

Equation~(\ref{logical_error_eqn}) illustrates that the error per virtual gate should be $\varepsilon_{\text{V}} < 0.2\varepsilon_{\text{thresh}}$; otherwise, the code distance, and hence size of the quantum computer, will be impractically large.  Table~\ref{QuDOS_L3_Size} provides an example of these calculations for the \mbox{QuDOS} quantum computer.  Error per virtual gate ($\varepsilon_{\text{V}}$) is also assumed, and the $K$ and $Q$ values are for Shor's algorithm factoring a 1024-bit integer (see Section~\ref{Shor}).  We require that $\varepsilon_{\text{L}} \le 10^{-2}/KQ$, so that the logical error probability of the quantum algorithm is less than $1\%$.


\begin{table}
  \centering
  \begin{tabular}{|m{5.3cm}|m{1.4cm}|m{1.4cm}|}
    \hline
    \textbf{Parameter} & \textbf{Symbol} & \textbf{Value} \\ \hline
    Threshold error per virtual gate~\cite{Fowler2011b} \rule{0Ex}{2.5Ex} &  $\varepsilon_{\text{thresh}}$ \rule{0Ex}{2.5Ex} & $9$$\times$$10^{-3}$  \\ \hline
    Error per virtual gate \rule{0Ex}{2.5Ex} & $\varepsilon_{\text{V}}$ \rule{0Ex}{2.5Ex} &  $1$$\times$$10^{-3}$  \\ \hline
    Circuit depth (lattice refresh cycles) \rule{0Ex}{2.5Ex} & $K$ &  $1.6$$\times$$10^{11}$ \\ \hline
    \rule{0Ex}{2.5Ex}Logical qubits (``Shor'', Section~\ref{Shor})  & $Q$ & $72708$ \\ \hline
    Error per lattice refresh cycle \rule{0Ex}{2.5Ex} & $\varepsilon_{\text{L}}$ \rule{0Ex}{2.5Ex} &  $2.6$$\times$$10^{-20}$ \\ \hline
    Surface code distance \rule{0Ex}{2.5Ex} & $d$ & $31$ \\ \hline
    Virtual qubits per logical qubit \rule{0Ex}{2.5Ex} & $n_V/n_L$ & $6240$ \\ \hline

  \end{tabular}
  \caption{Parameters determining the size of the surface code in \mbox{QuDOS} for an implementation of Shor's factoring algorithm.}
  \label{QuDOS_L3_Size}
\end{table}


Determining the necessary strength of error correction also indicates how large the quantum computer is in terms of qubits.  We can estimate the number of virtual qubits per logical qubit, or $n_V/n_L$, by considering the minimum area needed for the two lattice defects, which make up a logical qubit, separated by distance $d$ in the surface code~\cite{Fowler09}.  For a typical set of parameters, as might be required in a large-scale computing application such as Shor's factoring algorithm~\cite{Shor99}, 6240 virtual qubits are needed to construct a logical qubit.  This is a nontrivial overhead, because quantum algorithms require a substantial number of logical qubits, as we discuss in greater detail in subsequent sections.  For example, quantum simulation algorithms may require $\sim$1000--10,000 logical qubits~\cite{Aspuru2005,Kassal2008} and integer factoring may require 100,000 logical qubits or more, depending on the methods of calculating arithmetic~\cite{VanMeter05}; more detail on why so many logical qubits are necessary is given in Section~\ref{distillation}.  Combining the size of quantum computations with the requirements of error correction means that large-scale quantum computing architectures will require millions or billions of virtual qubits (and hence physical qubits).

\subsection{Pauli frames}
\label{Pauli_frames}
A Pauli frame~\cite{Knill2005,DiVincenzo07} is a simple and efficient classical computing technique to track the result of applying a series of Pauli gates (\texttt{X}, \texttt{Y}, or \texttt{Z}) to single qubits.  The Gottesman-Knill Theorem implies that tracking Pauli gates can be done efficiently on a classical computer~\cite{Simon06}.  Many quantum error correction codes, such as the surface code, project the encoded state into a perturbed codeword with erroneous single-qubit Pauli gates applied (relative to states within the codespace).  The syndrome reveals what these Pauli errors are, up to undetectable stabilizers and logical operators, and error correction is achieved by applying those same Pauli gates to the appropriate qubits (since Pauli gates are Hermitian and unitary).  However, quantum gates are faulty, and applying additional gates may introduce more errors into our system.

\begin{figure}
  \centering
  \includegraphics[width=8cm]{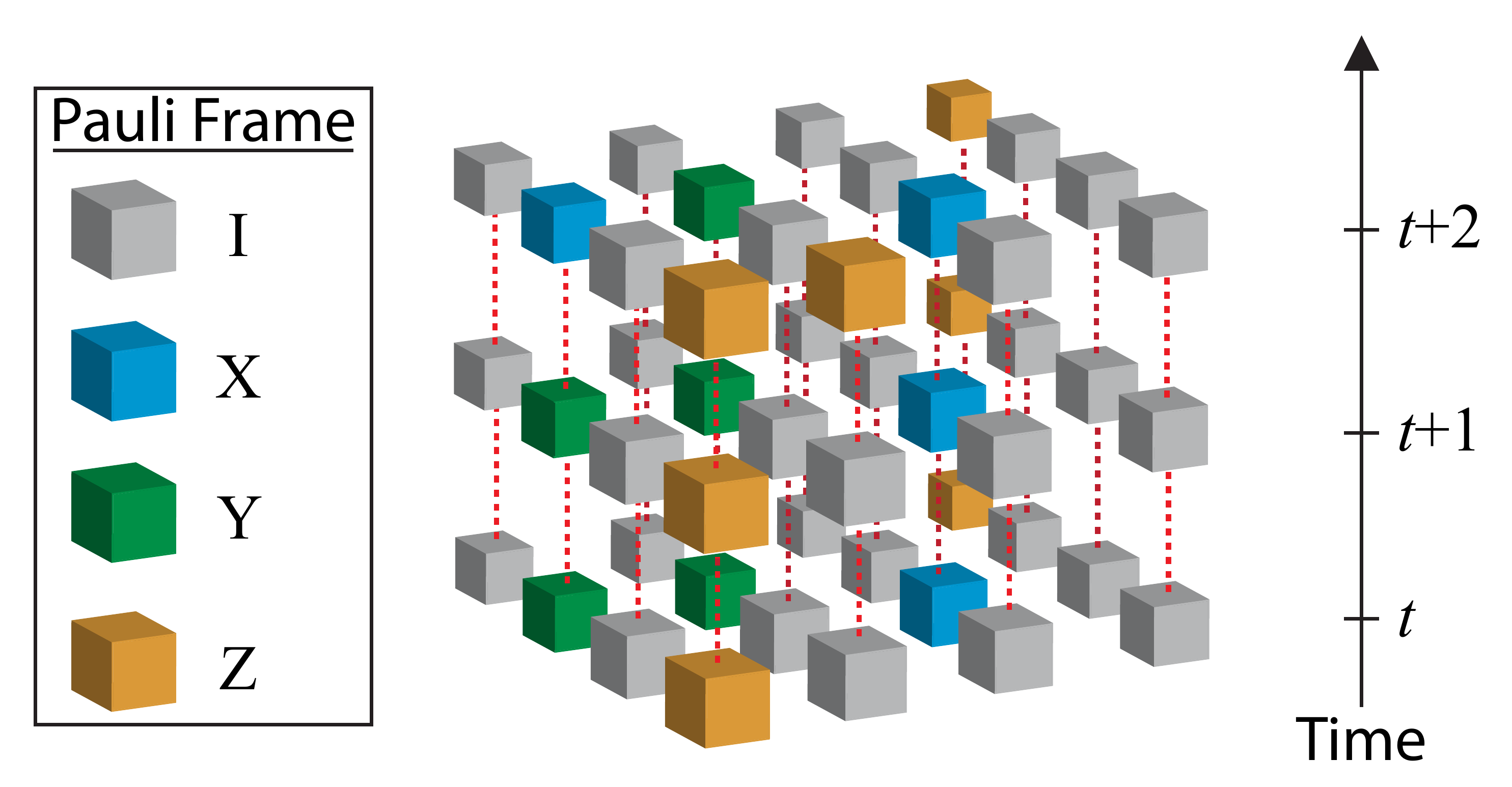}\\
  \caption{Color.  Example of a Pauli frame evolving over time with entries corresponding to virtual qubits forming a surface code.  Each horizontal slice is the Pauli frame at that time.  For example, if the qubit in the top-front-corner position is measured in the \texttt{X} basis, the interpreted result is the negation of the observed outcome, because the Pauli frame \texttt{Z} anticommutes with this measurement basis.}
  \label{PauliFrame}
\end{figure}

Rather than applying every correction operation, one can keep track of what Pauli correction operation \emph{would be applied}, and continue with the computation.  This is possible because the operations needed for error correction are in the Clifford group.  When a measurement in a Pauli \texttt{X}, \texttt{Y}, or \texttt{Z} basis is finally made on a qubit, the result is modified based on the corresponding Pauli gate which should have been applied earlier, as in Fig.~\ref{PauliFrame}.  This stored Pauli gate is called the Pauli frame~\cite{Knill2005,DiVincenzo07}, since instead of applying a Pauli gate, the quantum computer \emph{changes the reference frame for the qubit}, which can be understood by remapping the axes on the Bloch sphere, rather than moving the Bloch vector.

The Pauli frame is maintained as follows.  Denote the Pauli frame at time $t$ as $F_t$:
\begin{equation}
F_t = \bigotimes_j P_t(j),
\end{equation}
where $P_t(j) = \{\texttt{I},\texttt{X},\texttt{Y},\texttt{Z}\}$ is an element from the Pauli group corresponding to qubit $j$ at time $t$.  Any Pauli gate in the quantum circuit is multiplied into the Pauli frame and \emph{is not implemented} in hardware, so $F_{t+1} = \left(\bigotimes_j U_{\{\footnotesize{\texttt{I},\texttt{X},\texttt{Y},\texttt{Z}}\}}\right) F_t$ for all Pauli gates $U_{\{\footnotesize{\texttt{I},\texttt{X},\texttt{Y},\texttt{Z}}\}}$ in the circuit at time $t$.  Other gates $U_{\mathrm{C}}$ in the Clifford group \emph{are implemented}, but they will transform the Pauli frame by
\begin{equation}
\label{Pauli_Frame_Clifford}
F_{t+1} = U_{\mathrm{C}} F_t U_{\mathrm{C}}^{\dag}.
\end{equation}
The quantum computer operations proceed normally, with the only change being how the final measurement of that qubit is interpreted. The set of Clifford gates is sufficient for Layer 3, though the next section describes another Pauli frame for non-Clifford logical operations.


We emphasize that the Pauli frame is a \emph{classical object} stored in the digital circuitry that handles error correction. Pauli frames are nonetheless very important to the functioning of a surface code quantum computer. Layer~3 uses a Pauli frame with an entry for each virtual qubit in the error-correcting code.  As errors occur, the syndrome processing step identifies a most-likely pattern of Pauli errors. Instead of applying the recovery step directly, the Pauli frame is updated in classical memory.  The Pauli gates form a closed group under multiplication (and global phase of the quantum state is unimportant), so the Pauli frame only tracks one of four values (\texttt{X}, \texttt{Y}, \texttt{Z}, or \texttt{I}) for each virtual qubit in the lattice.

\section{Layer 4: Logical}
\label{Logical_layer}
\begin{figure*}
  \centering
  \includegraphics[width=15cm]{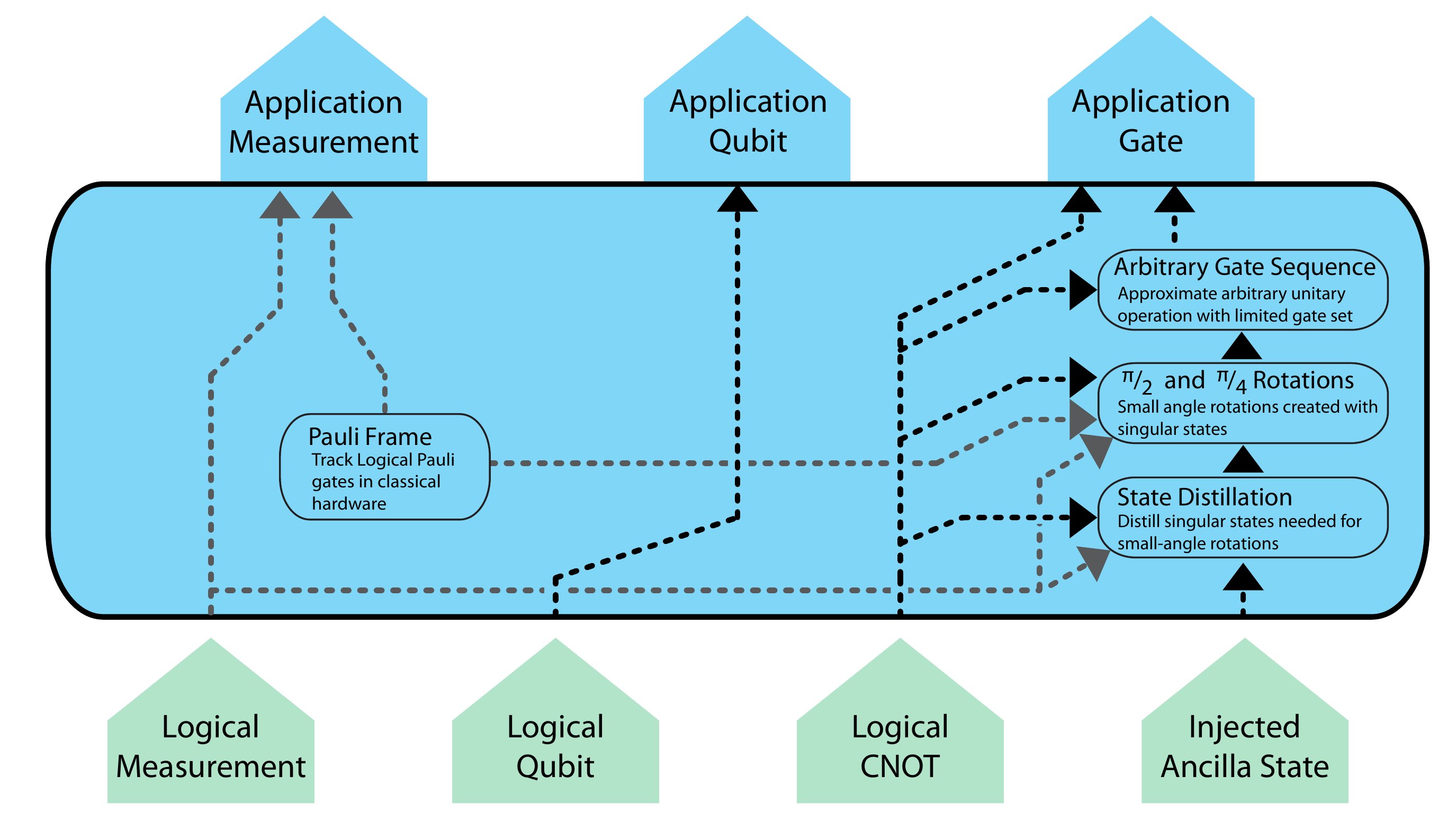}\\
  \caption{Color.  Organization of processes in the Logical layer. Logical qubits from Layer~3 are unaltered, but faulty singular states are distilled into high-fidelity states $\left| Y \right\rangle = \frac{1}{\sqrt{2}}\left(\left|0\right\rangle + i \left|1\right\rangle\right)$ and $\left| A \right\rangle = \frac{1}{\sqrt{2}}\left(\left|0\right\rangle + e^{i\frac{\pi}{4}} \left|1\right\rangle\right)$.  The distilled states are used to create arbitrary gates with specialized quantum circuits~\cite{Dawson05,Kitaev2002,Fowler2011,Jones2012Sim}.}
  \label{L4_ProcessTranslation}
\end{figure*}

The Logical layer takes the fault-tolerant resources from Layer~3 and creates a logical substrate for universal quantum computing.  This task requires additional processing of error-corrected gates and qubits to produce any arbitrary gate required in the Application layer, as shown in Fig.~\ref{L4_ProcessTranslation}.  Quantum error correction provides only a limited set of gates --- to see why, consider that no finite number of syndrome bits can distinguish arbitrarily small rotation gate errors.  A common set of gates provided by QEC is the Clifford group; although circuits from this set can be simulated efficiently on a classical computer by the Gottesman-Knill Theorem~\cite{Nielsen00}, the Clifford group forms the backbone of quantum circuits.  Still, some QEC schemes, such as the surface code, do not provide the full Clifford group without some sort of ancilla.  We identify the set of fault-tolerant gates generated by Layer~3 without the use of ancillas as the \emph{fundamental gates}.  The Logical layer then constructs arbitrary gates from circuits of fundamental gates and ancillas injected into the error-correcting code.  For example, surface code architectures inject and purify the ancillas $\ket{Y} = \frac{1}{\sqrt{2}}\left(\ket{0} + i\ket{1}\right)$ and $\ket{A} = \frac{1}{\sqrt{2}}\left(\ket{0} + e^{i\pi/4}\ket{1}\right)$; then the surface code consumes these ancillas in quantum circuits to produce $\texttt{S}=e^{i(\pi/4) \sigma_Z}$ and $\texttt{T}=e^{i(\pi/8) \sigma_Z}$ gates, respectively~\cite{Nielsen00,Fowler09}.  This section discusses the important functions of the Logical layer: implementing logical Pauli frames; distilling ancilla states like $\ket{Y}$ and $\ket{A}$; implementing the full Clifford group in the surface code without measurement; and approximating arbitrary quantum gates for the Application layer.


\subsection{Fundamental gates and logical Pauli frame}
\label{Fundamental_gates}
Fundamental gates are provided natively by the error-correcting code in Layer~3.  For example, Table~\ref{QuDOS_fundamental_gates} shows the fundamental gates used in \mbox{QuDOS}.  In practice, Pauli gates are implemented with a \emph{logical} Pauli frame, which is qualitatively the same as the Pauli frame in Layer~3 for virtual qubits (Section~\ref{Pauli_frames}).  However, in Layer 4 we may also need to apply gates $U_{\mathrm{NC}}$ outside the Clifford group.  The gate we actually implement, $U_{\mathrm{NC}}'$, results from a Pauli frame transform:
\begin{equation}
U_{\mathrm{NC}}' = F_t U_{\mathrm{NC}} F_t^{\dag}.
\end{equation}
Note the distinction between this expression and Eq.(\ref{Pauli_Frame_Clifford}), where the Pauli frame is changed by a Clifford gate.

The fundamental gate set in Table~\ref{QuDOS_fundamental_gates} is particular to the surface code, and it is not the full Clifford group because it is missing the phase gate \texttt{S}; Section~\ref{Phase_gate} illustrates a method to construct \texttt{S} using an ancilla, without measurement, which is efficient since it uses a small number of fundamental gates and the ancilla can be re-used.  The remaining logical gates to produce a universal set in the surface code will require ancilla states which are injected and distilled~\cite{Fowler09}.

\begin{table}
  \centering
  \begin{tabular}{|m{2.4cm}|m{3cm}|m{2.8cm}|}
    \hline
    \textbf{Gate} & \textbf{Implementation} & \textbf{Execution Time} \newline \footnotesize{(Lattice Steps)} \\ \hline
     \texttt{X}, \texttt{Y}, \texttt{Z} & Pauli frame & Instantaneous \\ \hline
     \texttt{CNOT} & Defect braiding & $13 \lceil d/4 \rceil$ \\ \hline
     \texttt{H} (Hadamard) & Shift lattice & $13 \lceil d/8 \rceil$ \\ \hline
     \texttt{MX}, \texttt{MZ} \newline (Measurement) & Measure stabilizers & 1 \\ \hline

  \end{tabular}
  \caption{Fundamental gates in \mbox{QuDOS} using surface code QEC (Ref.~\cite{Fowler09}).  The execution time here is one possible implementation, but in many cases the surface code computation can be deformed into other topologically equivalent circuits which yield faster execution at the expense of more spatial resources, or vice versa.}
  \label{QuDOS_fundamental_gates}
\end{table}

\subsection{Magic state distillation}
\label{distillation}
The conventional method for making a universal set of quantum gates in a \emph{fault-tolerant} manner is to produce a certain ancilla state and use it in a quantum circuit equivalent to the desired logical gate~\cite{Preskill97,Nielsen00,Aliferis2007}.  In some cases these circuits require measurement that consumes the ancilla, so that the number of ancilla states required is proportional to the number of gates in the quantum algorithm.  For example, the algorithms discussed in Section~\ref{application_layer} require around $10^{12}$ or more ancilla states, which are typically manufactured on an as-needed basis.

To complicate matters, ancillas must be produced by methods that are not fault-tolerant, such as initializing a virtual qubit and applying the appropriate virtual gate.  This ancilla state can then be \emph{injected} into a QEC code in Layer~3~\cite{Fowler09}, but it carries with it the errors in its production.  Fortunately, a few ``magic states'' can be distilled by using several low-fidelity ancillas and fundamental gates to produce one high-fidelity ancilla.  Once the ancilla fidelity is higher than the necessary logical gate fidelity, we may construct arbitrary fault-tolerant logical gates.  We examine here the resource costs for this process; each distillation is expensive, and very many ancillas must be distilled.  We characterize the performance of the magic state distillation because it will probably dominate the resource costs of any quantum computer that uses it.  Accordingly, this is an important area for future optimizations.

We focus first on distilling the ancilla state $\ket{A} = \frac{1}{\sqrt{2}}\left(\ket{0} + e^{i\frac{\pi}{4}} \ket{1}\right)$, which is used to construct the \texttt{T} or $\pi/8$ phase gate~\cite{Bravyi2005,Rauss07,Fowler09}.  In the next section, Fig.~\ref{Toffoli_decomposition} provides an illustration of why this process is important by showing the fault-tolerant construction of a Toffoli gate in Layer~5 using resources in Layer~4; specifically, ancilla distillation circuits constitute over 90\% of the computing effort for a single Toffoli gate.  As a result, the analysis in Appendix~\ref{Shor_details} contends that these distillation circuits account for the majority of resources in a surface code quantum computer executing Shor's algorithm.  In particular, for every qubit used by the algorithm, approximately 10 qubits are working in the background to generate the necessary distilled ancillas.  The ancilla distillation circuit in Fig.~\ref{Toffoli_decomposition} shows one level of $\ket{A}$ distillation, but a lengthy program like Shor's will typically require two levels (one concatenated on another).  Moreover, since perhaps trillions of distilled $\ket{A}$ ancillas will be needed for the algorithm, we create a ``distillation factory''~\cite{Steane1998,VanMeter09}, which is a dedicated region of the computer that continually produces these states as fast as possible.  Speed is important, because ancilla distillation can be the rate-limiting step in quantum circuits~\cite{Isailovic08}.

\begin{table}
  \centering
  \begin{tabular}{|m{3cm}|m{2.4cm}|m{2.8cm}|}
    \hline
    \textbf{Parameter}\rule{0Ex}{2.7Ex} & \textbf{Symbol}\rule{0Ex}{2.7Ex} & \textbf{Value}\rule{0Ex}{2.7Ex} \\ \hline
    Circuit depth\rule{0Ex}{2.7Ex}      & -\rule{0Ex}{2.7Ex}                    & 6 clock cycles\rule{0Ex}{2.7Ex} \\ \hline
    Circuit area\rule{0Ex}{2.7Ex}       & $A_{\text{distill}}$\rule{0Ex}{2.7Ex} & 12 logical qubits\rule{0Ex}{2.7Ex}\\ \hline
    Circuit volume\rule{0Ex}{2.7Ex}     & $V\left(\ket{A^{(1)}}\right)$\rule{0Ex}{2.5Ex} & 72 qubits$\times$cycles\rule{0Ex}{2.7Ex} \\ \hline
    Factory rate (level $n$) & $R_{\text{factory}}\left(\ket{A^{(n)}}\right)$ & $A_{\text{factory}}/V\left(\ket{A^{(n)}}\right)$\rule{0Ex}{2.7Ex} \newline ancillas/cycle \\ \hline

  \end{tabular}
  \caption{Resource analysis for a distillation factory.  These factories are crucial to quantum computers which require ancillas for universal gates.  Magic state distillation uses Clifford gates and measurement, so the circuit can be deformed to reduce depth and increase area, or vice versa, while keeping volume approximately constant.}
  \label{distillation_factory}
\end{table}

Each $\ket{A}$ distillation circuit will require 15 lower-level $\ket{A}$ states, but they are not all used at the same time.  For simplicity we will use a clock cycle for each gate equal to the time to implement a logical \texttt{CNOT} (see Section~\ref{Timing} for more on  this point), so that with initialization and measurement, the distillation circuit requires 6 cycles.  By only using $\ket{A}$ ancillas when they are needed, the circuit can be compacted to require at most 12 logical qubits at a given instant.  We characterize the computing effort by a ``circuit volume,'' which is the product of logical memory space (\emph{i.e.} area of the computer) and time. The circuit volume of $\ket{A}$ distillation is $V\left(\ket{A^{(1)}}\right) = (12 \textrm{logical qubits}) \times (6 \textrm{ clock cycles}) = 72$.  A two-level distillation will require 16 distillation circuits, or a circuit volume of $V\left(\ket{A^{(2)}}\right) = 1152$.  An efficient distillation factory with area $A_{\text{factory}}$ will produce on average $A_{\text{factory}}/V\left(\ket{A^{(2)}}\right)$ distilled ancillas per clock cycle.  Analysis of this problem in the context of Shor's algorithm is given in Appendix~\ref{Shor_details}, and Table~\ref{distillation_factory} lists a summary of these results.

\subsection{Logical phase gate without measurement}
\label{Phase_gate}
The \texttt{S} gate, or phase gate, is the final component of the Clifford group absent from the fundamental set of fault-tolerant gates in Section~\ref{Fundamental_gates}.  Previous implementations of the surface code presented a method for creating this gate by consuming a distilled $\left|Y\right\rangle$ state in a projective measurement-based circuit~\cite{Rauss07,Fowler09}.  However, this approach forces the quantum computer to distill a high-fidelity $\left|Y\right\rangle$ ancilla for each \texttt{S} gate, which can be very costly in both fundamental gates and qubits.

\begin{figure}
  \centering
  \includegraphics[width=7cm]{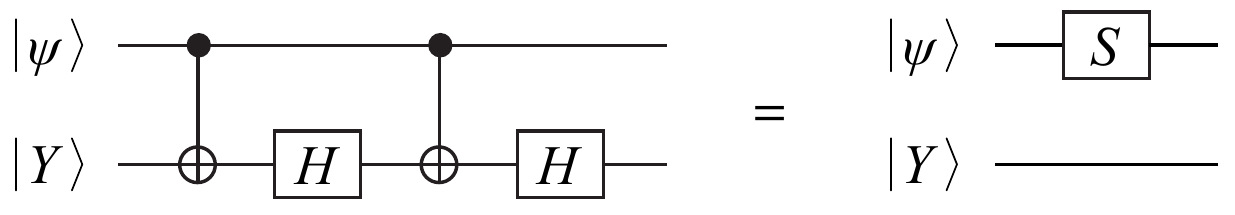}\\
  \caption{Circuit decomposition for a logical \texttt{S} gate which uses an ancilla $\ket{Y}$, but does not consume it. The inverse operation $\texttt{S}^{\dag}$ can be created by running this circuit in reverse.}\label{Sgate_circuit}
\end{figure}

We consider an alternative method that uses the $\left|Y\right\rangle$ ancilla without consuming it to make the \texttt{S} gate, which was originally presented in Ref.~\cite{Aliferis2007}.  The circuit uses only four fundamental gates and, unlike the previous technique, is deterministic because measurement is not needed.  Since $\left|Y\right\rangle$ ancillas are not consumed, one can distill a handful of such states when a quantum computer is turned on, then preserve them for later use.  The circuit in Fig.~\ref{Sgate_circuit} is equivalent to a simple \texttt{S} gate on the control qubit.  This is because $\ket{Y}$ is the $+i$ eigenstate of the operator $iY$, and so the controlled-$iY$ gate will impart a phase $+i$ only if the control qubit is in the state $\ket{1}$, which is identical to the \texttt{S} gate.  Note also that $\texttt{S}^{\dag}$ can be created by running the circuit in Fig.~\ref{Sgate_circuit} backwards.  This technique allows one to implement the entire Clifford group without measurement in the surface code.  Moreover, since \texttt{S} gates are used frequently in quantum algorithms, this improved gate construction substantially reduces the complexity of a quantum computer since fewer of the resource-intensive state distillations are necessary.

\subsection{Approximating arbitrary logical gates}
\begin{figure}
  \centering
  \includegraphics[width=8cm]{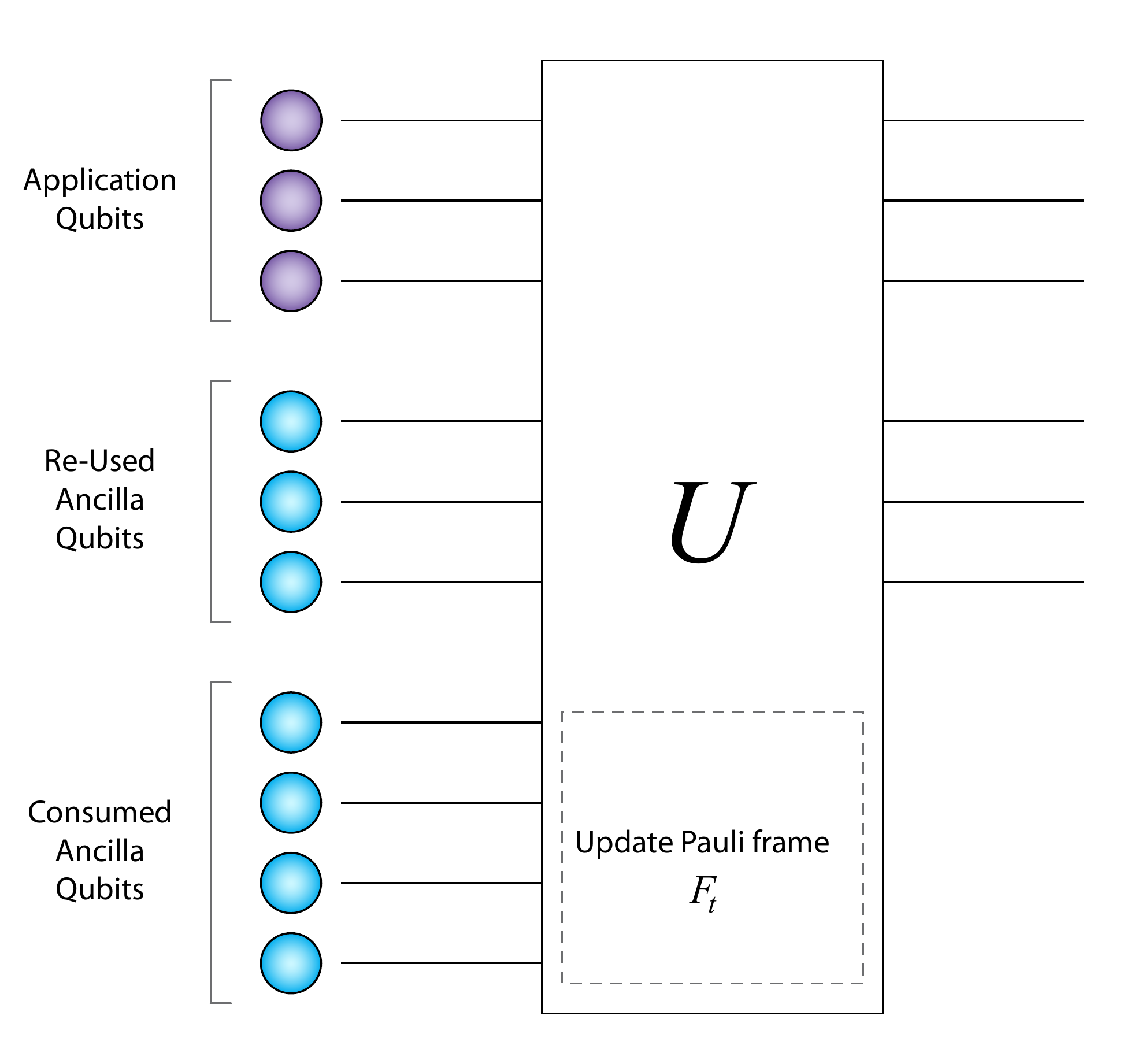}\\
  \caption{Color.  Constructing an arbitrary gate in the Logical layer.  Application qubits are visible to the quantum algorithm, while logical ancilla qubits facilitate universal quantum computation.  Some ancillas are re-used, such as $\ket{Y}$ for \texttt{S} gates, while other ancillas are consumed, such as $\ket{A}$ for \texttt{T} gates.  Often when an ancilla is consumed in a circuit that uses measurement, the circuit is probabilistic, and the Pauli frame is updated conditional on the measurement result.}
  \label{Logical_layer_gate}
\end{figure}

The primary function of the Logical layer is to decompose arbitrary unitary gates from the quantum algorithm into circuits containing fundamental gates available from the QEC layer.  The circuits in the Logical layer act on application qubits (used explicitly by the quantum algorithm) and ancilla logical qubits which facilitate universal quantum computation, as shown in Fig.~\ref{Logical_layer_gate}.  Since arbitrary quantum gates are not available directly, they must be approximated in some fashion, where the total resources required is a function of the approximation accuracy. We cover briefly some of the methods which can be employed to produce such arbitrary gates; a more comprehensive survey of techniques is given in Ref.~\cite{Jones2012Sim}.

When constructing approximations to a unitary operation in the Logical layer, one seeks to implement a quantum circuit that approximates the desired unitary with a minimal overhead in terms of gates and ancillas produced by Layer 3.  We denote approximation accuracy as
\begin{equation}
\varepsilon_{\mathrm{approx}} = \sqrt{\frac{d - \left| \mathrm{tr}(U^{\dag}U_{\mathrm{approx}})\right|}{d}},
\end{equation}
where $U_{\mathrm{approx}}$ is the fault-tolerant quantum circuit that approximates the desired unitary $U$, and $d$ is the dimensionality of these operators~\cite{Fowler2011}.  Several techniques exist for approximating arbitrary single-qubit gates, which can be generalized to arbitrary multi-qubit gates:
\begin{itemize}
\item Gate approximation sequences, such as those produced by the Solovay-Kitaev algorithm~\cite{Nielsen00,Dawson05} or Fowler's algorithm~\cite{Fowler2011}, generate a sequence of gates from the fault-tolerant set (\emph{e.g.} the set \{\texttt{X},\texttt{Y},\texttt{Z},\texttt{H},\texttt{S},\texttt{T}\}) that approximates the desired unitary $U$.  The depth of these sequences scales as $O(\log^c (\varepsilon_{\mathrm{approx}}))$ with $c \approx 4$ for Solovay-Kitaev sequences and $O(\log (\varepsilon_{\mathrm{approx}}))$ for Fowler sequences.
\item Phase kickback uses a special ancilla register and a quantum adder to produce fault-tolerant phase rotations~\cite{Kitaev1995,Cleve1998,Kitaev2002}.  The depth of phase kickback circuits is $O(\log (\varepsilon_{\mathrm{approx}}))$ or $O(\log \log (\varepsilon_{\mathrm{approx}}))$ depending on the quantum adder~\cite{Vedral1996,Cuccaro2008,Draper2006}.  The ancilla register, which is not consumed and can be reused, has size $m$ qubits to approximate a phase rotation to precision $\frac{\pi}{2^m}$ radians, which is also $O(\log (\varepsilon_{\mathrm{approx}}))$.
\item ``Teleportation gates''~\cite{Gottesman1999} can yield very fast quantum circuits, but typically a special purpose ancilla required for each such gate must be computed in advance, which demands a larger and more complex quantum computer; teleportation gates that increase performance in large-scale quantum computing are used extensively in the architecture of Ref.~\cite{Isailovic08} and in the simulation algorithms analyzed in Ref.~\cite{Jones2012Sim}.
\end{itemize}
Choosing among these methods depends on the capabilities of the quantum architecture, such as available logical qubits for parallel computation, and on the desired performance characteristics of the computer.

\section{Layer 5: Application}
\label{application_layer}
The Application layer is where quantum algorithms are executed.  The efforts of Layers~1 through 4 have produced a computing substrate that supplies any arbitrary gate needed.  The Application layer is therefore not concerned with the implementation details of the quantum computer---it is an ideal quantum programming environment.  We do not introduce any new algorithmic methods here, but rather we are interested in how to accurately estimate the quantum computing resources required for a target application.  This analysis can indicate the feasibility of a proposed quantum computer design, which is a worthwhile consideration when evaluating the long-term prospects of a quantum computing research program.

A quantum engineer could start here in Layer~5 with a specific application in mind and work down the layers to determine the system design necessary to achieve desired functionality.  We take this approach for \mbox{QuDOS} by examining two interesting quantum algorithms: Shor's factoring algorithm and simulation of quantum chemistry.  A rigorous system design is beyond the scope of the present work, but we consider the computing resources required for each application in sufficient detail that one may gauge the engineering effort necessary to design a quantum computer based on \mbox{QuDOS} technology.

\subsection{Elements of the Application Layer}
The Application layer is composed of \emph{application} qubits and gates that act on the qubits.  Application qubits are logical qubits used explicitly by a quantum algorithm (see Fig.~\ref{Logical_layer_gate}); as discussed in Section~\ref{Logical_layer}, many logical qubits are also used to distill ancilla states necessary to produce a universal set of gates, but these distillation logical qubits are not visible to the algorithm in Layer 5.  When an analysis of a quantum algorithm quotes a number of qubits without reference to fault-tolerant error correction, often this means the number of application qubits~\cite{Beauregard2003,Aspuru2005,Zalka06,Takahashi2006}.  Similarly, Application-layer gates are equivalent in most respects to logical gates; the distinction is made according to what resources are visible to the algorithm or deliberately hidden in the machinery of the Logical layer, which affords some discretion to the computer designer.

\begin{figure*}
  \centering
  \includegraphics[width=\textwidth]{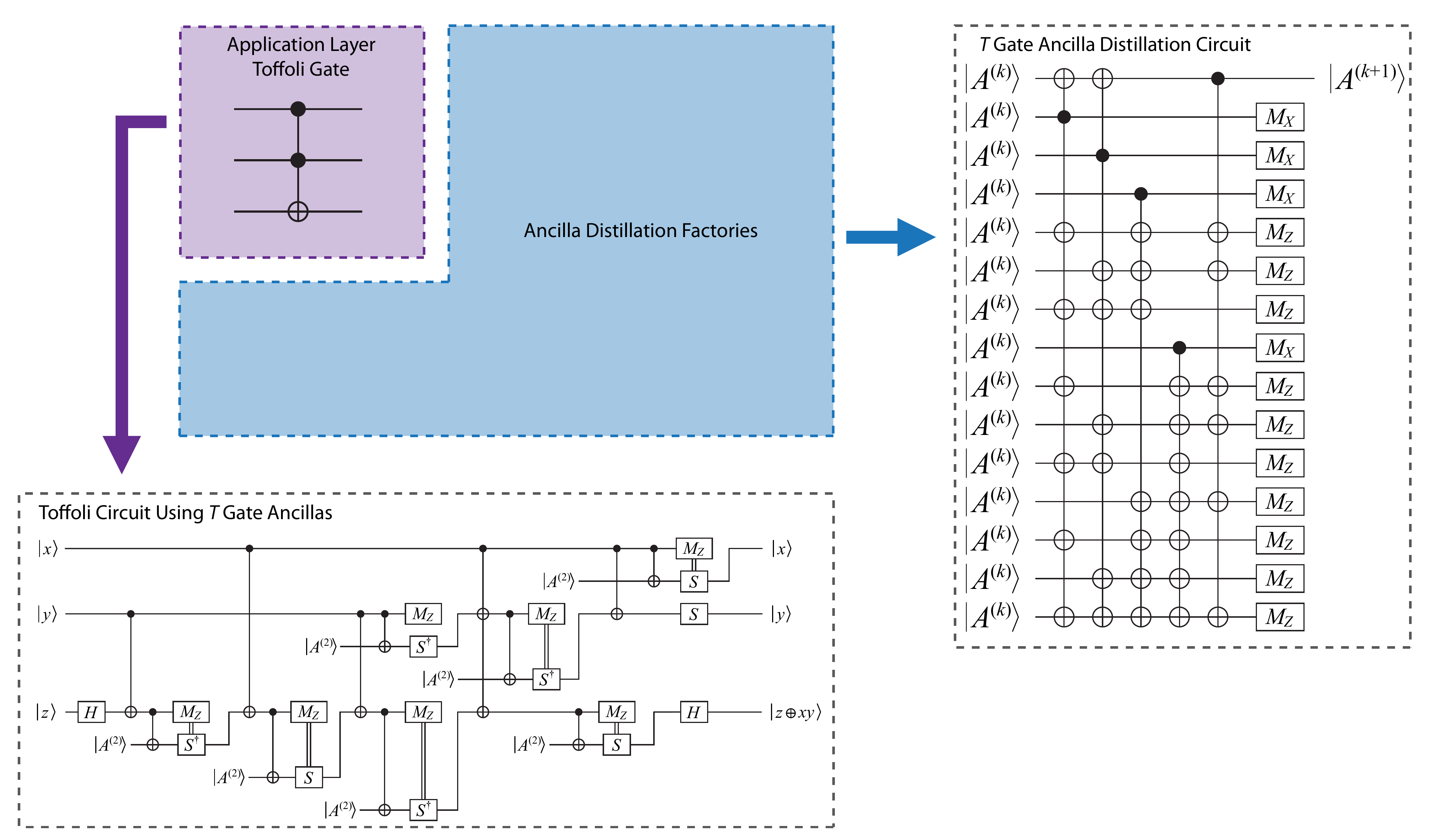}\\
  \caption{Color.  A Toffoli gate ($\ket{x,y,z} \rightarrow \ket{x,y,z \oplus xy}$) at the Application layer is constructed with assistance from the Logical layer, using the decomposition in Ref.~\cite{Nielsen00}.  There are only three application qubits, but substantially more logical qubits are needed for distillation circuits in Layer 4.  The $|A^{(2)}\rangle$ ancillas are the result of two levels of distillation ($|A^{(0)}\rangle$ is an injected state) on the ancilla required for \texttt{T}~gates.  Note that each time an ancilla is used with measurement, the Pauli frame may need to be updated.  The ancilla-based circuit for \texttt{S}~gates (see Fig.~\ref{Sgate_circuit}) is not shown here, for clarity.}
  \label{Toffoli_decomposition}
\end{figure*}

A quantum algorithm could request any arbitrary gate in Layer 5, but not all quantum gates are equal in terms of resource costs.  We saw in Section~\ref{distillation} that distilling $\ket{A}$ ancillas, which are needed for \texttt{T} gates, is a very expensive process.  For example, Fig.~\ref{Toffoli_decomposition} shows how Layers~4 and~5 coordinate to produce an Application-layer Toffoli gate, illustrating the extent to which ancilla distillation consumes resources in the computer.  When ancilla preparation is included, \texttt{T}~gates can account for over 90\% of the circuit complexity in a fault-tolerant quantum algorithm (\emph{cf.} Ref.~\cite{Isailovic08} as well).  For this reason, we count resources for applications in terms of Toffoli gates.  This is a natural choice, because the level of ancilla distillation, number of virtual qubits, \emph{etc.} depend on the choice of hardware, error correction, and many other design-specific parameters; by comparison, number of Toffoli gates is machine-independent since this quantity depends only on the algorithm (much like the number of application qubits mentioned above).  To determine error correction or hardware resources for a given algorithm, one can take the Layer~5 resource estimates and work down through Layers~4 to 1, which is an example of modularity in this architecture framework.  Using the analysis in the preceding sections, an Application-layer Toffoli gate in \mbox{QuDOS} has an execution time of 930 $\mu$s (31 logical gate cycles including the \texttt{S}~gate circuits, discussed in Section~\ref{Timing}).

\subsection{Shor's algorithm}
\label{Shor}
Perhaps the most well-known application of quantum computers is Shor's algorithm, which decomposes an integer into its prime factors~\cite{Shor99}.  Solving the factoring problem efficiently would compromise the RSA cryptosystem~\cite{RSA1978}.  Because of the prominence of Shor's algorithm in the field of large-scale, fault-tolerant quantum computing, we analyze the resources required to factor a number of size typical for RSA.


\begin{figure}
  \centering
  \includegraphics[width=8cm]{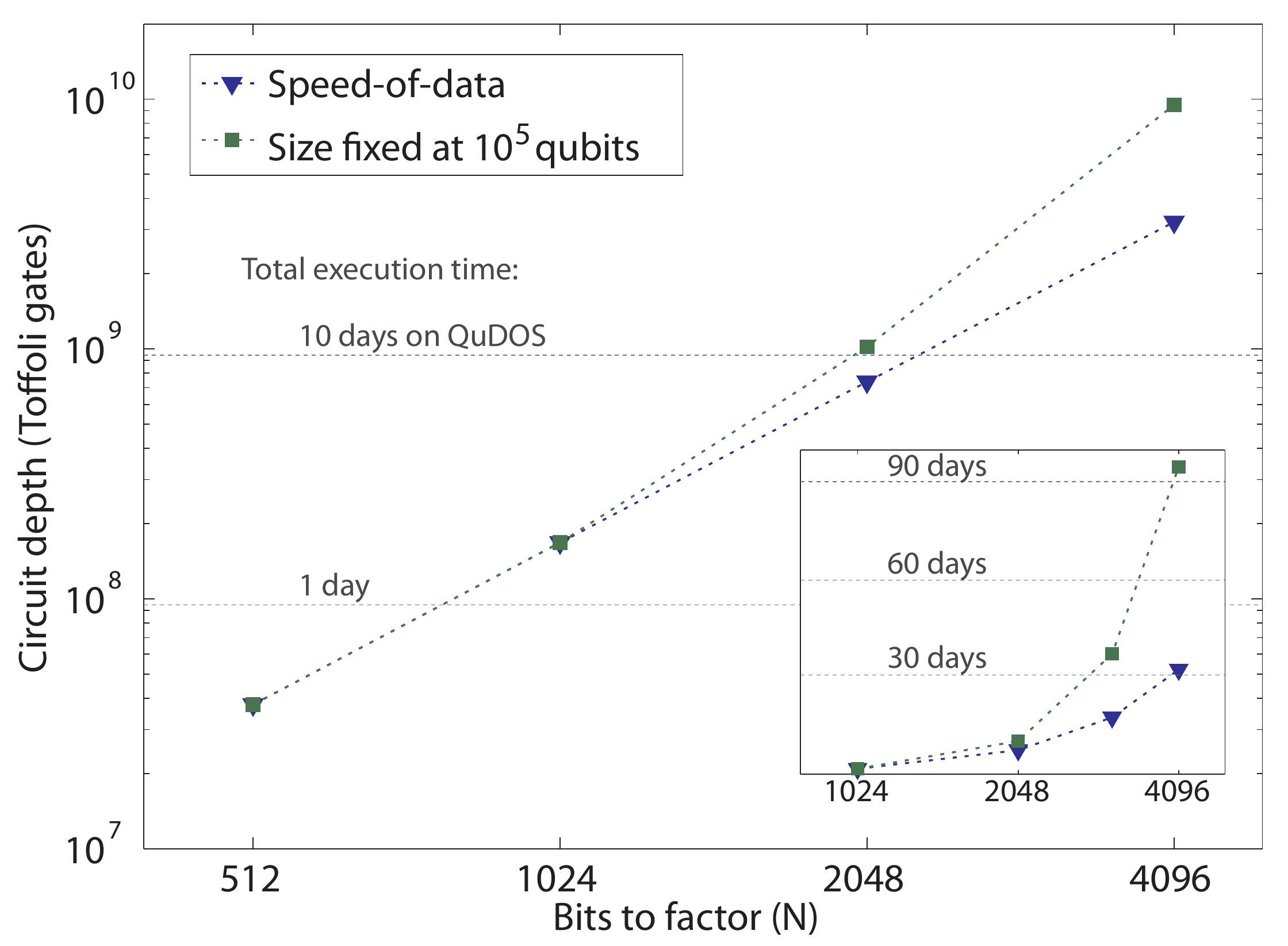}\\
  \caption{Color.  Execution time for Shor's algorithm, using the same circuit implementation as Ref.~\cite{VanMeter09}.  The vertical axis shows circuit depth, in terms of Toffoli gates, and the plot is labeled with estimated runtime on the \mbox{QuDOS} architecture.  The blue trace is a quantum computer whose size in logical qubits scales as necessary to compute at the speed of data (no delays).  The green trace is a machine with $10^5$ logical qubits, which experiences rapidly increasing delays as problem size increases beyond 2048 bits since insufficient resources are available to distill ancillas for \texttt{T}~gates, a necessary component of Shor's algorithm.  The inset shows the same data on a linear vertical scale, illustrating when the quantum computer experiences delays for lack of enough qubits.}
  \label{Shor_Toffoli}
\end{figure}


A common key length for RSA public-key cryptography is 1024 bits.  Factoring a number this large is not trivial, even on a quantum computer, as the following analysis shows.  Fig.~\ref{Shor_Toffoli} shows the expected run time on \mbox{QuDOS} for one iteration of Shor's algorithm versus key length in bits for two different quantum computers: one where system size increases with the problem size, and one where the system size is limited to $10^5$ logical qubits (including application qubits).  For the fixed-size quantum computer, the runtime begins to grow faster than the minimal circuit depth when factoring numbers 2048 bits and higher.  Fixing the machine size highlights the importance of the ancilla distillation factories.  For this instance of Shor's algorithm, about 90\% of the machine should be devoted to distillation; if insufficient resources are devoted to distillation, performance of the factoring algorithm plummets.  For example, the 4096-bit factorization devotes $\sim 75\%$ of the machine to distillation, but about $3\times$ as many factories would be needed to achieve maximum execution speed in the lower trace in Fig.~\ref{Shor_Toffoli}.  Many design parameters in an implementation of Shor's algorithm can be tuned as desired, and we collect the details of our analysis in Appendix \ref{Shor_details}.  We should also mention here that Shor's algorithm is probabilistic, so a few iterations may be required~\cite{Shor99}.

\subsection{Quantum simulation}
Quantum computers were inspired by the problem that simulating quantum systems on a classical computer is fundamentally difficult.  Feynman postulated that one quantum system could simulate another much more efficiently than a classical processor, and he proposed a quantum processor to perform this task~\cite{Feynman1982}.  Quantum simulation is one of the few known quantum algorithms that solves a useful problem believed to be intractable on classical computers, so we analyze the resource requirements for quantum simulation in the quantum architecture we propose.

We specifically consider fault-tolerant quantum simulation.  Other methods of simulation are under investigation~\cite{Buluta2009,Barreiro2011,Biamonte2010}, but they lie outside the scope of this work.  The particular example we select is simulating the Schr\"{o}dinger equation for time-independent Hamiltonians in first-quantized form, where each Hamiltonian represents the electron/nuclear configuration in a molecule~\cite{Zalka1998b,Kassal2008}.  An application of such a simulation is to determine ground- and excited-state energy levels in a molecule.  We select first-quantized instead of second-quantized form for better resource scaling at large problem sizes~\cite{Kassal2011}.


\begin{figure}
  \centering
  \includegraphics[width=8cm]{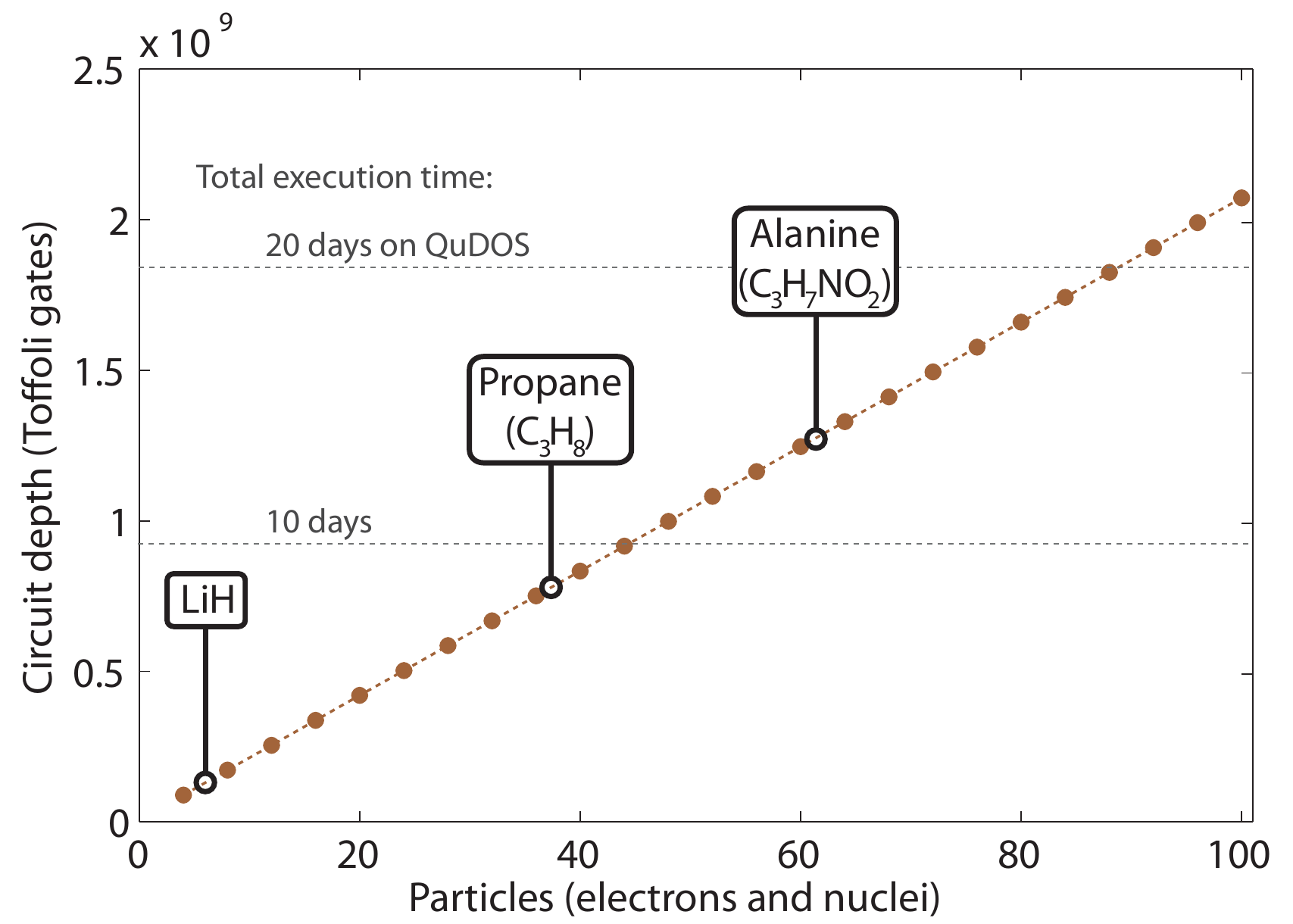}\\
  \caption{Color.  Execution time for simulation of a molecular Hamiltonian in first-quantized form, as a function of problem size.  The horizontal axis is number of particles being simulated, and the plot is labeled with some interesting examples from chemistry.  The vertical axis is circuit depth in Toffoli gates, and the plot is labeled with estimated runtime on \mbox{QuDOS}.  Each simulation uses 12-bit spatial precision in the wavefunction and $2^{10}$ time steps for 10-bit precision in readout, or at most $\sim 3$ significant figures.  The linear scaling in algorithm runtime versus problem size is due to two-body potential energy calculations, which constitute the majority of the quantum circuit.  The number of potential energy calculations increases quadratically with problem size, but through parallel computation they require linear execution time, as described in Appendix~\ref{simulation_details}.}
  \label{Simulation_depth}
\end{figure}


Fig.~\ref{Simulation_depth} shows the time necessary to execute the simulation algorithm for determining an energy eigenstate on the \mbox{QuDOS} computer as a function of the size of the simulation problem, expressed in number of electrons and nuclei.  First-quantized form stores the position-basis information for an electron wavefunction in a quantum register, and the complete Hamiltonian is a function of one- and two-body interactions between these registers, so this method does not depend on the particular molecular structure or arrangement; hence, the method is very general.  Note that the calculation time scales linearly in problem size, as opposed to the exponential scaling seen in classical methods.  The precision of the simulation scales with the number of time steps simulated~\cite{Aspuru2005}, and this example uses $2^{10}$ time steps for a maximum precision of about 3 significant figures.  Details of this simulation algorithm can be found in Appendix~\ref{simulation_details}.

\subsection{Large-scale quantum computing}
The factoring algorithm and quantum simulation represent interesting applications of large-scale quantum computing, and for each the computing resources required of a layered architecture based on \mbox{QuDOS} are listed in Table~\ref{Computing_Resources}.  The algorithms are comparable in total resource costs, as reflected by the fact that these two example problems require similar degrees of error correction (hence very similar $KQ$ product).  The simulation algorithm is more compact than Shor's, requiring in particular fewer logical qubits for distillation, which reflects the fact that this algorithm performs fewer arithmetic operations in parallel.  However, Shor's algorithm has a shorter execution time in this analysis.  Both algorithms can be accelerated through parallelism if the quantum computer has more logical qubits to work with~\cite{VanMeter05,Jones2012Sim}.

\begin{table}
  \centering
  \begin{tabular}{|m{1.2cm}|m{3.5cm}|m{1.7cm}|m{1.7cm}|}
    \hline
    \multicolumn{2}{|m{4.7cm}|}{\textbf{Computing Resource }}  & \textbf{Shor's \newline Algorithm \newline (1024-bit)} & \textbf{Molecular \newline Simulation \newline (alanine)}\\ \hline
    \multirow{2}{*}{Layer 5} & Application qubits\rule{0Ex}{2.5Ex}          & 6144\rule{0Ex}{2.5Ex}     & 6650\rule{0Ex}{2.5Ex} \\ \cline{2-4}
                             & Circuit depth (Toffoli)\rule{0Ex}{2.5Ex}     & $1.68$$\times$$10^8$\rule{0Ex}{2.5Ex}  & $1.27$$\times$$10^9$\rule{0Ex}{2.5Ex} \\ \hline
    \multirow{2}{*}{Layer 4} & Log. distillation qubits\rule{0Ex}{2.5Ex}    & 66564\rule{0Ex}{2.5Ex}    & 15860\rule{0Ex}{2.5Ex} \\ \cline{2-4}
                             & Logical clock cycles\rule{0Ex}{2.5Ex}        & $5.21$$\times$$10^9$\rule{0Ex}{2.5Ex}   & $3.94$$\times$$10^{10}$\rule{0Ex}{2.5Ex} \\ \hline
    \multirow{2}{*}{Layer 3} & Code distance\rule{0Ex}{2.5Ex}               & 31\rule{0Ex}{2.5Ex}       & 31\rule{0Ex}{2.5Ex} \\ \cline{2-4}
                             & Error per lattice cycle\rule{0Ex}{2.5Ex}     & $2.58$$\times$$10^{-20}$\rule{0Ex}{2.5Ex} & $2.58$$\times$$10^{-20}$\rule{0Ex}{2.5Ex} \\ \hline
    \multirow{2}{*}{Layer 2} & Virtual qubits             \rule{0Ex}{2.5Ex} & $4.54$$\times$$10^8$\rule{0Ex}{2.5Ex}    & $1.40$$\times$$10^8$\rule{0Ex}{2.5Ex} \\ \cline{2-4}
                             & Error per virtual gate     \rule{0Ex}{2.5Ex} & $1.00$$\times$$10^{-3}$\rule{0Ex}{2.5Ex}  & $1.00$$\times$$10^{-3}$\rule{0Ex}{2.5Ex} \\ \hline
    Layer 1                  & Quantum dots \rule{0Ex}{2.5Ex} \newline (area on chip)  & $4.54$$\times$$10^8$\rule{0Ex}{2.5Ex} \newline (4.54 $\mathrm{cm}^2$)  & $1.40$$\times$$10^8$\rule{0Ex}{2.5Ex} \newline (1.40 $\mathrm{cm}^2$) \\ \hline
    \multicolumn{2}{|m{4.7cm}|}{\textbf{Execution time (est.)}\rule{0Ex}{2.5Ex}} & \textbf{1.81 days}\rule{0Ex}{2.5Ex} & \textbf{13.7 days}\rule{0Ex}{2.5Ex} \\ \hline

  \end{tabular}
  \caption{Summary of the computing resources in a layered architecture based on the \mbox{QuDOS} platform, for Shor's algorithm factoring a 1024-bit number (same implementation as Ref.~\cite{VanMeter09}) and the ground state simulation of the molecule alanine ($\mathrm{C}_3\mathrm{H}_7\mathrm{N}\mathrm{O}_2$) using first-quantized representation.  Further details about the algorithms are provided in Appendix~\ref{Algorithm_details}.}
  \label{Computing_Resources}
\end{table}

\section{Timing Considerations}
\label{Timing}
\begin{figure*}
  \centering
  \includegraphics[width=\textwidth]{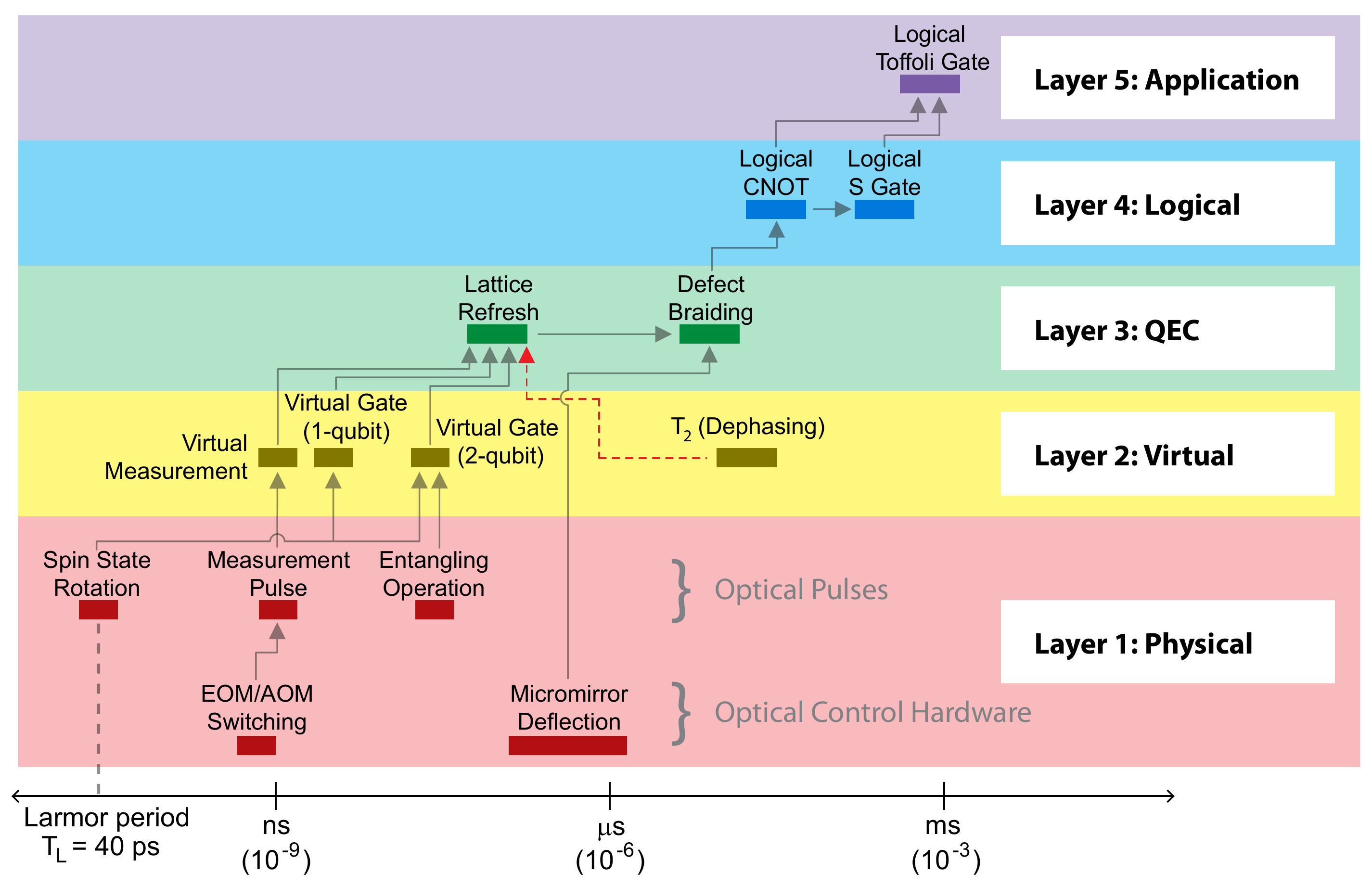}\\
  \caption{Color.  Relative timescales for critical operations in \mbox{QuDOS} within each layer.  Each bar indicates the approximate timescale of an operation, and the width indicates that some operation times may vary with improvements in technology.  The arrows indicate dependence of higher operations on lower layers. The red arrow signifies that the surface code lattice refresh must be 2--3 orders of magnitude faster than the dephasing time in order for error correction to function.  The Application layer is represented here with a Toffoli gate, which is a common building block of quantum algorithms.  Complete algorithm runtimes can vary significantly, depending on both the capabilities of the quantum computer and the specific way each algorithm is compiled, such as to what extent calculations are performed in parallel.  The optical control hardware is discussed in Appendix~\ref{App_Physical}.}
  \label{LayeredRelativeTimescales}
\end{figure*}

Precise timing and sequencing of operations are crucial to making an architecture efficient. In the framework we present here, an upper layer in the architecture depends on processes in the layer beneath, so that logical gate time is dictated by QEC operations, and so forth. This system of dependence of operation times is depicted for \mbox{QuDOS} in Fig.~\ref{LayeredRelativeTimescales}. The horizontal axis is a logarithmic scale in the time to execute an operation at a particular layer, while the arrows indicate fundamental dependence of one operation on other operations in lower layers.


Examining Fig.~\ref{LayeredRelativeTimescales}, we see that the timescales increase as one goes to higher layers. This is because a higher layer must often issue multiple commands to layers below. Using \mbox{QuDOS} as an example, the Virtual layer must construct a virtual 1-qubit gate from a sequence of spin-state rotations. This process includes the duration of the laser pulses and the delays between pulses, which all add together for the total duration of the virtual gate.  A crucial point shown in Fig.~\ref{LayeredRelativeTimescales} is that the time to implement a logical quantum gate can be orders of magnitude greater than the duration of each individual physical process, such as a laser pulse.  This increase in operation time is an important consideration for quantum computer designs which rely on comparatively slower physical processes.  At the same time, a quantum computer with a subset of very fast control mechanisms is limited by the slowest essential gate process, as \mbox{QuDOS} can only operate as fast as the 2-qubit entangling gate in Layer~1 permits.  For large-scale quantum computing, the speed of logical, error-corrected operations is the crucial figure of merit.

\begin{table}
  \centering
  \begin{tabular}{|m{2cm}|>{\centering}m{2.2cm}|m{3.9cm}|}
    \hline
    \textbf{Layer} & \textbf{Clock Cycle} & \textbf{Limiting Operation} \\ \hline
    (1) Physical & 8 ns & Laser repetition frequency \\ \hline
    (2) Virtual & 32 ns & Virtual 1-qubit gate \\ \hline
    (3) QEC & 256 ns & Lattice refresh \newline (syndrome circuit) \\ \hline
    (4) Logical & 30 $\mu$s & Logical \texttt{CNOT} \\ \hline

  \end{tabular}
  \caption{Clock cycle times for Layers 1 to 4 in our analysis of \mbox{QuDOS}.  The cycle time in each layer is determined by a fundamental control operation.  Many operations possess some flexibility that would permit tradeoffs in execution time and system size, and better methods may be discovered.}
  \label{Clock_frequency}
\end{table}

Fig.~\ref{LayeredRelativeTimescales} also highlights the fact that different control operations in the computer occur on substantially different timescales; achieving synchronization of these processes is an important function for a quantum computer architecture.  To facilitate this process, each layer in the architecture has an internal ``clock frequency,'' which is characteristic of the timescale of operations in that layer.  These clock cycle times for each layer in \mbox{QuDOS} are listed in Table~\ref{Clock_frequency}, along with the operations which define them.  Even within the same layer, some processes may take different lengths of time to execute, so setting a clock cycle synchronizes these operations.  Accordingly, as one layer builds on operations in a lower layer, the two layers are naturally synchronized.

Synchronization alone is not sufficient for a quantum computer to function.  Consider again the control cycle in Fig.~\ref{LayeredControlCycle_Abstract}.  Extracting and processing the error syndrome must be executed on timescales of the same order as the duration of a logical gate, or else errors will accumulate faster than they can be detected.  This function is performed by classical circuitry, but the required computing effort may not be trivial. Fast quantum operations can be a burden when error correction requires complex (classical) calculations, as is the case for the surface code.  \mbox{Devitt} \mbox{\emph{et al.}~\cite{Devitt10}} and \mbox{Fowler} \mbox{\emph{et al.}~\cite{Fowler2011b,Fowler2012}} examined this problem, finding that the processing requirements for surface code error correction are not trivial; performing these calculations ``live'' where the results may be needed within \emph{e.g.} 10 $\mu$s could be one of the more important problems for engineering a quantum computer.  Still, the recent progress in this area suggests that some combination of improved algorithm software and custom hardware can achieve the necessary performance~\cite{Fowler2012}.

\section{Discussion}
\label{discussion}
We have presented a layered framework for a quantum computer architecture.  The layered framework has two major strengths: it is \emph{modular}, and it facilitates \emph{fault tolerance}.  The layered nature of the architecture hints at modularity, but the defining characteristic of the layers we have chosen is encapsulation.  Each of the layers has a unique and important purpose, and that layer bundles the related operations to fulfill this purpose.  Additionally, each layer plays the role of resource manager, since often many operations in a lower layer are combined in a higher layer.  Since technologies in quantum computing will evolve over time, layers may need replacement in the future, and encapsulation makes integration of new processes a more straightforward task.

Fault tolerance is at present the biggest challenge for quantum computers, and the organization of layers is deliberately chosen to serve this need.  Arguably, Layers~1 and 5 define any quantum computer, but the layers in between are devoted exclusively to creating fault tolerance in an intelligent fashion.  Layer~2 uses simple control to mitigate systematic errors, so this layer is positioned close to the Physical layer where techniques like dynamical decoupling and decoherence-free subspaces are most effective.  Layer~3 hosts quantum error correction (QEC), which is essential for large-scale circuit-model quantum computing on any hardware, such as executing Shor's algorithm on a 1024-bit number.  There is a significant interplay between Layers~2 and 3, because Layer~2 enhances the effectiveness of Layer~3.  Finally, Layer~4 fills the gaps in the gate set provided by Layer~3 to form any desired unitary operation to arbitrary accuracy, thereby providing a complete substrate for universal quantum computation in Layer~5.

\mbox{QuDOS}, a specific hardware platform we introduce here, demonstrates the power of the layered architecture concept, but it also highlights a promising set of technologies for quantum computing, which are particularly noteworthy for the fast timescales of quantum operations, the high degree of integration possible with solid state fabrication, and the adoption of several mature technologies from other fields of engineering.  The execution times for fundamental quantum operations are discussed in Section~\ref{Layer1_summary}, but the importance of these fast processes becomes clear in Fig.~\ref{LayeredRelativeTimescales}, where the overhead resulting from virtual gates in Layer~2, QEC in Layer~3, and gate constructions in Layer 4 increases the time to implement quantum gates from nanoseconds in the Physical layer to milliseconds in the Application layer, or six orders of magnitude.  In this context, a quantum computer needs very fast physical operations.

One of our principal objectives is to better understand the resources required to construct a quantum computer that solves a problem intractable for classical computers.  Common figures of merit for evaluating quantum computing technology are gate fidelity, operation time, and qubit coherence time.  This investigation goes further to show how connectivity and classical
control performance are also crucial.  Designing a quantum computer requires viewing the system as a whole, such
that tradeoffs and compatibility between component choices must be addressed.  A holistic picture is equally
important for comparing different quantum computing technologies, such as ion traps or superconducting circuits. This work
illustrates how to approach the complete challenge of designing a quantum computer, so that one can adapt these techniques to develop architectures for other quantum computing technologies we have not considered here.  By doing so, differing system proposals can be compared within a common framework, which gives aspiring quantum engineers a common language for determining the best quantum computing technology for a desired application.

\begin{acknowledgments}
This work was supported by the National Science Foundation CCF-0829694, the Univ. of Tokyo Special Coordination Funds for Promoting Science and Technology, NICT, and the Japan Society for the Promotion of Science (JSPS) through its ``Funding Program for World-Leading Innovative R\&D on Science and Technology (FIRST  Program).''  NCJ was supported by the National Science Foundation Graduate Fellowship.  AGF acknowledges support from the Australian Research Council, the Australian Government, and the US National Security Agency (NSA) and the Army Research Office (ARO) under contract W911NF-08-1-0527.
\end{acknowledgments}

\appendix
\section{Parallel Control of Laser Pulses in \mbox{QuDOS}}
\label{App_Physical}
The \mbox{QuDOS} design depends on applying millions (or billions) of laser control pulses in parallel.  For completeness, we outline here a method for achieving this level of control, but a detailed analysis of the engineering problem lies outside the scope of this work.  Imagine that the 2D array of quantum dots in a cavity is an image plane, like a projector screen.  The challenge is to create a precisely controlled optical pattern on this screen.  We need two key elements for this scheme to work: the ability to modulate laser pulses to each point on the screen (\emph{i.e.} quantum dot), and the ability to focus laser signals near the diffraction limit.  Similar concepts were presented in Ref.~\cite{Steane2007}


\begin{figure*}
  \centering
  \includegraphics[width=12cm]{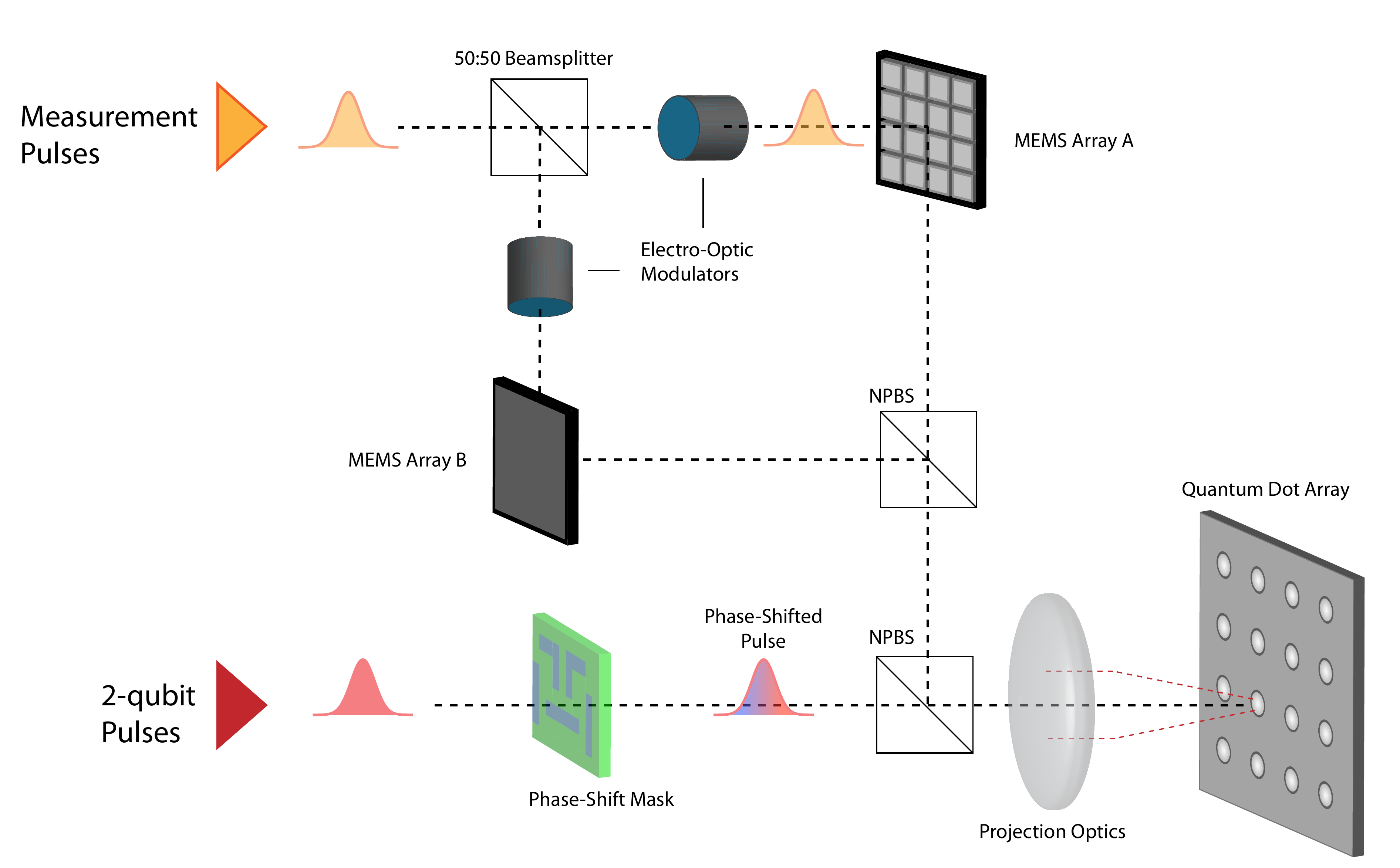}\\
  \caption{Color.  Optical setup in \mbox{QuDOS} which controls many quantum dots in parallel.  Phase-shift masks are used to produce diffraction-limited optical patterns which entangle the physical qubits (bottom).  Computation is achieved by measuring the resulting cluster-state; the measurement pattern is controlled by MEMS mirrors (top).  Two MEMS mirror arrays are multiplexed with electro-optic modulators, so that one can reposition mirrors while the other is ``active.''}
  \label{MEMS_AB}
\end{figure*}


To solve the first problem, we propose to use an array of MEMS mirrors.  This technology was developed for high-definition projectors and optical switches for telecommunications~\cite{TIDMD,Kim03}, but the same devices are being adapted for use in quantum information processing~\cite{Kim07,Knoer08}.  Since MEMS mirrors are based on the same fabrication techniques as integrated circuits, a controllable mirror array with millions of units has been demonstrated commercially~\cite{TIDMD}, and even larger arrays may be possible.

The 2-qubit gate mechanism in Section~\ref{2qubit} requires the quantum dots to be in relatively close proximity (1~$\mu$m), which is close to the wavelength of the laser light (920~nm).  Therefore, any optical patterns will have to compete with diffraction.  This is a familiar problem in photolithography, so we propose to adopt the method of phase-shift masking~\cite{Liu92,Aizenberg98} from this field.  In essence, a ``mask'' is defined by a transparent plate patterned in such a way that light passing through the mask receives a phase shift that is a function of position within the image plane.  This technique creates interference of light coming from different directions in such a manner that one can produce diffraction-limited patterns on the quantum dot array.

The MEMS mirrors and phase-shift masks work together as follows.  Operations on the surface code follow a highly regular pattern of virtual gates; more specifically, the surface code can be constructed by building a cluster state and performing measurement on selected virtual qubits to create defects in the lattice~\cite{Rauss07}.  The cluster state operations are decomposed into a sequence of laser pulses, which are in turn created by appropriately-designed phase-shift masks.  Separately, a pattern of measurement pulses for each virtual qubit are modulated by a MEMS array.  All of these optical signals are multiplexed together and sent to the quantum dot array, which is depicted in Fig.~\ref{MEMS_AB}.

The configuration of defects in the surface code changes more slowly than one cycle of the syndrome extraction circuit.  Because the defect boundaries must all be separated by the code distance $d$, the pattern of defects can be rearranged every $d/4$ lattice steps.  Therefore, the MEMS mirrors, which control where measurements are made, can be rearranged every 2 $\mu$s in \mbox{QuDOS}, which is compatible with current technology~\cite{NielsonCLEO2007}.  Still, one has to account for the time required to reposition a set of mirrors.  Two sets of mirrors are used in an alternating sequence: one is repositioning while the other is actively used, as shown in the top of Fig.~\ref{MEMS_AB}.  Electro-optic modulators can quickly multiplex laser pulses between the two mirror arrays.

\section{Application Layer Details}
\label{Algorithm_details}
We provide here a brief summary of the computational complexity for Shor's algorithm and quantum simulation in first-quantized form.  This analysis produces the resource estimates in Section~\ref{application_layer}.

\subsection{Shor's algorithm}
\label{Shor_details}
We adopt the same implementation of Shor's algorithm given in Van Meter \emph{et al.}~\cite{VanMeter09}.  In order to determine the performance of Shor's algorithm at Layer~5, we must look at how efficiently Layer~4 prepares Toffoli gates.  Let us suppose that the quantum computer has capacity for $10^5$ logical qubits; in general, one can interchange logical capacity and algorithm execution time.  To factor an $N$-bit number, approximately $6N$ application qubits are used by the algorithm itself, with the remainder of the logical qubits used to produce the crucial $\ket{A}$ ancillas.  Implementations with fewer application qubits are possible~\cite{Beauregard2003,Takahashi2006}, but the performance of such circuits is dramatically slower, especially if one is restricted to a limited set of gates.  As shown in Section~\ref{distillation}, one round of $\ket{A}$ distillation requires a volume of computing resources with cross-section 12 logical qubits and time 6 \texttt{CNOT} cycles.  We arrange the excess $(10^5 - 6N)$ qubits in ``factories'' which distill ancillas as fast as possible.

\begin{table}
  \centering
  \begin{tabular}{|m{1.2cm}|m{2.5cm}|m{2cm}|m{2.3cm}|}
    \hline
    \textbf{Bits to \newline Factor} & \textbf{Ancilla Factory \newline Cross-Section} \newline \footnotesize{(Logical Qubits)} & \textbf{Distillation \newline Rate} \newline \footnotesize{($\ket{A}$ per cycle)} & \textbf{Consumption \newline Rate (Max)} \newline \footnotesize{($\ket{A}$ per cycle)} \\ \hline
    512 & 96928 & 84.1 & 32.1\\ \hline
    1024 & 93856 & 81.5 & 57.8\\ \hline
    2048 & 87712 & 76.1 & 105.1\\ \hline
    4096 & 75424 & 65.5 & 192.7\\ \hline
    8192 & 50848 & 44.1 & 355.7\\ \hline
    16384 & 1696 & 1.5 & 660.6\\ \hline

  \end{tabular}
  \caption{Generation rates and maximal consumption rates for a $10^5$-qubit quantum computer running Shor's factoring algorithm.  When the speed-of-data consumption rate is higher than the distillation rate, Shor's algorithm experiences delays.}
  \label{ancilla_rates}
\end{table}

\begin{table}
  \centering
  \begin{tabular}{|m{2.6cm}|m{2.4cm}|m{3.2cm}|}
    \hline
    \textbf{Operator} & \textbf{Maximum Memory Size} \newline \footnotesize{(Logical Qubits)} & \textbf{Circuit Depth} \newline \footnotesize{(Layer 4 Clock Cycles)} \\ \hline
    Kinetic Energy \rule{0Ex}{2.5Ex} & $334 \times B$ & $1.55$$\times$$10^5$ \\ \hline
    Potential Energy \rule{0Ex}{2.5Ex} & $369 \times B$ & $6.26$$\times$$10^5 \times B$ \\ \hline
    QFT \rule{0Ex}{2.5Ex} & $272 \times B$ & $2.57$$\times$$10^4$\\ \hline

  \end{tabular}
  \caption{Resource requirements for the operators in first-quantized molecular simulation with $B$ particles and 12-bit spatial precision including ancilla distillation.}
  \label{simulation_operators}
\end{table}

As before, we define a Logical layer clock cycle as 1 \texttt{CNOT} gate.  We express the rate at which the factory generates ancillas by mean number of ancillas produced per clock cycle.  Two levels of distillation will require 16 distillation circuits (15 at the first level, 1 at the second level), which uses a circuit volume of $V_{\text{distill}} = 16 \times (12 \textrm{ log. qubits}) \times (6 \textrm{ clock cycles})$.  For a given cross-section area $A_{\text{factory}}$ of the quantum computer devoted to distillation, the maximal rate of ancilla production is given by $A_{\text{factory}}/V_{\text{distill}}$.  Calculations of these values are given in Table~\ref{ancilla_rates}.

We need to determine whether the quantum computer can run as fast as the circuit depth in Layer~5, or whether the distillation of $\ket{A}$ states limits performance.  Using a construction like Fig.~\ref{Toffoli_decomposition}, the depth of the Toffoli gate is 31 clock cycles, where each \texttt{S}~gate requires 4 cycles as shown in Fig.~\ref{Sgate_circuit}, and the circuit requires 7 distilled $\ket{A}$ ancillas.  The circuit uses the carry-lookahead adder construction in Ref.~\cite{Draper2006}, which requires $\sim 10N$ Toffoli gates in total with a circuit depth of $(\sim 4\log_2 N) t_{\text{Toffoli}}$, or $\sim 124\log_2 N$ cycles.  Using these figures, the maximal consumption rate of ancillas can be calculated, as shown in Table~\ref{ancilla_rates}.  As the size of the number to be factored increases, a fixed-size quantum computer is at some point unable to generate enough ancillas to run the algorithm at maximum speed; when this happens, execution time is limited by the distillation process.  One can make a crude estimate from Table~\ref{ancilla_rates} that an efficient quantum computer for Shor's algorithm must devote 90\% of its resources to distillation.  By similar arguments, a ``minimal'' size quantum computer that holds just the algorithm qubits and distills one $\ket{A}$ ancilla at a time will be very slow.

\subsection{Quantum simulation}
\label{simulation_details}
\begin{figure}
  \centering
  \includegraphics[width=7cm]{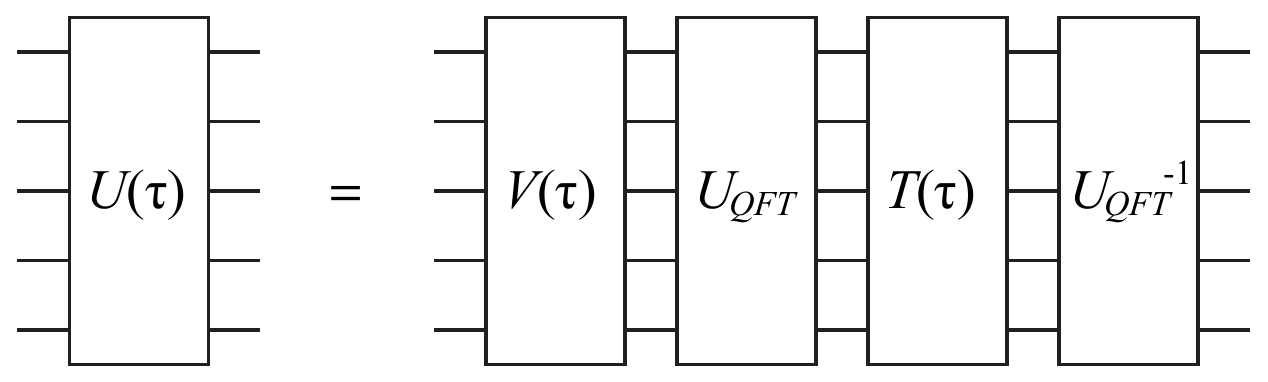}\\
  \caption{Circuit representation for one iteration of the Hamiltonian propagator in first-quantized form.  The QFT is performed on the wavefunction, transforming between position-basis and momentum-basis.}
  \label{propagator_circuit}
\end{figure}

We utilize the method in Ref.~\cite{Kassal2008} to perform simulation in first-quantized form.  Each electron wavefunction is represented on a 3-dimensional Cartesian grid with 12 bits of precision in each dimension, which requires a quantum register of 36 qubits per particle.  We elect to use a different set of adders and multipliers than Ref.~\cite{Kassal2008}, opting instead for simple ripple-carry adders which suffice for 12-bit precision~\cite{Cuccaro2008}.  First, the potential energy operator is calculated in the position-basis.  We transform the wavefunction representation from position-basis to momentum-basis with the quantum Fourier transform (QFT), allowing efficient evaluation of the kinetic energy operator.  The inverse QFT transforms our system back to position-basis.  The quantum circuit representation of the system propagator $\mathcal{U}$ is depicted in Fig.~\ref{propagator_circuit}.

\begin{figure}
  \centering
  \includegraphics[width=7cm]{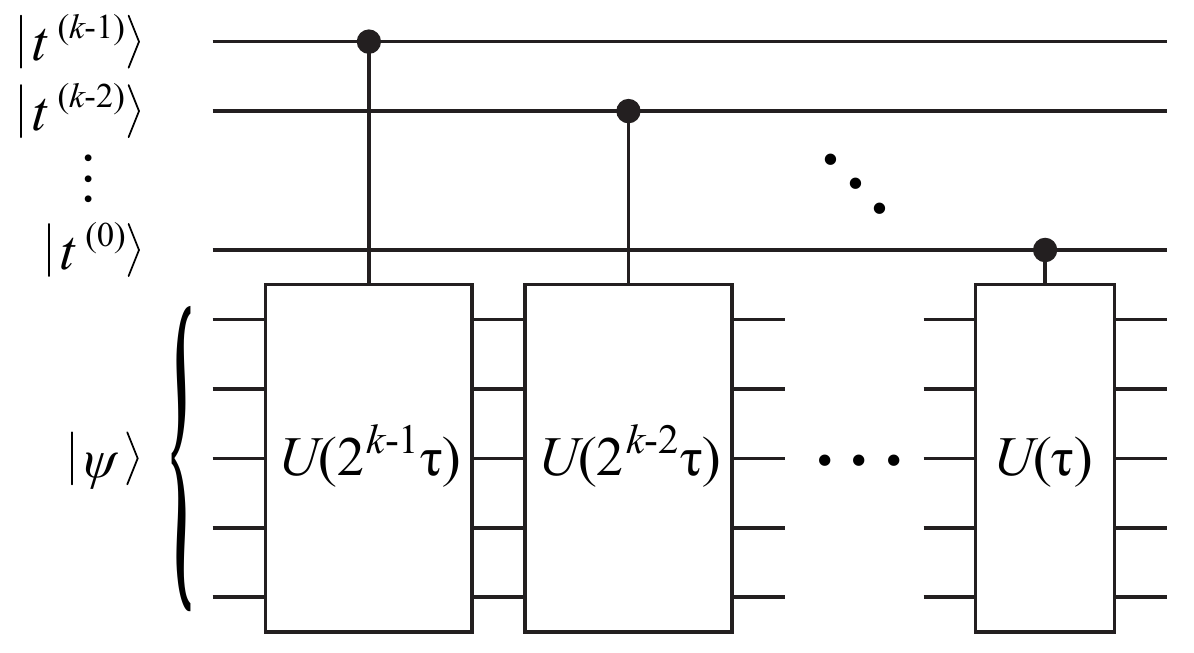}\\
  \caption{The time-evolution of the Hamiltonian is produced by iterating the system propagator over many time steps.  After evolution, a quantum Fourier transform of the time vector transforms the system into the energy eigenbasis, allowing readout of an energy eigenvalue.}
  \label{PEA}
\end{figure}

The resource requirements for each of the kinetic ($\mathcal{T}$), potential ($\mathcal{V}$), and QFT operators are summarized in Table~\ref{simulation_operators}.  The parameters in Table~\ref{simulation_operators} were derived assuming parallel calculation of commuting operator terms; for example, the Coulomb interaction between particles $\alpha$ and $\beta$ can be calculated simultaneously as $\gamma$ and $\delta$, because these terms in the Hamiltonian commute and the circuits are disjoint~\cite{Jones2012Sim}.  Moreover, we have used the preceding analysis in Appendix \ref{Shor_details} to include in these figures the size of ancilla factories, which is $\sim 260B$ logical qubits in order to simulate a system of $B$ particles.  For this parallel simulation algorithm, ancilla production consumes about 70\% of the quantum computer.

The circuit construction in Fig.~\ref{propagator_circuit} is just one iteration of the system propagator.  Estimating an energy eigenvalue requires simulation of the system at discrete time steps, so the propagator is repeated many times~\cite{Aspuru2005}, as shown in Fig.~\ref{PEA}.  After evolving the propagator along these time steps, the system is transformed to the energy eigenbasis via a QFT operation on the time vector $\ket{t}$~\cite{Nielsen00}.  The precision in the final answer is limited by the number of bits in $\ket{t}$, so for this analysis we assume the system is evolved for $2^{10}$ time steps, which offers at most $\sim 3$ decimal digits of precision.\\

\bibliography{References}

\begin{thebibliography}{130}%
\makeatletter
\providecommand \@ifxundefined [1]{%
 \@ifx{#1\undefined}
}%
\providecommand \@ifnum [1]{%
 \ifnum #1\expandafter \@firstoftwo
 \else \expandafter \@secondoftwo
 \fi
}%
\providecommand \@ifx [1]{%
 \ifx #1\expandafter \@firstoftwo
 \else \expandafter \@secondoftwo
 \fi
}%
\providecommand \natexlab [1]{#1}%
\providecommand \enquote  [1]{``#1''}%
\providecommand \bibnamefont  [1]{#1}%
\providecommand \bibfnamefont [1]{#1}%
\providecommand \citenamefont [1]{#1}%
\providecommand \href@noop [0]{\@secondoftwo}%
\providecommand \href [0]{\begingroup \@sanitize@url \@href}%
\providecommand \@href[1]{\@@startlink{#1}\@@href}%
\providecommand \@@href[1]{\endgroup#1\@@endlink}%
\providecommand \@sanitize@url [0]{\catcode `\\12\catcode `\$12\catcode
  `\&12\catcode `\#12\catcode `\^12\catcode `\_12\catcode `\%12\relax}%
\providecommand \@@startlink[1]{}%
\providecommand \@@endlink[0]{}%
\providecommand \url  [0]{\begingroup\@sanitize@url \@url }%
\providecommand \@url [1]{\endgroup\@href {#1}{\urlprefix }}%
\providecommand \urlprefix  [0]{URL }%
\providecommand \Eprint [0]{\href }%
\providecommand \doibase [0]{http://dx.doi.org/}%
\providecommand \selectlanguage [0]{\@gobble}%
\providecommand \bibinfo  [0]{\@secondoftwo}%
\providecommand \bibfield  [0]{\@secondoftwo}%
\providecommand \translation [1]{[#1]}%
\providecommand \BibitemOpen [0]{}%
\providecommand \bibitemStop [0]{}%
\providecommand \bibitemNoStop [0]{.\EOS\space}%
\providecommand \EOS [0]{\spacefactor3000\relax}%
\providecommand \BibitemShut  [1]{\csname bibitem#1\endcsname}%
\let\auto@bib@innerbib\@empty
\bibitem [{\citenamefont {Ladd}\ \emph {et~al.}(2010)\citenamefont {Ladd},
  \citenamefont {Jelezko}, \citenamefont {Laflamme}, \citenamefont {Nakamura},
  \citenamefont {Monroe},\ and\ \citenamefont {O'Brien}}]{Ladd10}%
  \BibitemOpen
  \bibfield  {author} {\bibinfo {author} {\bibfnamefont {T.D.}\ \bibnamefont
  {Ladd}}, \bibinfo {author} {\bibfnamefont {F.}~\bibnamefont {Jelezko}},
  \bibinfo {author} {\bibfnamefont {R.}~\bibnamefont {Laflamme}}, \bibinfo
  {author} {\bibfnamefont {Y.}~\bibnamefont {Nakamura}}, \bibinfo {author}
  {\bibfnamefont {C.}~\bibnamefont {Monroe}}, \ and\ \bibinfo {author}
  {\bibfnamefont {J.L.}\ \bibnamefont {O'Brien}},\ }\bibfield  {title}
  {\enquote {\bibinfo {title} {Quantum computers},}\ }\href@noop {} {\bibfield
  {journal} {\bibinfo  {journal} {Nature}\ }\textbf {\bibinfo {volume} {464}},\
  \bibinfo {pages} {45--53} (\bibinfo {year} {2010})}\BibitemShut {NoStop}%
\bibitem [{\citenamefont {Fowler}\ \emph {et~al.}(2009)\citenamefont {Fowler},
  \citenamefont {Stephens},\ and\ \citenamefont {Groszkowski}}]{Fowler09}%
  \BibitemOpen
  \bibfield  {author} {\bibinfo {author} {\bibfnamefont {Austin~G.}\
  \bibnamefont {Fowler}}, \bibinfo {author} {\bibfnamefont {Ashley~M.}\
  \bibnamefont {Stephens}}, \ and\ \bibinfo {author} {\bibfnamefont {Peter}\
  \bibnamefont {Groszkowski}},\ }\bibfield  {title} {\enquote {\bibinfo {title}
  {High-threshold universal quantum computation on the surface code},}\
  }\href@noop {} {\bibfield  {journal} {\bibinfo  {journal} {Phys. Rev. A}\
  }\textbf {\bibinfo {volume} {80}},\ \bibinfo {pages} {052312} (\bibinfo
  {year} {2009})}\BibitemShut {NoStop}%
\bibitem [{\citenamefont {DiVincenzo}(2000)}]{DiVincenzo00}%
  \BibitemOpen
  \bibfield  {author} {\bibinfo {author} {\bibfnamefont {David~P.}\
  \bibnamefont {DiVincenzo}},\ }\bibfield  {title} {\enquote {\bibinfo {title}
  {The physical implementation of quantum computation},}\ }\href@noop {}
  {\bibfield  {journal} {\bibinfo  {journal} {Fortschritte der Physik}\
  }\textbf {\bibinfo {volume} {48}},\ \bibinfo {pages} {771--783} (\bibinfo
  {year} {2000})}\BibitemShut {NoStop}%
\bibitem [{\citenamefont {Steane}(2002)}]{Steane02}%
  \BibitemOpen
  \bibfield  {author} {\bibinfo {author} {\bibfnamefont {Andrew~M.}\
  \bibnamefont {Steane}},\ }\bibfield  {title} {\enquote {\bibinfo {title}
  {Quantum computer architecture for fast entropy extraction},}\ }\href@noop {}
  {\bibfield  {journal} {\bibinfo  {journal} {Quantum Info. Comput.}\ }\textbf
  {\bibinfo {volume} {2}},\ \bibinfo {pages} {171--183} (\bibinfo {year}
  {2002})}\BibitemShut {NoStop}%
\bibitem [{\citenamefont {Steane}(2007)}]{Steane2007}%
  \BibitemOpen
  \bibfield  {author} {\bibinfo {author} {\bibfnamefont {Andrew~M.}\
  \bibnamefont {Steane}},\ }\bibfield  {title} {\enquote {\bibinfo {title} {How
  to build a 300 bit, 1 {G}iga-operation quantum computer},}\ }\href@noop {}
  {\bibfield  {journal} {\bibinfo  {journal} {Quantum Info. Comput.}\ }\textbf
  {\bibinfo {volume} {7}},\ \bibinfo {pages} {297--306} (\bibinfo {year}
  {2007})}\BibitemShut {NoStop}%
\bibitem [{\citenamefont {Spiller}\ \emph {et~al.}(2005)\citenamefont
  {Spiller}, \citenamefont {Munro}, \citenamefont {Barrett},\ and\
  \citenamefont {Kok}}]{Spiller05}%
  \BibitemOpen
  \bibfield  {author} {\bibinfo {author} {\bibfnamefont {Timothy~P.}\
  \bibnamefont {Spiller}}, \bibinfo {author} {\bibfnamefont {William~J.}\
  \bibnamefont {Munro}}, \bibinfo {author} {\bibfnamefont {Sean~D.}\
  \bibnamefont {Barrett}}, \ and\ \bibinfo {author} {\bibfnamefont {Pieter}\
  \bibnamefont {Kok}},\ }\bibfield  {title} {\enquote {\bibinfo {title} {An
  introduction to quantum information processing: applications and
  realizations},}\ }\href@noop {} {\bibfield  {journal} {\bibinfo  {journal}
  {Contemporary Physics}\ }\textbf {\bibinfo {volume} {46}},\ \bibinfo {pages}
  {407--436} (\bibinfo {year} {2005})}\BibitemShut {NoStop}%
\bibitem [{\citenamefont {Van{ }Meter}\ and\ \citenamefont
  {Oskin}(2006)}]{VanMeter06}%
  \BibitemOpen
  \bibfield  {author} {\bibinfo {author} {\bibfnamefont {Rodney}\ \bibnamefont
  {Van{ }Meter}}\ and\ \bibinfo {author} {\bibfnamefont {Mark}\ \bibnamefont
  {Oskin}},\ }\bibfield  {title} {\enquote {\bibinfo {title} {Architectural
  implications of quantum computing technologies},}\ }\href@noop {} {\bibfield
  {journal} {\bibinfo  {journal} {{ACM} Journal of Emerging Technologies in
  Computing Systems}\ }\textbf {\bibinfo {volume} {2}},\ \bibinfo {pages}
  {31--63} (\bibinfo {year} {2006})}\BibitemShut {NoStop}%
\bibitem [{\citenamefont {Taylor}\ \emph {et~al.}(2005)\citenamefont {Taylor},
  \citenamefont {Engel}, \citenamefont {D\"{u}r}, \citenamefont {Yacoby},
  \citenamefont {Marcus}, \citenamefont {Zoller},\ and\ \citenamefont
  {Lukin}}]{Taylor2005}%
  \BibitemOpen
  \bibfield  {author} {\bibinfo {author} {\bibfnamefont {J.M.}\ \bibnamefont
  {Taylor}}, \bibinfo {author} {\bibfnamefont {H.-A.}\ \bibnamefont {Engel}},
  \bibinfo {author} {\bibfnamefont {W.}~\bibnamefont {D\"{u}r}}, \bibinfo
  {author} {\bibfnamefont {A.}~\bibnamefont {Yacoby}}, \bibinfo {author}
  {\bibfnamefont {C.~M.}\ \bibnamefont {Marcus}}, \bibinfo {author}
  {\bibfnamefont {P.}~\bibnamefont {Zoller}}, \ and\ \bibinfo {author}
  {\bibfnamefont {M.~D.}\ \bibnamefont {Lukin}},\ }\bibfield  {title} {\enquote
  {\bibinfo {title} {Fault-tolerant architecture for quantum computation using
  electrically controlled semiconductor spins},}\ }\href@noop {} {\bibfield
  {journal} {\bibinfo  {journal} {Nature Physics}\ }\textbf {\bibinfo {volume}
  {1}},\ \bibinfo {pages} {177--183} (\bibinfo {year} {2005})}\BibitemShut
  {NoStop}%
\bibitem [{\citenamefont {Steane}(1998)}]{Steane1998}%
  \BibitemOpen
  \bibfield  {author} {\bibinfo {author} {\bibfnamefont {A.~M.}\ \bibnamefont
  {Steane}},\ }\bibfield  {title} {\enquote {\bibinfo {title} {Space, time,
  parallelism and noise requirements for reliable quantum computing},}\
  }\href@noop {} {\bibfield  {journal} {\bibinfo  {journal} {Fortschritte der
  Physik}\ }\textbf {\bibinfo {volume} {46}},\ \bibinfo {pages} {443--457}
  (\bibinfo {year} {1998})}\BibitemShut {NoStop}%
\bibitem [{\citenamefont {Isailovic}\ \emph {et~al.}(2008)\citenamefont
  {Isailovic}, \citenamefont {Whitney}, \citenamefont {Patel},\ and\
  \citenamefont {Kubiatowicz}}]{Isailovic08}%
  \BibitemOpen
  \bibfield  {author} {\bibinfo {author} {\bibfnamefont {N.}~\bibnamefont
  {Isailovic}}, \bibinfo {author} {\bibfnamefont {M.}~\bibnamefont {Whitney}},
  \bibinfo {author} {\bibfnamefont {Y.}~\bibnamefont {Patel}}, \ and\ \bibinfo
  {author} {\bibfnamefont {J.}~\bibnamefont {Kubiatowicz}},\ }\bibfield
  {title} {\enquote {\bibinfo {title} {Running a quantum circuit at the speed
  of data},}\ }in\ \href@noop {} {\emph {\bibinfo {booktitle} {35th
  International Symposium on Computer Architecture, 2008 (ISCA'08)}}}\
  (\bibinfo {year} {2008})\BibitemShut {NoStop}%
\bibitem [{\citenamefont {Metodi}\ \emph {et~al.}(2005)\citenamefont {Metodi},
  \citenamefont {Thaker}, \citenamefont {Cross}, \citenamefont {Chong},\ and\
  \citenamefont {Chuang}}]{Metodi2005}%
  \BibitemOpen
  \bibfield  {author} {\bibinfo {author} {\bibfnamefont {Tzvetan~S.}\
  \bibnamefont {Metodi}}, \bibinfo {author} {\bibfnamefont {Darshan~D.}\
  \bibnamefont {Thaker}}, \bibinfo {author} {\bibfnamefont {Andrew~W.}\
  \bibnamefont {Cross}}, \bibinfo {author} {\bibfnamefont {Frederic~T.}\
  \bibnamefont {Chong}}, \ and\ \bibinfo {author} {\bibfnamefont {Isaac~L.}\
  \bibnamefont {Chuang}},\ }\bibfield  {title} {\enquote {\bibinfo {title} {A
  quantum logic array microarchitecture: Scalable quantum data movement and
  computation},}\ }in\ \href@noop {} {\emph {\bibinfo {booktitle} {Proceedings
  of the 38th International Symposium on Microarchitecture MICRO-38}}}\
  (\bibinfo {year} {2005})\ pp.\ \bibinfo {pages} {305--318},\ \bibinfo {note}
  {\emph{Preprint} arXiv:quant-ph/0509051v1.}\BibitemShut {Stop}%
\bibitem [{\citenamefont {Kielpinski}\ \emph {et~al.}(2002)\citenamefont
  {Kielpinski}, \citenamefont {Monroe},\ and\ \citenamefont
  {Wineland}}]{Kielpinski02}%
  \BibitemOpen
  \bibfield  {author} {\bibinfo {author} {\bibfnamefont {D.}~\bibnamefont
  {Kielpinski}}, \bibinfo {author} {\bibfnamefont {C.}~\bibnamefont {Monroe}},
  \ and\ \bibinfo {author} {\bibfnamefont {D.}~\bibnamefont {Wineland}},\
  }\bibfield  {title} {\enquote {\bibinfo {title} {Architecture for a
  large-scale ion-trap quantum computer},}\ }\href@noop {} {\bibfield
  {journal} {\bibinfo  {journal} {Nature}\ }\textbf {\bibinfo {volume} {417}},\
  \bibinfo {pages} {709--711} (\bibinfo {year} {2002})}\BibitemShut {NoStop}%
\bibitem [{\citenamefont {Copsey}\ \emph {et~al.}(2003)\citenamefont {Copsey},
  \citenamefont {Oskin}, \citenamefont {Metodiev}, \citenamefont {Chong},
  \citenamefont {Chuang},\ and\ \citenamefont {Kubiatowicz}}]{Copsey03}%
  \BibitemOpen
  \bibfield  {author} {\bibinfo {author} {\bibfnamefont {Dean}\ \bibnamefont
  {Copsey}}, \bibinfo {author} {\bibfnamefont {Mark}\ \bibnamefont {Oskin}},
  \bibinfo {author} {\bibfnamefont {Tzvetan}\ \bibnamefont {Metodiev}},
  \bibinfo {author} {\bibfnamefont {Frederic~T.}\ \bibnamefont {Chong}},
  \bibinfo {author} {\bibfnamefont {Isaac}\ \bibnamefont {Chuang}}, \ and\
  \bibinfo {author} {\bibfnamefont {John}\ \bibnamefont {Kubiatowicz}},\
  }\bibfield  {title} {\enquote {\bibinfo {title} {The effect of communication
  costs in solid-state quantum computing architectures},}\ }in\ \href@noop {}
  {\emph {\bibinfo {booktitle} {Proceedings of the Fifteenth Annual ACM
  Symposium on Parallel Algorithms and Architectures (SPAA'03)}}}\ (\bibinfo
  {publisher} {ACM},\ \bibinfo {address} {New York, NY, USA},\ \bibinfo {year}
  {2003})\ pp.\ \bibinfo {pages} {65--74}\BibitemShut {NoStop}%
\bibitem [{\citenamefont {Svore}\ \emph {et~al.}(2006)\citenamefont {Svore},
  \citenamefont {Aho}, \citenamefont {Cross}, \citenamefont {Chuang},\ and\
  \citenamefont {Markov}}]{Svore06}%
  \BibitemOpen
  \bibfield  {author} {\bibinfo {author} {\bibfnamefont {K.M.}\ \bibnamefont
  {Svore}}, \bibinfo {author} {\bibfnamefont {A.V.}\ \bibnamefont {Aho}},
  \bibinfo {author} {\bibfnamefont {A.W.}\ \bibnamefont {Cross}}, \bibinfo
  {author} {\bibfnamefont {I.}~\bibnamefont {Chuang}}, \ and\ \bibinfo {author}
  {\bibfnamefont {I.L.}\ \bibnamefont {Markov}},\ }\bibfield  {title} {\enquote
  {\bibinfo {title} {A layered software architecture for quantum computing
  design tools},}\ }\href@noop {} {\bibfield  {journal} {\bibinfo  {journal}
  {Computer}\ }\textbf {\bibinfo {volume} {39}},\ \bibinfo {pages} {74--83}
  (\bibinfo {year} {2006})}\BibitemShut {NoStop}%
\bibitem [{\citenamefont {Oskin}\ \emph {et~al.}(2003)\citenamefont {Oskin},
  \citenamefont {Chong}, \citenamefont {Chuang},\ and\ \citenamefont
  {Kubiatowicz}}]{Oskin03}%
  \BibitemOpen
  \bibfield  {author} {\bibinfo {author} {\bibfnamefont {M.}~\bibnamefont
  {Oskin}}, \bibinfo {author} {\bibfnamefont {F.T.}\ \bibnamefont {Chong}},
  \bibinfo {author} {\bibfnamefont {I.L.}\ \bibnamefont {Chuang}}, \ and\
  \bibinfo {author} {\bibfnamefont {J.}~\bibnamefont {Kubiatowicz}},\
  }\bibfield  {title} {\enquote {\bibinfo {title} {Building quantum wires: the
  long and the short of it},}\ }in\ \href@noop {} {\emph {\bibinfo {booktitle}
  {30th International Symposium on Computer Architecture, 2003 (ISCA'03)}}}\
  (\bibinfo {year} {2003})\ pp.\ \bibinfo {pages} {374--385}\BibitemShut
  {NoStop}%
\bibitem [{\citenamefont {Kane}(1998)}]{Kane98}%
  \BibitemOpen
  \bibfield  {author} {\bibinfo {author} {\bibfnamefont {B.E.}\ \bibnamefont
  {Kane}},\ }\bibfield  {title} {\enquote {\bibinfo {title} {A silicon-based
  nuclear spin quantum computer},}\ }\href@noop {} {\bibfield  {journal}
  {\bibinfo  {journal} {Nature}\ }\textbf {\bibinfo {volume} {393}},\ \bibinfo
  {pages} {133--137} (\bibinfo {year} {1998})}\BibitemShut {NoStop}%
\bibitem [{\citenamefont {Mariantoni}\ \emph {et~al.}(2011)\citenamefont
  {Mariantoni}, \citenamefont {Wang}, \citenamefont {Yamamoto}, \citenamefont
  {Neeley}, \citenamefont {Bialczak}, \citenamefont {Chen}, \citenamefont
  {Lenander}, \citenamefont {Lucero}, \citenamefont {O'Connell}, \citenamefont
  {Sank}, \citenamefont {Weides}, \citenamefont {Wenner}, \citenamefont {Yin},
  \citenamefont {Zhao}, \citenamefont {Korotkov}, \citenamefont {Cleland},\
  and\ \citenamefont {Martinis}}]{Mariantoni2011}%
  \BibitemOpen
  \bibfield  {author} {\bibinfo {author} {\bibfnamefont {Matteo}\ \bibnamefont
  {Mariantoni}}, \bibinfo {author} {\bibfnamefont {H.}~\bibnamefont {Wang}},
  \bibinfo {author} {\bibfnamefont {T.}~\bibnamefont {Yamamoto}}, \bibinfo
  {author} {\bibfnamefont {M.}~\bibnamefont {Neeley}}, \bibinfo {author}
  {\bibfnamefont {Radoslaw~C.}\ \bibnamefont {Bialczak}}, \bibinfo {author}
  {\bibfnamefont {Y.}~\bibnamefont {Chen}}, \bibinfo {author} {\bibfnamefont
  {M.}~\bibnamefont {Lenander}}, \bibinfo {author} {\bibfnamefont {Erik}\
  \bibnamefont {Lucero}}, \bibinfo {author} {\bibfnamefont {A.D.}\ \bibnamefont
  {O'Connell}}, \bibinfo {author} {\bibfnamefont {D.}~\bibnamefont {Sank}},
  \bibinfo {author} {\bibfnamefont {M.}~\bibnamefont {Weides}}, \bibinfo
  {author} {\bibfnamefont {J.}~\bibnamefont {Wenner}}, \bibinfo {author}
  {\bibfnamefont {Y.}~\bibnamefont {Yin}}, \bibinfo {author} {\bibfnamefont
  {J.}~\bibnamefont {Zhao}}, \bibinfo {author} {\bibfnamefont {A.N.}\
  \bibnamefont {Korotkov}}, \bibinfo {author} {\bibfnamefont {A.N.}\
  \bibnamefont {Cleland}}, \ and\ \bibinfo {author} {\bibfnamefont {John~M.}\
  \bibnamefont {Martinis}},\ }\bibfield  {title} {\enquote {\bibinfo {title}
  {{Implementing the Quantum von Neumann Architecture with Superconducting
  Circuits}},}\ }\href@noop {} {\bibfield  {journal} {\bibinfo  {journal}
  {Science}\ }\textbf {\bibinfo {volume} {334}},\ \bibinfo {pages} {61--65}
  (\bibinfo {year} {2011})}\BibitemShut {NoStop}%
\bibitem [{\citenamefont {Cirac}\ \emph {et~al.}(1997)\citenamefont {Cirac},
  \citenamefont {Zoller}, \citenamefont {Kimble},\ and\ \citenamefont
  {Mabuchi}}]{Cirac1997}%
  \BibitemOpen
  \bibfield  {author} {\bibinfo {author} {\bibfnamefont {J.I.}\ \bibnamefont
  {Cirac}}, \bibinfo {author} {\bibfnamefont {P.}~\bibnamefont {Zoller}},
  \bibinfo {author} {\bibfnamefont {H.J.}\ \bibnamefont {Kimble}}, \ and\
  \bibinfo {author} {\bibfnamefont {H.}~\bibnamefont {Mabuchi}},\ }\bibfield
  {title} {\enquote {\bibinfo {title} {Quantum state transfer and entanglement
  distribution among distant nodes in a quantum network},}\ }\href@noop {}
  {\bibfield  {journal} {\bibinfo  {journal} {Phys. Rev. Lett.}\ }\textbf
  {\bibinfo {volume} {78}},\ \bibinfo {pages} {3221--3224} (\bibinfo {year}
  {1997})}\BibitemShut {NoStop}%
\bibitem [{\citenamefont {van Enk}\ \emph {et~al.}(1999)\citenamefont {van
  Enk}, \citenamefont {Kimble}, \citenamefont {Cirac},\ and\ \citenamefont
  {Zoller}}]{vanEnk1999}%
  \BibitemOpen
  \bibfield  {author} {\bibinfo {author} {\bibfnamefont {S.~J.}\ \bibnamefont
  {van Enk}}, \bibinfo {author} {\bibfnamefont {H.J.}\ \bibnamefont {Kimble}},
  \bibinfo {author} {\bibfnamefont {J.I.}\ \bibnamefont {Cirac}}, \ and\
  \bibinfo {author} {\bibfnamefont {P.}~\bibnamefont {Zoller}},\ }\bibfield
  {title} {\enquote {\bibinfo {title} {Quantum communication with dark
  photons},}\ }\href@noop {} {\bibfield  {journal} {\bibinfo  {journal} {Phys.
  Rev. A}\ }\textbf {\bibinfo {volume} {59}},\ \bibinfo {pages} {2659--2664}
  (\bibinfo {year} {1999})}\BibitemShut {NoStop}%
\bibitem [{\citenamefont {Steane}\ and\ \citenamefont
  {Lucas}(2000)}]{Steane2000}%
  \BibitemOpen
  \bibfield  {author} {\bibinfo {author} {\bibfnamefont {A.M.}\ \bibnamefont
  {Steane}}\ and\ \bibinfo {author} {\bibfnamefont {D.M.}\ \bibnamefont
  {Lucas}},\ }\bibfield  {title} {\enquote {\bibinfo {title} {Quantum computing
  with trapped ions, atoms and light},}\ }\href@noop {} {\bibfield  {journal}
  {\bibinfo  {journal} {Fortschritte der Physik}\ }\textbf {\bibinfo {volume}
  {48}},\ \bibinfo {pages} {839--858} (\bibinfo {year} {2000})}\BibitemShut
  {NoStop}%
\bibitem [{\citenamefont {Duan}\ \emph {et~al.}(2001)\citenamefont {Duan},
  \citenamefont {Lukin}, \citenamefont {Cirac},\ and\ \citenamefont
  {Zoller}}]{Duan01}%
  \BibitemOpen
  \bibfield  {author} {\bibinfo {author} {\bibfnamefont {L.-M.}\ \bibnamefont
  {Duan}}, \bibinfo {author} {\bibfnamefont {M.D.}\ \bibnamefont {Lukin}},
  \bibinfo {author} {\bibfnamefont {J.I.}\ \bibnamefont {Cirac}}, \ and\
  \bibinfo {author} {\bibfnamefont {P.}~\bibnamefont {Zoller}},\ }\bibfield
  {title} {\enquote {\bibinfo {title} {Long-distance quantum communication with
  atomic ensembles and linear optics},}\ }\href@noop {} {\bibfield  {journal}
  {\bibinfo  {journal} {Nature}\ }\textbf {\bibinfo {volume} {414}},\ \bibinfo
  {pages} {413--418} (\bibinfo {year} {2001})}\BibitemShut {NoStop}%
\bibitem [{\citenamefont {Van~Meter}\ \emph {et~al.}(2007)\citenamefont
  {Van~Meter}, \citenamefont {Nemoto},\ and\ \citenamefont
  {Munro}}]{VanMeter2007}%
  \BibitemOpen
  \bibfield  {author} {\bibinfo {author} {\bibfnamefont {R.}~\bibnamefont
  {Van~Meter}}, \bibinfo {author} {\bibfnamefont {K.}~\bibnamefont {Nemoto}}, \
  and\ \bibinfo {author} {\bibfnamefont {W.J.}\ \bibnamefont {Munro}},\
  }\bibfield  {title} {\enquote {\bibinfo {title} {Communication links for
  distributed quantum computation},}\ }\href@noop {} {\bibfield  {journal}
  {\bibinfo  {journal} {Computers, IEEE Transactions on}\ }\textbf {\bibinfo
  {volume} {56}},\ \bibinfo {pages} {1643--1653} (\bibinfo {year}
  {2007})}\BibitemShut {NoStop}%
\bibitem [{\citenamefont {Duan}\ and\ \citenamefont
  {Monroe}(2010)}]{DuanMonroe}%
  \BibitemOpen
  \bibfield  {author} {\bibinfo {author} {\bibfnamefont {L.-M.}\ \bibnamefont
  {Duan}}\ and\ \bibinfo {author} {\bibfnamefont {C.}~\bibnamefont {Monroe}},\
  }\bibfield  {title} {\enquote {\bibinfo {title} {Colloquium: Quantum networks
  with trapped ions},}\ }\href@noop {} {\bibfield  {journal} {\bibinfo
  {journal} {Rev. Mod. Phys.}\ }\textbf {\bibinfo {volume} {82}},\ \bibinfo
  {pages} {1209--1224} (\bibinfo {year} {2010})}\BibitemShut {NoStop}%
\bibitem [{\citenamefont {Kim}\ and\ \citenamefont {Kim}(2009)}]{KimKim}%
  \BibitemOpen
  \bibfield  {author} {\bibinfo {author} {\bibfnamefont {Jungsang}\
  \bibnamefont {Kim}}\ and\ \bibinfo {author} {\bibfnamefont {Changsoon}\
  \bibnamefont {Kim}},\ }\bibfield  {title} {\enquote {\bibinfo {title}
  {Integrated optical approach to trapped ion quantum computation},}\
  }\href@noop {} {\bibfield  {journal} {\bibinfo  {journal} {Quantum Info.
  Comput.}\ }\textbf {\bibinfo {volume} {9}},\ \bibinfo {pages} {181--202}
  (\bibinfo {year} {2009})}\BibitemShut {NoStop}%
\bibitem [{\citenamefont {Fowler}\ \emph {et~al.}(2007)\citenamefont {Fowler},
  \citenamefont {Thompson}, \citenamefont {Yan}, \citenamefont {Stephens},
  \citenamefont {Plourde},\ and\ \citenamefont {Wilhelm}}]{Fowler07}%
  \BibitemOpen
  \bibfield  {author} {\bibinfo {author} {\bibfnamefont {Austin~G.}\
  \bibnamefont {Fowler}}, \bibinfo {author} {\bibfnamefont {William~F.}\
  \bibnamefont {Thompson}}, \bibinfo {author} {\bibfnamefont {Zhizhong}\
  \bibnamefont {Yan}}, \bibinfo {author} {\bibfnamefont {Ashley~M.}\
  \bibnamefont {Stephens}}, \bibinfo {author} {\bibfnamefont {B.L.T.}\
  \bibnamefont {Plourde}}, \ and\ \bibinfo {author} {\bibfnamefont {Frank~K.}\
  \bibnamefont {Wilhelm}},\ }\bibfield  {title} {\enquote {\bibinfo {title}
  {Long-range coupling and scalable architecture for superconducting flux
  qubits},}\ }\href@noop {} {\bibfield  {journal} {\bibinfo  {journal} {Phys.
  Rev. B}\ }\textbf {\bibinfo {volume} {76}},\ \bibinfo {pages} {174507}
  (\bibinfo {year} {2007})}\BibitemShut {NoStop}%
\bibitem [{\citenamefont {Whitney}\ \emph {et~al.}(2007)\citenamefont
  {Whitney}, \citenamefont {Isailovic}, \citenamefont {Patel},\ and\
  \citenamefont {Kubiatowicz}}]{Whitney07}%
  \BibitemOpen
  \bibfield  {author} {\bibinfo {author} {\bibfnamefont {M.}~\bibnamefont
  {Whitney}}, \bibinfo {author} {\bibfnamefont {N.}~\bibnamefont {Isailovic}},
  \bibinfo {author} {\bibfnamefont {Y.}~\bibnamefont {Patel}}, \ and\ \bibinfo
  {author} {\bibfnamefont {J.}~\bibnamefont {Kubiatowicz}},\ }\bibfield
  {title} {\enquote {\bibinfo {title} {Automated generation of layout and
  control for quantum circuits},}\ }in\ \href@noop {} {\emph {\bibinfo
  {booktitle} {Proceedings of the 4th International Conference on Computing
  Frontiers}}}\ (\bibinfo  {publisher} {ACM Press New York, NY, USA},\ \bibinfo
  {year} {2007})\ pp.\ \bibinfo {pages} {83--94}\BibitemShut {NoStop}%
\bibitem [{\citenamefont {Whitney}\ \emph {et~al.}(2009)\citenamefont
  {Whitney}, \citenamefont {Isailovic}, \citenamefont {Patel},\ and\
  \citenamefont {Kubiatowicz}}]{Whitney09}%
  \BibitemOpen
  \bibfield  {author} {\bibinfo {author} {\bibfnamefont {M.}~\bibnamefont
  {Whitney}}, \bibinfo {author} {\bibfnamefont {N.}~\bibnamefont {Isailovic}},
  \bibinfo {author} {\bibfnamefont {Y.}~\bibnamefont {Patel}}, \ and\ \bibinfo
  {author} {\bibfnamefont {J.}~\bibnamefont {Kubiatowicz}},\ }\bibfield
  {title} {\enquote {\bibinfo {title} {A fault tolerant, area efficient
  architecture for {Shor's} factoring algorithm},}\ }in\ \href@noop {} {\emph
  {\bibinfo {booktitle} {36th International Symposium on Computer Architecture,
  2009 (ISCA'09)}}}\ (\bibinfo {year} {2009})\BibitemShut {NoStop}%
\bibitem [{\citenamefont {Isailovic}\ \emph {et~al.}(2006)\citenamefont
  {Isailovic}, \citenamefont {Patel}, \citenamefont {Whitney},\ and\
  \citenamefont {Kubiatowicz}}]{Isailovic06}%
  \BibitemOpen
  \bibfield  {author} {\bibinfo {author} {\bibfnamefont {N.}~\bibnamefont
  {Isailovic}}, \bibinfo {author} {\bibfnamefont {Y.}~\bibnamefont {Patel}},
  \bibinfo {author} {\bibfnamefont {M.}~\bibnamefont {Whitney}}, \ and\
  \bibinfo {author} {\bibfnamefont {J.}~\bibnamefont {Kubiatowicz}},\
  }\bibfield  {title} {\enquote {\bibinfo {title} {Interconnection networks for
  scalable quantum computers},}\ }in\ \href@noop {} {\emph {\bibinfo
  {booktitle} {33rd International Symposium on Computer Architecture, 2006
  (ISCA'06)}}}\ (\bibinfo {year} {2006})\ pp.\ \bibinfo {pages}
  {366--377}\BibitemShut {NoStop}%
\bibitem [{\citenamefont {Gottesman}\ and\ \citenamefont
  {Chuang}(1999)}]{Gottesman1999}%
  \BibitemOpen
  \bibfield  {author} {\bibinfo {author} {\bibfnamefont {Daniel}\ \bibnamefont
  {Gottesman}}\ and\ \bibinfo {author} {\bibfnamefont {Isaac~L.}\ \bibnamefont
  {Chuang}},\ }\bibfield  {title} {\enquote {\bibinfo {title} {Demonstrating
  the viability of universal quantum computation using teleportation and
  single-qubit operations},}\ }\href@noop {} {\bibfield  {journal} {\bibinfo
  {journal} {Nature}\ }\textbf {\bibinfo {volume} {402}},\ \bibinfo {pages}
  {390--393} (\bibinfo {year} {1999})}\BibitemShut {NoStop}%
\bibitem [{\citenamefont {Levy}(2001)}]{Levy2001}%
  \BibitemOpen
  \bibfield  {author} {\bibinfo {author} {\bibfnamefont {Jeremy}\ \bibnamefont
  {Levy}},\ }\bibfield  {title} {\enquote {\bibinfo {title}
  {Quantum-information processing with ferroelectrically coupled quantum
  dots},}\ }\href@noop {} {\bibfield  {journal} {\bibinfo  {journal} {Phys.
  Rev. A}\ }\textbf {\bibinfo {volume} {64}},\ \bibinfo {pages} {052306}
  (\bibinfo {year} {2001})}\BibitemShut {NoStop}%
\bibitem [{\citenamefont {Fowler}\ \emph {et~al.}(2004)\citenamefont {Fowler},
  \citenamefont {Devitt},\ and\ \citenamefont {Hollenberg}}]{Fowler2004}%
  \BibitemOpen
  \bibfield  {author} {\bibinfo {author} {\bibfnamefont {Austin~G.}\
  \bibnamefont {Fowler}}, \bibinfo {author} {\bibfnamefont {Simon~J.}\
  \bibnamefont {Devitt}}, \ and\ \bibinfo {author} {\bibfnamefont {Lloyd
  C.~L.}\ \bibnamefont {Hollenberg}},\ }\bibfield  {title} {\enquote {\bibinfo
  {title} {{Implementation of Shor's Algorithm on a Linear Nearest Neighbour
  Qubit Array}},}\ }\href@noop {} {\bibfield  {journal} {\bibinfo  {journal}
  {Quantum Info. Comput.}\ }\textbf {\bibinfo {volume} {4}},\ \bibinfo {pages}
  {237--251} (\bibinfo {year} {2004})}\BibitemShut {NoStop}%
\bibitem [{\citenamefont {Aliferis}\ and\ \citenamefont
  {Cross}(2007)}]{Aliferis2007b}%
  \BibitemOpen
  \bibfield  {author} {\bibinfo {author} {\bibfnamefont {Panos}\ \bibnamefont
  {Aliferis}}\ and\ \bibinfo {author} {\bibfnamefont {Andrew~W.}\ \bibnamefont
  {Cross}},\ }\bibfield  {title} {\enquote {\bibinfo {title} {{Subsystem Fault
  Tolerance with the Bacon-Shor Code}},}\ }\href@noop {} {\bibfield  {journal}
  {\bibinfo  {journal} {Phys. Rev. Lett.}\ }\textbf {\bibinfo {volume} {98}},\
  \bibinfo {pages} {220502} (\bibinfo {year} {2007})}\BibitemShut {NoStop}%
\bibitem [{\citenamefont {Levy}\ \emph {et~al.}(2009)\citenamefont {Levy},
  \citenamefont {Ganti}, \citenamefont {Phillips}, \citenamefont {Hamlet},
  \citenamefont {Landahl}, \citenamefont {Gurrieri}, \citenamefont {Carr},\
  and\ \citenamefont {Carroll}}]{Levy2009}%
  \BibitemOpen
  \bibfield  {author} {\bibinfo {author} {\bibfnamefont {James~E.}\
  \bibnamefont {Levy}}, \bibinfo {author} {\bibfnamefont {Anand}\ \bibnamefont
  {Ganti}}, \bibinfo {author} {\bibfnamefont {Cynthia~A.}\ \bibnamefont
  {Phillips}}, \bibinfo {author} {\bibfnamefont {Benjamin~R.}\ \bibnamefont
  {Hamlet}}, \bibinfo {author} {\bibfnamefont {Andrew~J.}\ \bibnamefont
  {Landahl}}, \bibinfo {author} {\bibfnamefont {Thomas~M.}\ \bibnamefont
  {Gurrieri}}, \bibinfo {author} {\bibfnamefont {Robert~D.}\ \bibnamefont
  {Carr}}, \ and\ \bibinfo {author} {\bibfnamefont {Malcolm~S.}\ \bibnamefont
  {Carroll}},\ }\href@noop {} {\enquote {\bibinfo {title} {The impact of
  classical electronics constraints on a solid-state logical qubit memory},}\ }
  (\bibinfo {year} {2009}),\ \bibinfo {note} {\emph{Preprint}
  arXiv:0904.0003v1}\BibitemShut {NoStop}%
\bibitem [{\citenamefont {Levy}\ \emph {et~al.}(2011)\citenamefont {Levy},
  \citenamefont {Carroll}, \citenamefont {Ganti}, \citenamefont {Phillips},
  \citenamefont {Landahl}, \citenamefont {Gurrieri}, \citenamefont {Carr},
  \citenamefont {Stalford},\ and\ \citenamefont {Nielsen}}]{Levy2011}%
  \BibitemOpen
  \bibfield  {author} {\bibinfo {author} {\bibfnamefont {James~E.}\
  \bibnamefont {Levy}}, \bibinfo {author} {\bibfnamefont {Malcolm~S.}\
  \bibnamefont {Carroll}}, \bibinfo {author} {\bibfnamefont {Anand}\
  \bibnamefont {Ganti}}, \bibinfo {author} {\bibfnamefont {Cynthia~A.}\
  \bibnamefont {Phillips}}, \bibinfo {author} {\bibfnamefont {Andrew~J.}\
  \bibnamefont {Landahl}}, \bibinfo {author} {\bibfnamefont {Thomas~M.}\
  \bibnamefont {Gurrieri}}, \bibinfo {author} {\bibfnamefont {Robert~D.}\
  \bibnamefont {Carr}}, \bibinfo {author} {\bibfnamefont {Harold~L.}\
  \bibnamefont {Stalford}}, \ and\ \bibinfo {author} {\bibfnamefont {Erik}\
  \bibnamefont {Nielsen}},\ }\bibfield  {title} {\enquote {\bibinfo {title}
  {Implications of electronics constraints for solid-state quantum error
  correction and quantum circuit failure probability},}\ }\href@noop {}
  {\bibfield  {journal} {\bibinfo  {journal} {New J. Phys.}\ }\textbf {\bibinfo
  {volume} {13}},\ \bibinfo {pages} {083021} (\bibinfo {year}
  {2011})}\BibitemShut {NoStop}%
\bibitem [{\citenamefont {Weinstein}\ \emph {et~al.}(2005)\citenamefont
  {Weinstein}, \citenamefont {Hellberg},\ and\ \citenamefont
  {Levy}}]{Weinstein2005}%
  \BibitemOpen
  \bibfield  {author} {\bibinfo {author} {\bibfnamefont {Yaakov~S.}\
  \bibnamefont {Weinstein}}, \bibinfo {author} {\bibfnamefont {C.~Stephen}\
  \bibnamefont {Hellberg}}, \ and\ \bibinfo {author} {\bibfnamefont {Jeremy}\
  \bibnamefont {Levy}},\ }\bibfield  {title} {\enquote {\bibinfo {title}
  {Quantum-dot cluster-state computing with encoded qubits},}\ }\href@noop {}
  {\bibfield  {journal} {\bibinfo  {journal} {Phys. Rev. A}\ }\textbf {\bibinfo
  {volume} {72}},\ \bibinfo {pages} {020304} (\bibinfo {year}
  {2005})}\BibitemShut {NoStop}%
\bibitem [{\citenamefont {Stock}\ and\ \citenamefont {James}(2009)}]{Stock09}%
  \BibitemOpen
  \bibfield  {author} {\bibinfo {author} {\bibfnamefont {Ren\'e}\ \bibnamefont
  {Stock}}\ and\ \bibinfo {author} {\bibfnamefont {Daniel F.~V.}\ \bibnamefont
  {James}},\ }\bibfield  {title} {\enquote {\bibinfo {title} {Scalable,
  high-speed measurement-based quantum computer using trapped ions},}\
  }\href@noop {} {\bibfield  {journal} {\bibinfo  {journal} {Phys. Rev. Lett.}\
  }\textbf {\bibinfo {volume} {102}},\ \bibinfo {pages} {170501} (\bibinfo
  {year} {2009})}\BibitemShut {NoStop}%
\bibitem [{\citenamefont {Devitt}\ \emph {et~al.}(2010)\citenamefont {Devitt},
  \citenamefont {Fowler}, \citenamefont {Tilma}, \citenamefont {Munro},\ and\
  \citenamefont {Nemoto}}]{Devitt10}%
  \BibitemOpen
  \bibfield  {author} {\bibinfo {author} {\bibfnamefont {Simon~J.}\
  \bibnamefont {Devitt}}, \bibinfo {author} {\bibfnamefont {Austin~G.}\
  \bibnamefont {Fowler}}, \bibinfo {author} {\bibfnamefont {Todd}\ \bibnamefont
  {Tilma}}, \bibinfo {author} {\bibfnamefont {W.~J.}\ \bibnamefont {Munro}}, \
  and\ \bibinfo {author} {\bibfnamefont {Kae}\ \bibnamefont {Nemoto}},\
  }\bibfield  {title} {\enquote {\bibinfo {title} {Classical processing
  requirements for a topological quantum computing system},}\ }\href@noop {}
  {\bibfield  {journal} {\bibinfo  {journal} {International Journal of Quantum
  Information}\ }\textbf {\bibinfo {volume} {8}},\ \bibinfo {pages} {121--147}
  (\bibinfo {year} {2010})}\BibitemShut {NoStop}%
\bibitem [{\citenamefont {Devitt}\ \emph
  {et~al.}(2011{\natexlab{a}})\citenamefont {Devitt}, \citenamefont {Stephens},
  \citenamefont {Munro},\ and\ \citenamefont {Nemoto}}]{Devitt2011}%
  \BibitemOpen
  \bibfield  {author} {\bibinfo {author} {\bibfnamefont {Simon~J}\ \bibnamefont
  {Devitt}}, \bibinfo {author} {\bibfnamefont {Ashley~M}\ \bibnamefont
  {Stephens}}, \bibinfo {author} {\bibfnamefont {William~J}\ \bibnamefont
  {Munro}}, \ and\ \bibinfo {author} {\bibfnamefont {Kae}\ \bibnamefont
  {Nemoto}},\ }\bibfield  {title} {\enquote {\bibinfo {title} {Integration of
  highly probabilistic sources into optical quantum architectures: perpetual
  quantum computation},}\ }\href@noop {} {\bibfield  {journal} {\bibinfo
  {journal} {New J. Phys.}\ }\textbf {\bibinfo {volume} {13}},\ \bibinfo
  {pages} {095001} (\bibinfo {year} {2011}{\natexlab{a}})}\BibitemShut
  {NoStop}%
\bibitem [{\citenamefont {Kitaev}\ \emph {et~al.}(2002)\citenamefont {Kitaev},
  \citenamefont {Shen},\ and\ \citenamefont {Vyalyi}}]{Kitaev2002}%
  \BibitemOpen
  \bibfield  {author} {\bibinfo {author} {\bibfnamefont {Alexei~Yu.}\
  \bibnamefont {Kitaev}}, \bibinfo {author} {\bibfnamefont {Alexander~H.}\
  \bibnamefont {Shen}}, \ and\ \bibinfo {author} {\bibfnamefont {Mikhail~N.}\
  \bibnamefont {Vyalyi}},\ }\href@noop {} {\emph {\bibinfo {title} {Classical
  and Quantum Computation}}},\ \bibinfo {edition} {1st}\ ed.\ (\bibinfo
  {publisher} {American Mathematical Society},\ \bibinfo {year}
  {2002})\BibitemShut {NoStop}%
\bibitem [{\citenamefont {Oskin}\ \emph {et~al.}(2002)\citenamefont {Oskin},
  \citenamefont {Chong},\ and\ \citenamefont {Chuang}}]{Oskin2002}%
  \BibitemOpen
  \bibfield  {author} {\bibinfo {author} {\bibfnamefont {Mark}\ \bibnamefont
  {Oskin}}, \bibinfo {author} {\bibfnamefont {Frederic~T.}\ \bibnamefont
  {Chong}}, \ and\ \bibinfo {author} {\bibfnamefont {Isaac~L.}\ \bibnamefont
  {Chuang}},\ }\bibfield  {title} {\enquote {\bibinfo {title} {A practical
  architecture for reliable quantum computers},}\ }\href@noop {} {\bibfield
  {journal} {\bibinfo  {journal} {IEEE Computer}\ }\textbf {\bibinfo {volume}
  {35}},\ \bibinfo {pages} {79--87} (\bibinfo {year} {2002})}\BibitemShut
  {NoStop}%
\bibitem [{\citenamefont {Devitt}\ \emph
  {et~al.}(2011{\natexlab{b}})\citenamefont {Devitt}, \citenamefont {Munro},\
  and\ \citenamefont {Nemoto}}]{Devitt2011b}%
  \BibitemOpen
  \bibfield  {author} {\bibinfo {author} {\bibfnamefont {Simon~J.}\
  \bibnamefont {Devitt}}, \bibinfo {author} {\bibfnamefont {William~J.}\
  \bibnamefont {Munro}}, \ and\ \bibinfo {author} {\bibfnamefont {Kae}\
  \bibnamefont {Nemoto}},\ }\bibfield  {title} {\enquote {\bibinfo {title}
  {High performance quantum computing},}\ }\href@noop {} {\bibfield  {journal}
  {\bibinfo  {journal} {Progress in Informatics}\ }\textbf {\bibinfo {volume}
  {8}},\ \bibinfo {pages} {1--7} (\bibinfo {year} {2011}{\natexlab{b}})},\
  \bibinfo {note} {\emph{Preprint} arXiv:0810.2444v1.}\BibitemShut {Stop}%
\bibitem [{\citenamefont {Van~Meter}\ \emph {et~al.}(2006)\citenamefont
  {Van~Meter}, \citenamefont {Nemoto}, \citenamefont {Munro},\ and\
  \citenamefont {Itoh}}]{VanMeter06b}%
  \BibitemOpen
  \bibfield  {author} {\bibinfo {author} {\bibfnamefont {Rodney}\ \bibnamefont
  {Van~Meter}}, \bibinfo {author} {\bibfnamefont {Kae}\ \bibnamefont {Nemoto}},
  \bibinfo {author} {\bibfnamefont {W.~J.}\ \bibnamefont {Munro}}, \ and\
  \bibinfo {author} {\bibfnamefont {Kohei~M.}\ \bibnamefont {Itoh}},\
  }\bibfield  {title} {\enquote {\bibinfo {title} {Distributed arithmetic on a
  quantum multicomputer},}\ }\href@noop {} {\bibfield  {journal} {\bibinfo
  {journal} {SIGARCH Comput. Archit. News}\ }\textbf {\bibinfo {volume} {34}},\
  \bibinfo {pages} {354--365} (\bibinfo {year} {2006})}\BibitemShut {NoStop}%
\bibitem [{\citenamefont {Van{ }Meter}\ \emph {et~al.}(2010)\citenamefont {Van{
  }Meter}, \citenamefont {Ladd}, \citenamefont {Fowler},\ and\ \citenamefont
  {Yamamoto}}]{VanMeter09}%
  \BibitemOpen
  \bibfield  {author} {\bibinfo {author} {\bibfnamefont {Rodney}\ \bibnamefont
  {Van{ }Meter}}, \bibinfo {author} {\bibfnamefont {Thaddeus~D.}\ \bibnamefont
  {Ladd}}, \bibinfo {author} {\bibfnamefont {Austin~G.}\ \bibnamefont
  {Fowler}}, \ and\ \bibinfo {author} {\bibfnamefont {Yoshihisa}\ \bibnamefont
  {Yamamoto}},\ }\bibfield  {title} {\enquote {\bibinfo {title} {Distributed
  quantum computation architecture using semiconductor nanophotonics},}\
  }\href@noop {} {\bibfield  {journal} {\bibinfo  {journal} {International
  Journal of Quantum Information}\ }\textbf {\bibinfo {volume} {8}},\ \bibinfo
  {pages} {295--323} (\bibinfo {year} {2010})},\ \bibinfo {note}
  {\emph{Preprint} arXiv:quant-ph/0906.2686v2.}\BibitemShut {Stop}%
\bibitem [{\citenamefont {Preskill}(1997)}]{Preskill97}%
  \BibitemOpen
  \bibfield  {author} {\bibinfo {author} {\bibfnamefont {John}\ \bibnamefont
  {Preskill}},\ }\href@noop {} {\enquote {\bibinfo {title} {Fault-tolerant
  quantum computation},}\ } (\bibinfo {year} {1997}),\ \bibinfo {note}
  {\emph{Preprint} arXiv:quant-ph/9712048.}\BibitemShut {Stop}%
\bibitem [{\citenamefont {Shen}\ and\ \citenamefont {Lipasti}(2005)}]{Shen05}%
  \BibitemOpen
  \bibfield  {author} {\bibinfo {author} {\bibfnamefont {John~Paul}\
  \bibnamefont {Shen}}\ and\ \bibinfo {author} {\bibfnamefont {Mikko~H.}\
  \bibnamefont {Lipasti}},\ }\href@noop {} {\emph {\bibinfo {title} {Modern
  Processor Design: Fundamentals of Superscalar Processors}}}\ (\bibinfo
  {publisher} {McGraw-Hill Higher Education},\ \bibinfo {year}
  {2005})\BibitemShut {NoStop}%
\bibitem [{\citenamefont {Bj\"ork}\ \emph {et~al.}(1994)\citenamefont
  {Bj\"ork}, \citenamefont {Pau}, \citenamefont {Jacobson},\ and\ \citenamefont
  {Yamamoto}}]{Bjork94}%
  \BibitemOpen
  \bibfield  {author} {\bibinfo {author} {\bibfnamefont {Gunnar}\ \bibnamefont
  {Bj\"ork}}, \bibinfo {author} {\bibfnamefont {Stanley}\ \bibnamefont {Pau}},
  \bibinfo {author} {\bibfnamefont {Joseph}\ \bibnamefont {Jacobson}}, \ and\
  \bibinfo {author} {\bibfnamefont {Yoshihisa}\ \bibnamefont {Yamamoto}},\
  }\bibfield  {title} {\enquote {\bibinfo {title} {Wannier exciton
  superradiance in a quantum-well microcavity},}\ }\href@noop {} {\bibfield
  {journal} {\bibinfo  {journal} {Phys. Rev. B}\ }\textbf {\bibinfo {volume}
  {50}},\ \bibinfo {pages} {17336--17348} (\bibinfo {year} {1994})}\BibitemShut
  {NoStop}%
\bibitem [{\citenamefont {{Imamo\={g}lu}}\ \emph {et~al.}(1999)\citenamefont
  {{Imamo\={g}lu}}, \citenamefont {Awschalom}, \citenamefont {Burkard},
  \citenamefont {DiVincenzo}, \citenamefont {Loss}, \citenamefont {Sherwin},\
  and\ \citenamefont {Small}}]{Imamoglu99}%
  \BibitemOpen
  \bibfield  {author} {\bibinfo {author} {\bibfnamefont {A.}~\bibnamefont
  {{Imamo\={g}lu}}}, \bibinfo {author} {\bibfnamefont {D.D.}\ \bibnamefont
  {Awschalom}}, \bibinfo {author} {\bibfnamefont {G.}~\bibnamefont {Burkard}},
  \bibinfo {author} {\bibfnamefont {D.P.}\ \bibnamefont {DiVincenzo}}, \bibinfo
  {author} {\bibfnamefont {D.}~\bibnamefont {Loss}}, \bibinfo {author}
  {\bibfnamefont {M.}~\bibnamefont {Sherwin}}, \ and\ \bibinfo {author}
  {\bibfnamefont {A.}~\bibnamefont {Small}},\ }\bibfield  {title} {\enquote
  {\bibinfo {title} {{Quantum Information Processing Using Quantum Dot Spins
  and Cavity QED}},}\ }\href@noop {} {\bibfield  {journal} {\bibinfo  {journal}
  {Phys. Rev. Lett.}\ }\textbf {\bibinfo {volume} {83}},\ \bibinfo {pages}
  {4204--4207} (\bibinfo {year} {1999})}\BibitemShut {NoStop}%
\bibitem [{\citenamefont {Bonadeo}\ \emph {et~al.}(2000)\citenamefont
  {Bonadeo}, \citenamefont {Chen}, \citenamefont {Gammon},\ and\ \citenamefont
  {Steel}}]{Bonadeo00}%
  \BibitemOpen
  \bibfield  {author} {\bibinfo {author} {\bibfnamefont {N.H.}\ \bibnamefont
  {Bonadeo}}, \bibinfo {author} {\bibfnamefont {Gang}\ \bibnamefont {Chen}},
  \bibinfo {author} {\bibfnamefont {D.}~\bibnamefont {Gammon}}, \ and\ \bibinfo
  {author} {\bibfnamefont {D.G.}\ \bibnamefont {Steel}},\ }\bibfield  {title}
  {\enquote {\bibinfo {title} {Single quantum dot nonlinear optical
  spectroscopy},}\ }\href@noop {} {\bibfield  {journal} {\bibinfo  {journal}
  {Physica Status Solidi B}\ }\textbf {\bibinfo {volume} {221}},\ \bibinfo
  {pages} {5--18} (\bibinfo {year} {2000})}\BibitemShut {NoStop}%
\bibitem [{\citenamefont {Guest}\ \emph {et~al.}(2002)\citenamefont {Guest},
  \citenamefont {Stievater}, \citenamefont {Li}, \citenamefont {Cheng},
  \citenamefont {Steel}, \citenamefont {Gammon}, \citenamefont {Katzer},
  \citenamefont {Park}, \citenamefont {Ell}, \citenamefont {Thr\"anhardt},
  \citenamefont {Khitrova},\ and\ \citenamefont {Gibbs}}]{Guest02}%
  \BibitemOpen
  \bibfield  {author} {\bibinfo {author} {\bibfnamefont {J.R.}\ \bibnamefont
  {Guest}}, \bibinfo {author} {\bibfnamefont {T.H.}\ \bibnamefont {Stievater}},
  \bibinfo {author} {\bibfnamefont {Xiaoqin}\ \bibnamefont {Li}}, \bibinfo
  {author} {\bibfnamefont {Jun}\ \bibnamefont {Cheng}}, \bibinfo {author}
  {\bibfnamefont {D.G.}\ \bibnamefont {Steel}}, \bibinfo {author}
  {\bibfnamefont {D.}~\bibnamefont {Gammon}}, \bibinfo {author} {\bibfnamefont
  {D.S.}\ \bibnamefont {Katzer}}, \bibinfo {author} {\bibfnamefont
  {D.}~\bibnamefont {Park}}, \bibinfo {author} {\bibfnamefont {C.}~\bibnamefont
  {Ell}}, \bibinfo {author} {\bibfnamefont {A.}~\bibnamefont {Thr\"anhardt}},
  \bibinfo {author} {\bibfnamefont {G.}~\bibnamefont {Khitrova}}, \ and\
  \bibinfo {author} {\bibfnamefont {H.M.}\ \bibnamefont {Gibbs}},\ }\bibfield
  {title} {\enquote {\bibinfo {title} {Measurement of optical absorption by a
  single quantum dot exciton},}\ }\href@noop {} {\bibfield  {journal} {\bibinfo
   {journal} {Phys. Rev. B}\ }\textbf {\bibinfo {volume} {65}},\ \bibinfo
  {pages} {241310} (\bibinfo {year} {2002})}\BibitemShut {NoStop}%
\bibitem [{\citenamefont {Hours}\ \emph {et~al.}(2005)\citenamefont {Hours},
  \citenamefont {Senellart}, \citenamefont {Peter}, \citenamefont {Cavanna},\
  and\ \citenamefont {Bloch}}]{Hours05}%
  \BibitemOpen
  \bibfield  {author} {\bibinfo {author} {\bibfnamefont {J.}~\bibnamefont
  {Hours}}, \bibinfo {author} {\bibfnamefont {P.}~\bibnamefont {Senellart}},
  \bibinfo {author} {\bibfnamefont {E.}~\bibnamefont {Peter}}, \bibinfo
  {author} {\bibfnamefont {A.}~\bibnamefont {Cavanna}}, \ and\ \bibinfo
  {author} {\bibfnamefont {J.}~\bibnamefont {Bloch}},\ }\bibfield  {title}
  {\enquote {\bibinfo {title} {Exciton radiative lifetime controlled by the
  lateral confinement energy in a single quantum dot},}\ }\href@noop {}
  {\bibfield  {journal} {\bibinfo  {journal} {Phys. Rev. B}\ }\textbf {\bibinfo
  {volume} {71}},\ \bibinfo {pages} {161306} (\bibinfo {year}
  {2005})}\BibitemShut {NoStop}%
\bibitem [{\citenamefont {Yamamoto}\ \emph {et~al.}(2009)\citenamefont
  {Yamamoto}, \citenamefont {Ladd}, \citenamefont {Press}, \citenamefont
  {Clark}, \citenamefont {Sanaka}, \citenamefont {Santori}, \citenamefont
  {Fattal}, \citenamefont {Fu}, \citenamefont {H\"{o}fling}, \citenamefont
  {Reitzenstein},\ and\ \citenamefont {Forchel}}]{Yamamoto09}%
  \BibitemOpen
  \bibfield  {author} {\bibinfo {author} {\bibfnamefont {Y.}~\bibnamefont
  {Yamamoto}}, \bibinfo {author} {\bibfnamefont {T.D.}\ \bibnamefont {Ladd}},
  \bibinfo {author} {\bibfnamefont {D.}~\bibnamefont {Press}}, \bibinfo
  {author} {\bibfnamefont {S.}~\bibnamefont {Clark}}, \bibinfo {author}
  {\bibfnamefont {K.}~\bibnamefont {Sanaka}}, \bibinfo {author} {\bibfnamefont
  {C.}~\bibnamefont {Santori}}, \bibinfo {author} {\bibfnamefont
  {D.}~\bibnamefont {Fattal}}, \bibinfo {author} {\bibfnamefont {K.-M.}\
  \bibnamefont {Fu}}, \bibinfo {author} {\bibfnamefont {S.}~\bibnamefont
  {H\"{o}fling}}, \bibinfo {author} {\bibfnamefont {S.}~\bibnamefont
  {Reitzenstein}}, \ and\ \bibinfo {author} {\bibfnamefont {A.}~\bibnamefont
  {Forchel}},\ }\bibfield  {title} {\enquote {\bibinfo {title} {Optically
  controlled semiconductor spin qubits for quantum information processing},}\
  }\href@noop {} {\bibfield  {journal} {\bibinfo  {journal} {Physica Scripta}\
  }\textbf {\bibinfo {volume} {2009}},\ \bibinfo {pages} {014010} (\bibinfo
  {year} {2009})}\BibitemShut {NoStop}%
\bibitem [{\citenamefont {Bayer}\ \emph {et~al.}(2002)\citenamefont {Bayer},
  \citenamefont {Ortner}, \citenamefont {Stern}, \citenamefont {Kuther},
  \citenamefont {Gorbunov}, \citenamefont {Forchel}, \citenamefont {Hawrylak},
  \citenamefont {Fafard}, \citenamefont {Hinzer}, \citenamefont {Reinecke},
  \citenamefont {Walck}, \citenamefont {Reithmaier}, \citenamefont {Klopf},\
  and\ \citenamefont {Sch\"afer}}]{Bayer02}%
  \BibitemOpen
  \bibfield  {author} {\bibinfo {author} {\bibfnamefont {M.}~\bibnamefont
  {Bayer}}, \bibinfo {author} {\bibfnamefont {G.}~\bibnamefont {Ortner}},
  \bibinfo {author} {\bibfnamefont {O.}~\bibnamefont {Stern}}, \bibinfo
  {author} {\bibfnamefont {A.}~\bibnamefont {Kuther}}, \bibinfo {author}
  {\bibfnamefont {A.~A.}\ \bibnamefont {Gorbunov}}, \bibinfo {author}
  {\bibfnamefont {A.}~\bibnamefont {Forchel}}, \bibinfo {author} {\bibfnamefont
  {P.}~\bibnamefont {Hawrylak}}, \bibinfo {author} {\bibfnamefont
  {S.}~\bibnamefont {Fafard}}, \bibinfo {author} {\bibfnamefont
  {K.}~\bibnamefont {Hinzer}}, \bibinfo {author} {\bibfnamefont {T.~L.}\
  \bibnamefont {Reinecke}}, \bibinfo {author} {\bibfnamefont {S.~N.}\
  \bibnamefont {Walck}}, \bibinfo {author} {\bibfnamefont {J.~P.}\ \bibnamefont
  {Reithmaier}}, \bibinfo {author} {\bibfnamefont {F.}~\bibnamefont {Klopf}}, \
  and\ \bibinfo {author} {\bibfnamefont {F.}~\bibnamefont {Sch\"afer}},\
  }\bibfield  {title} {\enquote {\bibinfo {title} {Fine structure of neutral
  and charged excitons in self-assembled {In(Ga)As/(Al)GaAs} quantum dots},}\
  }\href@noop {} {\bibfield  {journal} {\bibinfo  {journal} {Phys. Rev. B}\
  }\textbf {\bibinfo {volume} {65}},\ \bibinfo {pages} {195315} (\bibinfo
  {year} {2002})}\BibitemShut {NoStop}%
\bibitem [{\citenamefont {Reitzenstein}\ \emph {et~al.}(2007)\citenamefont
  {Reitzenstein}, \citenamefont {Hofmann}, \citenamefont {Gorbunov},
  \citenamefont {Strau\ss{}}, \citenamefont {Kwon}, \citenamefont {Schneider},
  \citenamefont {L\"offler}, \citenamefont {H\"ofling}, \citenamefont {Kamp},\
  and\ \citenamefont {Forchel}}]{Reitz07}%
  \BibitemOpen
  \bibfield  {author} {\bibinfo {author} {\bibfnamefont {S.}~\bibnamefont
  {Reitzenstein}}, \bibinfo {author} {\bibfnamefont {C.}~\bibnamefont
  {Hofmann}}, \bibinfo {author} {\bibfnamefont {A.}~\bibnamefont {Gorbunov}},
  \bibinfo {author} {\bibfnamefont {M.}~\bibnamefont {Strau\ss{}}}, \bibinfo
  {author} {\bibfnamefont {S.H.}\ \bibnamefont {Kwon}}, \bibinfo {author}
  {\bibfnamefont {C.}~\bibnamefont {Schneider}}, \bibinfo {author}
  {\bibfnamefont {A.}~\bibnamefont {L\"offler}}, \bibinfo {author}
  {\bibfnamefont {S.}~\bibnamefont {H\"ofling}}, \bibinfo {author}
  {\bibfnamefont {M.}~\bibnamefont {Kamp}}, \ and\ \bibinfo {author}
  {\bibfnamefont {A.}~\bibnamefont {Forchel}},\ }\bibfield  {title} {\enquote
  {\bibinfo {title} {{AlAs/GaAs} micropillar cavities with quality factors
  exceeding 150.000},}\ }\href@noop {} {\bibfield  {journal} {\bibinfo
  {journal} {Appl. Phys. Lett.}\ }\textbf {\bibinfo {volume} {90}},\ \bibinfo
  {pages} {251109} (\bibinfo {year} {2007})}\BibitemShut {NoStop}%
\bibitem [{\citenamefont {Nielsen}\ and\ \citenamefont
  {Chuang}(2000)}]{Nielsen00}%
  \BibitemOpen
  \bibfield  {author} {\bibinfo {author} {\bibfnamefont {Michael~A.}\
  \bibnamefont {Nielsen}}\ and\ \bibinfo {author} {\bibfnamefont {Isaac~L.}\
  \bibnamefont {Chuang}},\ }\href@noop {} {\emph {\bibinfo {title} {Quantum
  Computation and Quantum Information}}},\ \bibinfo {edition} {1st}\ ed.\
  (\bibinfo  {publisher} {Cambridge University Press},\ \bibinfo {year}
  {2000})\BibitemShut {NoStop}%
\bibitem [{\citenamefont {Press}\ \emph {et~al.}(2008)\citenamefont {Press},
  \citenamefont {Ladd}, \citenamefont {Zhang},\ and\ \citenamefont
  {Yamamoto}}]{Press08}%
  \BibitemOpen
  \bibfield  {author} {\bibinfo {author} {\bibfnamefont {David}\ \bibnamefont
  {Press}}, \bibinfo {author} {\bibfnamefont {Thaddeus~D.}\ \bibnamefont
  {Ladd}}, \bibinfo {author} {\bibfnamefont {Bingyang}\ \bibnamefont {Zhang}},
  \ and\ \bibinfo {author} {\bibfnamefont {Yoshihisa}\ \bibnamefont
  {Yamamoto}},\ }\bibfield  {title} {\enquote {\bibinfo {title} {Complete
  quantum control of a single quantum dot spin using ultrafast optical
  pulses},}\ }\href@noop {} {\bibfield  {journal} {\bibinfo  {journal}
  {Nature}\ }\textbf {\bibinfo {volume} {456}},\ \bibinfo {pages} {218--221}
  (\bibinfo {year} {2008})}\BibitemShut {NoStop}%
\bibitem [{\citenamefont {Greve}\ \emph {et~al.}()\citenamefont {Greve},
  \citenamefont {McMahon}, \citenamefont {Jones} \emph {et~al.}}]{DeGreve2012}%
  \BibitemOpen
  \bibfield  {author} {\bibinfo {author} {\bibfnamefont {Kristiaan~De}\
  \bibnamefont {Greve}}, \bibinfo {author} {\bibfnamefont {Peter~L.}\
  \bibnamefont {McMahon}}, \bibinfo {author} {\bibfnamefont {N.~Cody}\
  \bibnamefont {Jones}},  \emph {et~al.},\ }\href@noop {} {}\bibinfo {note}
  {\emph{In preparation.}}\BibitemShut {Stop}%
\bibitem [{\citenamefont {Economou}\ \emph {et~al.}(2006)\citenamefont
  {Economou}, \citenamefont {Sham}, \citenamefont {Wu},\ and\ \citenamefont
  {Steel}}]{Economou2006}%
  \BibitemOpen
  \bibfield  {author} {\bibinfo {author} {\bibfnamefont {Sophia~E.}\
  \bibnamefont {Economou}}, \bibinfo {author} {\bibfnamefont {L.J.}\
  \bibnamefont {Sham}}, \bibinfo {author} {\bibfnamefont {Yanwen}\ \bibnamefont
  {Wu}}, \ and\ \bibinfo {author} {\bibfnamefont {D.G.}\ \bibnamefont
  {Steel}},\ }\bibfield  {title} {\enquote {\bibinfo {title} {Proposal for
  optical {U(1)} rotations of electron spin trapped in a quantum dot},}\
  }\href@noop {} {\bibfield  {journal} {\bibinfo  {journal} {Phys. Rev. B}\
  }\textbf {\bibinfo {volume} {74}},\ \bibinfo {pages} {205415} (\bibinfo
  {year} {2006})}\BibitemShut {NoStop}%
\bibitem [{\citenamefont {Clark}\ \emph {et~al.}(2007)\citenamefont {Clark},
  \citenamefont {Fu}, \citenamefont {Ladd},\ and\ \citenamefont
  {Yamamoto}}]{Clark07}%
  \BibitemOpen
  \bibfield  {author} {\bibinfo {author} {\bibfnamefont {Susan~M.}\
  \bibnamefont {Clark}}, \bibinfo {author} {\bibfnamefont {Kai-Mei~C.}\
  \bibnamefont {Fu}}, \bibinfo {author} {\bibfnamefont {Thaddeus~D.}\
  \bibnamefont {Ladd}}, \ and\ \bibinfo {author} {\bibfnamefont {Yoshihisa}\
  \bibnamefont {Yamamoto}},\ }\bibfield  {title} {\enquote {\bibinfo {title}
  {Quantum computers based on electron spins controlled by ultrafast
  off-resonant single optical pulses},}\ }\href@noop {} {\bibfield  {journal}
  {\bibinfo  {journal} {Phys. Rev. Lett.}\ }\textbf {\bibinfo {volume} {99}},\
  \bibinfo {pages} {040501} (\bibinfo {year} {2007})}\BibitemShut {NoStop}%
\bibitem [{\citenamefont {Ladd}\ and\ \citenamefont
  {Yamamoto}(2011)}]{Ladd2011}%
  \BibitemOpen
  \bibfield  {author} {\bibinfo {author} {\bibfnamefont {T.~D.}\ \bibnamefont
  {Ladd}}\ and\ \bibinfo {author} {\bibfnamefont {Y.}~\bibnamefont
  {Yamamoto}},\ }\bibfield  {title} {\enquote {\bibinfo {title} {Simple quantum
  logic gate with quantum dot cavity {QED} systems},}\ }\href@noop {}
  {\bibfield  {journal} {\bibinfo  {journal} {Phys. Rev. B}\ }\textbf {\bibinfo
  {volume} {84}},\ \bibinfo {pages} {235307} (\bibinfo {year}
  {2011})}\BibitemShut {NoStop}%
\bibitem [{\citenamefont {Quinteiro}\ \emph {et~al.}(2006)\citenamefont
  {Quinteiro}, \citenamefont {Fern\'andez-Rossier},\ and\ \citenamefont
  {Piermarocchi}}]{polariton_mediate}%
  \BibitemOpen
  \bibfield  {author} {\bibinfo {author} {\bibfnamefont {G.~F.}\ \bibnamefont
  {Quinteiro}}, \bibinfo {author} {\bibfnamefont {J.}~\bibnamefont
  {Fern\'andez-Rossier}}, \ and\ \bibinfo {author} {\bibfnamefont
  {C.}~\bibnamefont {Piermarocchi}},\ }\bibfield  {title} {\enquote {\bibinfo
  {title} {Long-range spin-qubit interaction mediated by microcavity
  polaritons},}\ }\href@noop {} {\bibfield  {journal} {\bibinfo  {journal}
  {Phys. Rev. Lett.}\ }\textbf {\bibinfo {volume} {97}},\ \bibinfo {pages}
  {097401} (\bibinfo {year} {2006})}\BibitemShut {NoStop}%
\bibitem [{\citenamefont {Berezovsky}\ \emph {et~al.}(2006)\citenamefont
  {Berezovsky}, \citenamefont {Mikkelsen}, \citenamefont {Gywat}, \citenamefont
  {Stoltz}, \citenamefont {Coldren},\ and\ \citenamefont
  {Awschalom}}]{Berez06}%
  \BibitemOpen
  \bibfield  {author} {\bibinfo {author} {\bibfnamefont {J.}~\bibnamefont
  {Berezovsky}}, \bibinfo {author} {\bibfnamefont {M.H.}\ \bibnamefont
  {Mikkelsen}}, \bibinfo {author} {\bibfnamefont {O.}~\bibnamefont {Gywat}},
  \bibinfo {author} {\bibfnamefont {N.G.}\ \bibnamefont {Stoltz}}, \bibinfo
  {author} {\bibfnamefont {L.A.}\ \bibnamefont {Coldren}}, \ and\ \bibinfo
  {author} {\bibfnamefont {D.D.}\ \bibnamefont {Awschalom}},\ }\bibfield
  {title} {\enquote {\bibinfo {title} {Nondestructive optical measurements of a
  single electron spin in a quantum dot},}\ }\href@noop {} {\bibfield
  {journal} {\bibinfo  {journal} {Science}\ }\textbf {\bibinfo {volume}
  {314}},\ \bibinfo {pages} {1916--1920} (\bibinfo {year} {2006})}\BibitemShut
  {NoStop}%
\bibitem [{\citenamefont {Atat\"ure}\ \emph {et~al.}(2007)\citenamefont
  {Atat\"ure}, \citenamefont {Dreiser}, \citenamefont {Badolato},\ and\
  \citenamefont {{Imamo\={g}lu}}}]{Atature07}%
  \BibitemOpen
  \bibfield  {author} {\bibinfo {author} {\bibfnamefont {Mete}\ \bibnamefont
  {Atat\"ure}}, \bibinfo {author} {\bibfnamefont {Jan}\ \bibnamefont
  {Dreiser}}, \bibinfo {author} {\bibfnamefont {Antonio}\ \bibnamefont
  {Badolato}}, \ and\ \bibinfo {author} {\bibfnamefont {Atac}\ \bibnamefont
  {{Imamo\={g}lu}}},\ }\bibfield  {title} {\enquote {\bibinfo {title}
  {{Observation of Faraday rotation from a single confined spin}},}\
  }\href@noop {} {\bibfield  {journal} {\bibinfo  {journal} {Nature Physics}\
  }\textbf {\bibinfo {volume} {3}},\ \bibinfo {pages} {101--106} (\bibinfo
  {year} {2007})}\BibitemShut {NoStop}%
\bibitem [{\citenamefont {Fushman}\ \emph {et~al.}(2008)\citenamefont
  {Fushman}, \citenamefont {Englund}, \citenamefont {Faraon}, \citenamefont
  {Stoltz}, \citenamefont {Petroff},\ and\ \citenamefont
  {Vuckovic}}]{Fushman08}%
  \BibitemOpen
  \bibfield  {author} {\bibinfo {author} {\bibfnamefont {Ilya}\ \bibnamefont
  {Fushman}}, \bibinfo {author} {\bibfnamefont {Dirk}\ \bibnamefont {Englund}},
  \bibinfo {author} {\bibfnamefont {Andrei}\ \bibnamefont {Faraon}}, \bibinfo
  {author} {\bibfnamefont {Nick}\ \bibnamefont {Stoltz}}, \bibinfo {author}
  {\bibfnamefont {Pierre}\ \bibnamefont {Petroff}}, \ and\ \bibinfo {author}
  {\bibfnamefont {Jelena}\ \bibnamefont {Vuckovic}},\ }\bibfield  {title}
  {\enquote {\bibinfo {title} {{Controlled Phase Shifts with a Single Quantum
  Dot}},}\ }\href@noop {} {\bibfield  {journal} {\bibinfo  {journal} {Science}\
  }\textbf {\bibinfo {volume} {320}},\ \bibinfo {pages} {769--772} (\bibinfo
  {year} {2008})}\BibitemShut {NoStop}%
\bibitem [{\citenamefont {Young}\ \emph {et~al.}(2011)\citenamefont {Young},
  \citenamefont {Oulton}, \citenamefont {Hu}, \citenamefont {Thijssen},
  \citenamefont {Schneider}, \citenamefont {Reitzenstein}, \citenamefont
  {Kamp}, \citenamefont {H\"ofling}, \citenamefont {Worschech}, \citenamefont
  {Forchel},\ and\ \citenamefont {Rarity}}]{Young2011}%
  \BibitemOpen
  \bibfield  {author} {\bibinfo {author} {\bibfnamefont {A.B.}\ \bibnamefont
  {Young}}, \bibinfo {author} {\bibfnamefont {R.}~\bibnamefont {Oulton}},
  \bibinfo {author} {\bibfnamefont {C.Y.}\ \bibnamefont {Hu}}, \bibinfo
  {author} {\bibfnamefont {A.C.T.}\ \bibnamefont {Thijssen}}, \bibinfo {author}
  {\bibfnamefont {C.}~\bibnamefont {Schneider}}, \bibinfo {author}
  {\bibfnamefont {S.}~\bibnamefont {Reitzenstein}}, \bibinfo {author}
  {\bibfnamefont {M.}~\bibnamefont {Kamp}}, \bibinfo {author} {\bibfnamefont
  {S.}~\bibnamefont {H\"ofling}}, \bibinfo {author} {\bibfnamefont
  {L.}~\bibnamefont {Worschech}}, \bibinfo {author} {\bibfnamefont
  {A.}~\bibnamefont {Forchel}}, \ and\ \bibinfo {author} {\bibfnamefont {J.G.}\
  \bibnamefont {Rarity}},\ }\bibfield  {title} {\enquote {\bibinfo {title}
  {Quantum-dot-induced phase shift in a pillar microcavity},}\ }\href@noop {}
  {\bibfield  {journal} {\bibinfo  {journal} {Phys. Rev. A}\ }\textbf {\bibinfo
  {volume} {84}},\ \bibinfo {pages} {011803} (\bibinfo {year}
  {2011})}\BibitemShut {NoStop}%
\bibitem [{\citenamefont {Press}\ \emph {et~al.}(2010)\citenamefont {Press},
  \citenamefont {De{ }Greve}, \citenamefont {McMahon}, \citenamefont {Ladd},
  \citenamefont {Friess}, \citenamefont {Schneider}, \citenamefont {Kamp},
  \citenamefont {H\"ofling}, \citenamefont {Forchel},\ and\ \citenamefont
  {Yamamoto}}]{Press10}%
  \BibitemOpen
  \bibfield  {author} {\bibinfo {author} {\bibfnamefont {David}\ \bibnamefont
  {Press}}, \bibinfo {author} {\bibfnamefont {Kristiaan}\ \bibnamefont {De{
  }Greve}}, \bibinfo {author} {\bibfnamefont {Peter~L.}\ \bibnamefont
  {McMahon}}, \bibinfo {author} {\bibfnamefont {Thaddeus~D.}\ \bibnamefont
  {Ladd}}, \bibinfo {author} {\bibfnamefont {Benedikt}\ \bibnamefont {Friess}},
  \bibinfo {author} {\bibfnamefont {Christian}\ \bibnamefont {Schneider}},
  \bibinfo {author} {\bibfnamefont {Martin}\ \bibnamefont {Kamp}}, \bibinfo
  {author} {\bibfnamefont {Sven}\ \bibnamefont {H\"ofling}}, \bibinfo {author}
  {\bibfnamefont {Alfred}\ \bibnamefont {Forchel}}, \ and\ \bibinfo {author}
  {\bibfnamefont {Yoshihisa}\ \bibnamefont {Yamamoto}},\ }\bibfield  {title}
  {\enquote {\bibinfo {title} {Ultrafast optical spin echo in a single quantum
  dot},}\ }\href@noop {} {\bibfield  {journal} {\bibinfo  {journal} {Nature
  Photonics}\ }\textbf {\bibinfo {volume} {4}},\ \bibinfo {pages} {367--370}
  (\bibinfo {year} {2010})}\BibitemShut {NoStop}%
\bibitem [{\citenamefont {Aliferis}\ and\ \citenamefont
  {Preskill}(2008)}]{Aliferis2008}%
  \BibitemOpen
  \bibfield  {author} {\bibinfo {author} {\bibfnamefont {Panos}\ \bibnamefont
  {Aliferis}}\ and\ \bibinfo {author} {\bibfnamefont {John}\ \bibnamefont
  {Preskill}},\ }\bibfield  {title} {\enquote {\bibinfo {title} {Fault-tolerant
  quantum computation against biased noise},}\ }\href@noop {} {\bibfield
  {journal} {\bibinfo  {journal} {Phys. Rev. A}\ }\textbf {\bibinfo {volume}
  {78}},\ \bibinfo {pages} {052331} (\bibinfo {year} {2008})}\BibitemShut
  {NoStop}%
\bibitem [{\citenamefont {Lidar}\ \emph {et~al.}(1998)\citenamefont {Lidar},
  \citenamefont {Chuang},\ and\ \citenamefont {Whaley}}]{Lidar1998}%
  \BibitemOpen
  \bibfield  {author} {\bibinfo {author} {\bibfnamefont {D.A.}\ \bibnamefont
  {Lidar}}, \bibinfo {author} {\bibfnamefont {I.L.}\ \bibnamefont {Chuang}}, \
  and\ \bibinfo {author} {\bibfnamefont {K.B.}\ \bibnamefont {Whaley}},\
  }\bibfield  {title} {\enquote {\bibinfo {title} {Decoherence-free subspaces
  for quantum computation},}\ }\href@noop {} {\bibfield  {journal} {\bibinfo
  {journal} {Phys. Rev. Lett.}\ }\textbf {\bibinfo {volume} {81}},\ \bibinfo
  {pages} {2594--2597} (\bibinfo {year} {1998})}\BibitemShut {NoStop}%
\bibitem [{\citenamefont {Lidar}\ and\ \citenamefont {Whaley}(2003)}]{Lidar03}%
  \BibitemOpen
  \bibfield  {author} {\bibinfo {author} {\bibfnamefont {Daniel~A.}\
  \bibnamefont {Lidar}}\ and\ \bibinfo {author} {\bibfnamefont {K.~Birgitta}\
  \bibnamefont {Whaley}},\ }\enquote {\bibinfo {title} {Irreversible quantum
  dynamics},}\ \ (\bibinfo  {publisher} {Springer},\ \bibinfo {address}
  {Berlin},\ \bibinfo {year} {2003})\ Chap.\ \bibinfo {chapter}
  {Decoherence-Free Subspaces and Subsystems}, pp.\ \bibinfo {pages}
  {83--120},\ \bibinfo {note} {\emph{Preprint}
  arXiv:quant-ph/0301032.}\BibitemShut {Stop}%
\bibitem [{\citenamefont {Grace}\ \emph {et~al.}(2006)\citenamefont {Grace},
  \citenamefont {Brif}, \citenamefont {Rabitz}, \citenamefont {Walmsley},
  \citenamefont {Kosut},\ and\ \citenamefont {Lidar}}]{Grace2006}%
  \BibitemOpen
  \bibfield  {author} {\bibinfo {author} {\bibfnamefont {Matthew}\ \bibnamefont
  {Grace}}, \bibinfo {author} {\bibfnamefont {Constantin}\ \bibnamefont
  {Brif}}, \bibinfo {author} {\bibfnamefont {Herschel}\ \bibnamefont {Rabitz}},
  \bibinfo {author} {\bibfnamefont {Ian}\ \bibnamefont {Walmsley}}, \bibinfo
  {author} {\bibfnamefont {Robert}\ \bibnamefont {Kosut}}, \ and\ \bibinfo
  {author} {\bibfnamefont {Daniel}\ \bibnamefont {Lidar}},\ }\bibfield  {title}
  {\enquote {\bibinfo {title} {Encoding a qubit into multilevel subspaces},}\
  }\href@noop {} {\bibfield  {journal} {\bibinfo  {journal} {New J. Phys.}\
  }\textbf {\bibinfo {volume} {8}},\ \bibinfo {pages} {35} (\bibinfo {year}
  {2006})}\BibitemShut {NoStop}%
\bibitem [{\citenamefont {Hahn}(1950)}]{Hahn50}%
  \BibitemOpen
  \bibfield  {author} {\bibinfo {author} {\bibfnamefont {E.L.}\ \bibnamefont
  {Hahn}},\ }\bibfield  {title} {\enquote {\bibinfo {title} {Spin echoes},}\
  }\href@noop {} {\bibfield  {journal} {\bibinfo  {journal} {Phys. Rev.}\
  }\textbf {\bibinfo {volume} {80}},\ \bibinfo {pages} {580--594} (\bibinfo
  {year} {1950})}\BibitemShut {NoStop}%
\bibitem [{\citenamefont {Viola}\ \emph {et~al.}(1999)\citenamefont {Viola},
  \citenamefont {Knill},\ and\ \citenamefont {Lloyd}}]{Viola1999}%
  \BibitemOpen
  \bibfield  {author} {\bibinfo {author} {\bibfnamefont {Lorenza}\ \bibnamefont
  {Viola}}, \bibinfo {author} {\bibfnamefont {Emanuel}\ \bibnamefont {Knill}},
  \ and\ \bibinfo {author} {\bibfnamefont {Seth}\ \bibnamefont {Lloyd}},\
  }\bibfield  {title} {\enquote {\bibinfo {title} {Dynamical decoupling of open
  quantum systems},}\ }\href@noop {} {\bibfield  {journal} {\bibinfo  {journal}
  {Phys. Rev. Lett.}\ }\textbf {\bibinfo {volume} {82}},\ \bibinfo {pages}
  {2417--2421} (\bibinfo {year} {1999})}\BibitemShut {NoStop}%
\bibitem [{\citenamefont {Khodjasteh}\ and\ \citenamefont
  {Lidar}(2005)}]{Khodjasteh2005}%
  \BibitemOpen
  \bibfield  {author} {\bibinfo {author} {\bibfnamefont {K.}~\bibnamefont
  {Khodjasteh}}\ and\ \bibinfo {author} {\bibfnamefont {D.~A.}\ \bibnamefont
  {Lidar}},\ }\bibfield  {title} {\enquote {\bibinfo {title} {Fault-tolerant
  quantum dynamical decoupling},}\ }\href@noop {} {\bibfield  {journal}
  {\bibinfo  {journal} {Phys. Rev. Lett.}\ }\textbf {\bibinfo {volume} {95}},\
  \bibinfo {pages} {180501} (\bibinfo {year} {2005})}\BibitemShut {NoStop}%
\bibitem [{\citenamefont {Biercuk}\ \emph {et~al.}(2009)\citenamefont
  {Biercuk}, \citenamefont {Uys}, \citenamefont {{VanDevender}}, \citenamefont
  {Shiga}, \citenamefont {Itano},\ and\ \citenamefont
  {Bollinger}}]{Biercuk2009}%
  \BibitemOpen
  \bibfield  {author} {\bibinfo {author} {\bibfnamefont {Michael~J.}\
  \bibnamefont {Biercuk}}, \bibinfo {author} {\bibfnamefont {Hermann}\
  \bibnamefont {Uys}}, \bibinfo {author} {\bibfnamefont {Aaron~P.}\
  \bibnamefont {{VanDevender}}}, \bibinfo {author} {\bibfnamefont {Nobuyasu}\
  \bibnamefont {Shiga}}, \bibinfo {author} {\bibfnamefont {Wayne~M.}\
  \bibnamefont {Itano}}, \ and\ \bibinfo {author} {\bibfnamefont {John~J.}\
  \bibnamefont {Bollinger}},\ }\bibfield  {title} {\enquote {\bibinfo {title}
  {Optimized dynamical decoupling in a model quantum memory},}\ }\href@noop {}
  {\bibfield  {journal} {\bibinfo  {journal} {Nature}\ }\textbf {\bibinfo
  {volume} {458}},\ \bibinfo {pages} {996--1000} (\bibinfo {year}
  {2009})}\BibitemShut {NoStop}%
\bibitem [{\citenamefont {Viola}\ and\ \citenamefont {Knill}(2003)}]{Viola03}%
  \BibitemOpen
  \bibfield  {author} {\bibinfo {author} {\bibfnamefont {Lorenza}\ \bibnamefont
  {Viola}}\ and\ \bibinfo {author} {\bibfnamefont {Emanuel}\ \bibnamefont
  {Knill}},\ }\bibfield  {title} {\enquote {\bibinfo {title} {Robust dynamical
  decoupling of quantum systems with bounded controls},}\ }\href@noop {}
  {\bibfield  {journal} {\bibinfo  {journal} {Phys. Rev. Lett.}\ }\textbf
  {\bibinfo {volume} {90}},\ \bibinfo {pages} {037901} (\bibinfo {year}
  {2003})}\BibitemShut {NoStop}%
\bibitem [{\citenamefont {Ng}\ \emph {et~al.}(2011)\citenamefont {Ng},
  \citenamefont {Lidar},\ and\ \citenamefont {Preskill}}]{Ng09}%
  \BibitemOpen
  \bibfield  {author} {\bibinfo {author} {\bibfnamefont {Hui~Khoon}\
  \bibnamefont {Ng}}, \bibinfo {author} {\bibfnamefont {Daniel~A.}\
  \bibnamefont {Lidar}}, \ and\ \bibinfo {author} {\bibfnamefont {John}\
  \bibnamefont {Preskill}},\ }\bibfield  {title} {\enquote {\bibinfo {title}
  {Combining dynamical decoupling with fault-tolerant quantum computation},}\
  }\href@noop {} {\bibfield  {journal} {\bibinfo  {journal} {Phys. Rev. A}\
  }\textbf {\bibinfo {volume} {84}},\ \bibinfo {pages} {012305} (\bibinfo
  {year} {2011})}\BibitemShut {NoStop}%
\bibitem [{\citenamefont {Carr}\ and\ \citenamefont {Purcell}(1954)}]{Carr54}%
  \BibitemOpen
  \bibfield  {author} {\bibinfo {author} {\bibfnamefont {H.Y.}\ \bibnamefont
  {Carr}}\ and\ \bibinfo {author} {\bibfnamefont {E.M.}\ \bibnamefont
  {Purcell}},\ }\bibfield  {title} {\enquote {\bibinfo {title} {Effects of
  diffusion on free precession in nuclear magnetic resonance experiments},}\
  }\href@noop {} {\bibfield  {journal} {\bibinfo  {journal} {Phys. Rev.}\
  }\textbf {\bibinfo {volume} {94}},\ \bibinfo {pages} {630--638} (\bibinfo
  {year} {1954})}\BibitemShut {NoStop}%
\bibitem [{\citenamefont {Haeberlen}\ and\ \citenamefont
  {Waugh}(1968)}]{Haeberlen68}%
  \BibitemOpen
  \bibfield  {author} {\bibinfo {author} {\bibfnamefont {U.}~\bibnamefont
  {Haeberlen}}\ and\ \bibinfo {author} {\bibfnamefont {J.S.}\ \bibnamefont
  {Waugh}},\ }\bibfield  {title} {\enquote {\bibinfo {title} {Coherent
  averaging effects in magnetic resonance},}\ }\href@noop {} {\bibfield
  {journal} {\bibinfo  {journal} {Phys. Rev.}\ }\textbf {\bibinfo {volume}
  {175}},\ \bibinfo {pages} {453--467} (\bibinfo {year} {1968})}\BibitemShut
  {NoStop}%
\bibitem [{\citenamefont {Uhrig}(2007)}]{Uhrig07}%
  \BibitemOpen
  \bibfield  {author} {\bibinfo {author} {\bibfnamefont {G\"otz~S.}\
  \bibnamefont {Uhrig}},\ }\bibfield  {title} {\enquote {\bibinfo {title}
  {Keeping a quantum bit alive by optimized $\pi{}$-pulse sequences},}\
  }\href@noop {} {\bibfield  {journal} {\bibinfo  {journal} {Phys. Rev. Lett.}\
  }\textbf {\bibinfo {volume} {98}},\ \bibinfo {pages} {100504} (\bibinfo
  {year} {2007})}\BibitemShut {NoStop}%
\bibitem [{\citenamefont {Brown}\ \emph {et~al.}(2004)\citenamefont {Brown},
  \citenamefont {Harrow},\ and\ \citenamefont {Chuang}}]{Brown04}%
  \BibitemOpen
  \bibfield  {author} {\bibinfo {author} {\bibfnamefont {Kenneth~R.}\
  \bibnamefont {Brown}}, \bibinfo {author} {\bibfnamefont {Aram~W.}\
  \bibnamefont {Harrow}}, \ and\ \bibinfo {author} {\bibfnamefont {Isaac~L.}\
  \bibnamefont {Chuang}},\ }\bibfield  {title} {\enquote {\bibinfo {title}
  {Arbitrarily accurate composite pulse sequences},}\ }\href@noop {} {\bibfield
   {journal} {\bibinfo  {journal} {Phys. Rev. A}\ }\textbf {\bibinfo {volume}
  {70}} (\bibinfo {year} {2004})}\BibitemShut {NoStop}%
\bibitem [{\citenamefont {Tomita}\ \emph {et~al.}(2010)\citenamefont {Tomita},
  \citenamefont {Merrill},\ and\ \citenamefont {Brown}}]{Tomita10}%
  \BibitemOpen
  \bibfield  {author} {\bibinfo {author} {\bibfnamefont {Y.}~\bibnamefont
  {Tomita}}, \bibinfo {author} {\bibfnamefont {J.T.}\ \bibnamefont {Merrill}},
  \ and\ \bibinfo {author} {\bibfnamefont {K.R.}\ \bibnamefont {Brown}},\
  }\bibfield  {title} {\enquote {\bibinfo {title} {Multi-qubit compensation
  sequences},}\ }\href@noop {} {\bibfield  {journal} {\bibinfo  {journal} {New
  J. Phys.}\ }\textbf {\bibinfo {volume} {12}},\ \bibinfo {pages} {015002}
  (\bibinfo {year} {2010})}\BibitemShut {NoStop}%
\bibitem [{\citenamefont {Khodjasteh}\ and\ \citenamefont
  {Viola}(2009)}]{Khodjasteh2009}%
  \BibitemOpen
  \bibfield  {author} {\bibinfo {author} {\bibfnamefont {Kaveh}\ \bibnamefont
  {Khodjasteh}}\ and\ \bibinfo {author} {\bibfnamefont {Lorenza}\ \bibnamefont
  {Viola}},\ }\bibfield  {title} {\enquote {\bibinfo {title} {Dynamically
  error-corrected gates for universal quantum computation},}\ }\href@noop {}
  {\bibfield  {journal} {\bibinfo  {journal} {Phys. Rev. Lett.}\ }\textbf
  {\bibinfo {volume} {102}},\ \bibinfo {pages} {080501} (\bibinfo {year}
  {2009})}\BibitemShut {NoStop}%
\bibitem [{\citenamefont {Cappellaro}\ \emph {et~al.}(2005)\citenamefont
  {Cappellaro}, \citenamefont {Emerson}, \citenamefont {Boulant}, \citenamefont
  {Ramanathan}, \citenamefont {Lloyd},\ and\ \citenamefont
  {Cory}}]{Capellaro2005}%
  \BibitemOpen
  \bibfield  {author} {\bibinfo {author} {\bibfnamefont {P.}~\bibnamefont
  {Cappellaro}}, \bibinfo {author} {\bibfnamefont {J.}~\bibnamefont {Emerson}},
  \bibinfo {author} {\bibfnamefont {N.}~\bibnamefont {Boulant}}, \bibinfo
  {author} {\bibfnamefont {C.}~\bibnamefont {Ramanathan}}, \bibinfo {author}
  {\bibfnamefont {S.}~\bibnamefont {Lloyd}}, \ and\ \bibinfo {author}
  {\bibfnamefont {D.~G.}\ \bibnamefont {Cory}},\ }\bibfield  {title} {\enquote
  {\bibinfo {title} {Entanglement assisted metrology},}\ }\href@noop {}
  {\bibfield  {journal} {\bibinfo  {journal} {Phys. Rev. Lett.}\ }\textbf
  {\bibinfo {volume} {94}},\ \bibinfo {pages} {020502} (\bibinfo {year}
  {2005})}\BibitemShut {NoStop}%
\bibitem [{\citenamefont {Elzerman}\ \emph {et~al.}(2004)\citenamefont
  {Elzerman}, \citenamefont {Hanson}, \citenamefont {van Beveren},
  \citenamefont {Vandersypen},\ and\ \citenamefont {Kouwenhoven}}]{Elzerman04}%
  \BibitemOpen
  \bibfield  {author} {\bibinfo {author} {\bibfnamefont {J.~M.}\ \bibnamefont
  {Elzerman}}, \bibinfo {author} {\bibfnamefont {R.}~\bibnamefont {Hanson}},
  \bibinfo {author} {\bibfnamefont {L.~H.~Willems}\ \bibnamefont {van
  Beveren}}, \bibinfo {author} {\bibfnamefont {B.~Witkamp L. M.~K.}\
  \bibnamefont {Vandersypen}}, \ and\ \bibinfo {author} {\bibfnamefont {L.~P.}\
  \bibnamefont {Kouwenhoven}},\ }\bibfield  {title} {\enquote {\bibinfo {title}
  {Single-shot read-out of an individual electron spin in a quantum dot},}\
  }\href@noop {} {\bibfield  {journal} {\bibinfo  {journal} {Nature}\ }\textbf
  {\bibinfo {volume} {430}},\ \bibinfo {pages} {431--435} (\bibinfo {year}
  {2004})}\BibitemShut {NoStop}%
\bibitem [{\citenamefont {Kroutvar}\ \emph {et~al.}(2004)\citenamefont
  {Kroutvar}, \citenamefont {Ducommun}, \citenamefont {Heiss}, \citenamefont
  {Bichler}, \citenamefont {Schuh}, \citenamefont {Abstreiter},\ and\
  \citenamefont {Finley}}]{Kroutvar04}%
  \BibitemOpen
  \bibfield  {author} {\bibinfo {author} {\bibfnamefont {Miro}\ \bibnamefont
  {Kroutvar}}, \bibinfo {author} {\bibfnamefont {Yann}\ \bibnamefont
  {Ducommun}}, \bibinfo {author} {\bibfnamefont {Dominik}\ \bibnamefont
  {Heiss}}, \bibinfo {author} {\bibfnamefont {Max}\ \bibnamefont {Bichler}},
  \bibinfo {author} {\bibfnamefont {Dieter}\ \bibnamefont {Schuh}}, \bibinfo
  {author} {\bibfnamefont {Gerhard}\ \bibnamefont {Abstreiter}}, \ and\
  \bibinfo {author} {\bibfnamefont {Jonathan~J.}\ \bibnamefont {Finley}},\
  }\bibfield  {title} {\enquote {\bibinfo {title} {Optically programmable
  electron spin memory using semiconductor quantum dots},}\ }\href@noop {}
  {\bibfield  {journal} {\bibinfo  {journal} {Nature}\ }\textbf {\bibinfo
  {volume} {432}},\ \bibinfo {pages} {81--84} (\bibinfo {year}
  {2004})}\BibitemShut {NoStop}%
\bibitem [{\citenamefont {Aharonov}\ and\ \citenamefont
  {Ben-Or}(1997)}]{Aharonov1997}%
  \BibitemOpen
  \bibfield  {author} {\bibinfo {author} {\bibfnamefont {D.}~\bibnamefont
  {Aharonov}}\ and\ \bibinfo {author} {\bibfnamefont {M.}~\bibnamefont
  {Ben-Or}},\ }\bibfield  {title} {\enquote {\bibinfo {title} {Fault-tolerant
  quantum computation with constant error},}\ }in\ \href@noop {} {\emph
  {\bibinfo {booktitle} {Proceedings of the Twenty-Ninth Annual ACM Symposium
  on Theory of Computing (STOC '97)}}}\ (\bibinfo  {publisher} {ACM},\ \bibinfo
  {address} {New York, NY, USA},\ \bibinfo {year} {1997})\ pp.\ \bibinfo
  {pages} {176--188}\BibitemShut {NoStop}%
\bibitem [{\citenamefont {Preskill}(1998)}]{Preskill1998}%
  \BibitemOpen
  \bibfield  {author} {\bibinfo {author} {\bibfnamefont {John}\ \bibnamefont
  {Preskill}},\ }\bibfield  {title} {\enquote {\bibinfo {title} {Reliable
  quantum computers},}\ }\href@noop {} {\bibfield  {journal} {\bibinfo
  {journal} {P. Roy. Soc. A-Math. Phy.}\ }\textbf {\bibinfo {volume} {454}},\
  \bibinfo {pages} {385--410} (\bibinfo {year} {1998})},\ \bibinfo {note}
  {\emph{Preprint} arXiv:quant-ph/9705031.}\BibitemShut {Stop}%
\bibitem [{\citenamefont {Devitt}\ \emph {et~al.}(2009)\citenamefont {Devitt},
  \citenamefont {Nemoto},\ and\ \citenamefont {Munro}}]{Devitt2009}%
  \BibitemOpen
  \bibfield  {author} {\bibinfo {author} {\bibfnamefont {Simon~J.}\
  \bibnamefont {Devitt}}, \bibinfo {author} {\bibfnamefont {Kae}\ \bibnamefont
  {Nemoto}}, \ and\ \bibinfo {author} {\bibfnamefont {William~J.}\ \bibnamefont
  {Munro}},\ }\href@noop {} {\enquote {\bibinfo {title} {The idiots guide to
  quantum error correction},}\ } (\bibinfo {year} {2009}),\ \bibinfo {note}
  {\emph{Preprint} arXiv:0905.2794.}\BibitemShut {Stop}%
\bibitem [{\citenamefont {Gottesman}(1997)}]{Gottesman97}%
  \BibitemOpen
  \bibfield  {author} {\bibinfo {author} {\bibfnamefont {D.}~\bibnamefont
  {Gottesman}},\ }\emph {\bibinfo {title} {Stabilizer Codes and Quantum Error
  Correction}},\ \href@noop {} {Ph.D. thesis},\ \bibinfo  {school} {California
  Institute of Technology}, \bibinfo {address} {Pasadena, CA} (\bibinfo {year}
  {1997})\BibitemShut {NoStop}%
\bibitem [{\citenamefont {Raussendorf}\ and\ \citenamefont
  {Harrington}(2007)}]{Rauss07b}%
  \BibitemOpen
  \bibfield  {author} {\bibinfo {author} {\bibfnamefont {Robert}\ \bibnamefont
  {Raussendorf}}\ and\ \bibinfo {author} {\bibfnamefont {Jim}\ \bibnamefont
  {Harrington}},\ }\bibfield  {title} {\enquote {\bibinfo {title}
  {Fault-tolerant quantum computation with high threshold in two dimensions},}\
  }\href@noop {} {\bibfield  {journal} {\bibinfo  {journal} {Phys. Rev. Lett.}\
  }\textbf {\bibinfo {volume} {98}},\ \bibinfo {pages} {190504} (\bibinfo
  {year} {2007})}\BibitemShut {NoStop}%
\bibitem [{\citenamefont {Raussendorf}\ \emph {et~al.}(2007)\citenamefont
  {Raussendorf}, \citenamefont {Harrington},\ and\ \citenamefont
  {Goyal}}]{Rauss07}%
  \BibitemOpen
  \bibfield  {author} {\bibinfo {author} {\bibfnamefont {R}~\bibnamefont
  {Raussendorf}}, \bibinfo {author} {\bibfnamefont {J}~\bibnamefont
  {Harrington}}, \ and\ \bibinfo {author} {\bibfnamefont {K}~\bibnamefont
  {Goyal}},\ }\bibfield  {title} {\enquote {\bibinfo {title} {Topological
  fault-tolerance in cluster state quantum computation},}\ }\href@noop {}
  {\bibfield  {journal} {\bibinfo  {journal} {New J. Phys.}\ }\textbf {\bibinfo
  {volume} {9}},\ \bibinfo {pages} {199} (\bibinfo {year} {2007})}\BibitemShut
  {NoStop}%
\bibitem [{\citenamefont {Knill}(2005)}]{Knill2005}%
  \BibitemOpen
  \bibfield  {author} {\bibinfo {author} {\bibfnamefont {E.}~\bibnamefont
  {Knill}},\ }\bibfield  {title} {\enquote {\bibinfo {title} {Quantum computing
  with realistically noisy devices},}\ }\href@noop {} {\bibfield  {journal}
  {\bibinfo  {journal} {Nature}\ }\textbf {\bibinfo {volume} {434}},\ \bibinfo
  {pages} {39--44} (\bibinfo {year} {2005})}\BibitemShut {NoStop}%
\bibitem [{\citenamefont {Aliferis}\ \emph {et~al.}(2008)\citenamefont
  {Aliferis}, \citenamefont {Gottesman},\ and\ \citenamefont
  {Preskill}}]{Aliferis2008b}%
  \BibitemOpen
  \bibfield  {author} {\bibinfo {author} {\bibfnamefont {Panos}\ \bibnamefont
  {Aliferis}}, \bibinfo {author} {\bibfnamefont {Daniel}\ \bibnamefont
  {Gottesman}}, \ and\ \bibinfo {author} {\bibfnamefont {John}\ \bibnamefont
  {Preskill}},\ }\bibfield  {title} {\enquote {\bibinfo {title} {Accuracy
  threshold for postselected quantum computation},}\ }\href@noop {} {\bibfield
  {journal} {\bibinfo  {journal} {Quantum Info. Comput.}\ }\textbf {\bibinfo
  {volume} {8}},\ \bibinfo {pages} {181--244} (\bibinfo {year}
  {2008})}\BibitemShut {NoStop}%
\bibitem [{\citenamefont {Bacon}(2006)}]{Bacon2006}%
  \BibitemOpen
  \bibfield  {author} {\bibinfo {author} {\bibfnamefont {Dave}\ \bibnamefont
  {Bacon}},\ }\bibfield  {title} {\enquote {\bibinfo {title} {Operator quantum
  error-correcting subsystems for self-correcting quantum memories},}\
  }\href@noop {} {\bibfield  {journal} {\bibinfo  {journal} {Phys. Rev. A}\
  }\textbf {\bibinfo {volume} {73}},\ \bibinfo {pages} {012340} (\bibinfo
  {year} {2006})}\BibitemShut {NoStop}%
\bibitem [{\citenamefont {Fowler}\ \emph
  {et~al.}(2011{\natexlab{a}})\citenamefont {Fowler}, \citenamefont {Wang},\
  and\ \citenamefont {Hollenberg}}]{Fowler10}%
  \BibitemOpen
  \bibfield  {author} {\bibinfo {author} {\bibfnamefont {Austin~G.}\
  \bibnamefont {Fowler}}, \bibinfo {author} {\bibfnamefont {David~S.}\
  \bibnamefont {Wang}}, \ and\ \bibinfo {author} {\bibfnamefont {Lloyd C.~L.}\
  \bibnamefont {Hollenberg}},\ }\bibfield  {title} {\enquote {\bibinfo {title}
  {Surface code quantum error correction incorporating accurate error
  propagation},}\ }\href@noop {} {\bibfield  {journal} {\bibinfo  {journal}
  {Quantum Info. Comput.}\ }\textbf {\bibinfo {volume} {11}},\ \bibinfo {pages}
  {8} (\bibinfo {year} {2011}{\natexlab{a}})}\BibitemShut {NoStop}%
\bibitem [{\citenamefont {Wang}\ \emph {et~al.}(2011)\citenamefont {Wang},
  \citenamefont {Fowler},\ and\ \citenamefont {Hollenberg}}]{Wang10}%
  \BibitemOpen
  \bibfield  {author} {\bibinfo {author} {\bibfnamefont {David~S.}\
  \bibnamefont {Wang}}, \bibinfo {author} {\bibfnamefont {Austin~G.}\
  \bibnamefont {Fowler}}, \ and\ \bibinfo {author} {\bibfnamefont {Lloyd
  C.~L.}\ \bibnamefont {Hollenberg}},\ }\bibfield  {title} {\enquote {\bibinfo
  {title} {Surface code quantum computing with error rates over 1\%},}\
  }\href@noop {} {\bibfield  {journal} {\bibinfo  {journal} {Phys. Rev. A}\
  }\textbf {\bibinfo {volume} {83}},\ \bibinfo {pages} {020302} (\bibinfo
  {year} {2011})}\BibitemShut {NoStop}%
\bibitem [{\citenamefont {Fowler}\ \emph
  {et~al.}(2011{\natexlab{b}})\citenamefont {Fowler}, \citenamefont
  {Whiteside},\ and\ \citenamefont {Hollenberg}}]{Fowler2011b}%
  \BibitemOpen
  \bibfield  {author} {\bibinfo {author} {\bibfnamefont {Austin~G.}\
  \bibnamefont {Fowler}}, \bibinfo {author} {\bibfnamefont {Adam~C.}\
  \bibnamefont {Whiteside}}, \ and\ \bibinfo {author} {\bibfnamefont {Lloyd
  C.~L.}\ \bibnamefont {Hollenberg}},\ }\href@noop {} {\enquote {\bibinfo
  {title} {Towards practical classical processing for the surface code},}\ }
  (\bibinfo {year} {2011}{\natexlab{b}}),\ \bibinfo {note} {\emph{Preprint}
  arXiv:1110.5133}\BibitemShut {NoStop}%
\bibitem [{\citenamefont {Aliferis}(2007)}]{Aliferis2007}%
  \BibitemOpen
  \bibfield  {author} {\bibinfo {author} {\bibfnamefont {Panos}\ \bibnamefont
  {Aliferis}},\ }\emph {\bibinfo {title} {Level Reduction and the Quantum
  Threshold Theorem}},\ \href@noop {} {Ph.D. thesis},\ \bibinfo  {school}
  {California Institute of Technology}, \bibinfo {address} {Pasadena, CA}
  (\bibinfo {year} {2007})\BibitemShut {NoStop}%
\bibitem [{\citenamefont {Shor}(1997)}]{Shor99}%
  \BibitemOpen
  \bibfield  {author} {\bibinfo {author} {\bibfnamefont {Peter~W}\ \bibnamefont
  {Shor}},\ }\bibfield  {title} {\enquote {\bibinfo {title} {Polynomial-time
  algorithms for prime factorization and discrete logarithms on a quantum
  computer},}\ }\href@noop {} {\bibfield  {journal} {\bibinfo  {journal} {SIAM
  J. Comput.}\ }\textbf {\bibinfo {volume} {26}},\ \bibinfo {pages}
  {1484--1509} (\bibinfo {year} {1997})}\BibitemShut {NoStop}%
\bibitem [{\citenamefont {Aspuru-Guzik}\ \emph {et~al.}(2005)\citenamefont
  {Aspuru-Guzik}, \citenamefont {Dutoi}, \citenamefont {Love},\ and\
  \citenamefont {Head-Gordon}}]{Aspuru2005}%
  \BibitemOpen
  \bibfield  {author} {\bibinfo {author} {\bibfnamefont {Al\'{a}n}\
  \bibnamefont {Aspuru-Guzik}}, \bibinfo {author} {\bibfnamefont {Anthony~D.}\
  \bibnamefont {Dutoi}}, \bibinfo {author} {\bibfnamefont {Peter~J.}\
  \bibnamefont {Love}}, \ and\ \bibinfo {author} {\bibfnamefont {Martin}\
  \bibnamefont {Head-Gordon}},\ }\bibfield  {title} {\enquote {\bibinfo {title}
  {Simulated quantum computation of molecular energies},}\ }\href@noop {}
  {\bibfield  {journal} {\bibinfo  {journal} {Science}\ }\textbf {\bibinfo
  {volume} {309}},\ \bibinfo {pages} {1704--1707} (\bibinfo {year}
  {2005})}\BibitemShut {NoStop}%
\bibitem [{\citenamefont {Kassal}\ \emph {et~al.}(2008)\citenamefont {Kassal},
  \citenamefont {Jordan}, \citenamefont {Love}, \citenamefont {Mohseni},\ and\
  \citenamefont {Aspuru-Guzik}}]{Kassal2008}%
  \BibitemOpen
  \bibfield  {author} {\bibinfo {author} {\bibfnamefont {Ivan}\ \bibnamefont
  {Kassal}}, \bibinfo {author} {\bibfnamefont {Stephen~P.}\ \bibnamefont
  {Jordan}}, \bibinfo {author} {\bibfnamefont {Peter~J.}\ \bibnamefont {Love}},
  \bibinfo {author} {\bibfnamefont {Masoud}\ \bibnamefont {Mohseni}}, \ and\
  \bibinfo {author} {\bibfnamefont {Al\'{a}n}\ \bibnamefont {Aspuru-Guzik}},\
  }\bibfield  {title} {\enquote {\bibinfo {title} {Polynomial-time quantum
  algorithm for the simulation of chemical dynamics},}\ }\href@noop {}
  {\bibfield  {journal} {\bibinfo  {journal} {P. Natl. Acad. Sci. USA}\
  }\textbf {\bibinfo {volume} {105}},\ \bibinfo {pages} {18681--18686}
  (\bibinfo {year} {2008})}\BibitemShut {NoStop}%
\bibitem [{\citenamefont {Van~Meter}\ and\ \citenamefont
  {Itoh}(2005)}]{VanMeter05}%
  \BibitemOpen
  \bibfield  {author} {\bibinfo {author} {\bibfnamefont {Rodney}\ \bibnamefont
  {Van~Meter}}\ and\ \bibinfo {author} {\bibfnamefont {Kohei~M.}\ \bibnamefont
  {Itoh}},\ }\bibfield  {title} {\enquote {\bibinfo {title} {Fast quantum
  modular exponentiation},}\ }\href@noop {} {\bibfield  {journal} {\bibinfo
  {journal} {Phys. Rev. A}\ }\textbf {\bibinfo {volume} {71}},\ \bibinfo
  {pages} {052320} (\bibinfo {year} {2005})}\BibitemShut {NoStop}%
\bibitem [{\citenamefont {DiVincenzo}\ and\ \citenamefont
  {Aliferis}(2007)}]{DiVincenzo07}%
  \BibitemOpen
  \bibfield  {author} {\bibinfo {author} {\bibfnamefont {David~P.}\
  \bibnamefont {DiVincenzo}}\ and\ \bibinfo {author} {\bibfnamefont {Panos}\
  \bibnamefont {Aliferis}},\ }\bibfield  {title} {\enquote {\bibinfo {title}
  {Effective fault-tolerant quantum computation with slow measurements},}\
  }\href@noop {} {\bibfield  {journal} {\bibinfo  {journal} {Phys. Rev. Lett.}\
  }\textbf {\bibinfo {volume} {98}},\ \bibinfo {pages} {020501} (\bibinfo
  {year} {2007})}\BibitemShut {NoStop}%
\bibitem [{\citenamefont {Anders}\ and\ \citenamefont
  {Briegel}(2006)}]{Simon06}%
  \BibitemOpen
  \bibfield  {author} {\bibinfo {author} {\bibfnamefont {Simon}\ \bibnamefont
  {Anders}}\ and\ \bibinfo {author} {\bibfnamefont {Hans~J.}\ \bibnamefont
  {Briegel}},\ }\bibfield  {title} {\enquote {\bibinfo {title} {Fast simulation
  of stabilizer circuits using a graph-state representation},}\ }\href@noop {}
  {\bibfield  {journal} {\bibinfo  {journal} {Phys. Rev. A}\ }\textbf {\bibinfo
  {volume} {73}},\ \bibinfo {pages} {022334} (\bibinfo {year}
  {2006})}\BibitemShut {NoStop}%
\bibitem [{\citenamefont {Dawson}\ and\ \citenamefont
  {Nielsen}(2006)}]{Dawson05}%
  \BibitemOpen
  \bibfield  {author} {\bibinfo {author} {\bibfnamefont {Christopher~M.}\
  \bibnamefont {Dawson}}\ and\ \bibinfo {author} {\bibfnamefont {Michael~A.}\
  \bibnamefont {Nielsen}},\ }\bibfield  {title} {\enquote {\bibinfo {title}
  {{The Solovay-Kitaev Algorithm}},}\ }\href@noop {} {\bibfield  {journal}
  {\bibinfo  {journal} {Quantum Info. Comput.}\ }\textbf {\bibinfo {volume}
  {6}},\ \bibinfo {pages} {81} (\bibinfo {year} {2006})}\BibitemShut {NoStop}%
\bibitem [{\citenamefont {Fowler}(2011)}]{Fowler2011}%
  \BibitemOpen
  \bibfield  {author} {\bibinfo {author} {\bibfnamefont {Austin~G.}\
  \bibnamefont {Fowler}},\ }\bibfield  {title} {\enquote {\bibinfo {title}
  {{Constructing arbitrary Steane code single logical qubit fault-tolerant
  gates}},}\ }\href@noop {} {\bibfield  {journal} {\bibinfo  {journal} {Quantum
  Info. Comput.}\ }\textbf {\bibinfo {volume} {11}},\ \bibinfo {pages}
  {867--873} (\bibinfo {year} {2011})}\BibitemShut {NoStop}%
\bibitem [{\citenamefont {Jones}\ \emph {et~al.}(2012)\citenamefont {Jones},
  \citenamefont {Whitfield}, \citenamefont {McMahon}, \citenamefont {Yung},
  \citenamefont {Van{ }Meter}, \citenamefont {Aspuru-Guzik},\ and\
  \citenamefont {Yamamoto}}]{Jones2012Sim}%
  \BibitemOpen
  \bibfield  {author} {\bibinfo {author} {\bibfnamefont {N.~Cody}\ \bibnamefont
  {Jones}}, \bibinfo {author} {\bibfnamefont {James~D.}\ \bibnamefont
  {Whitfield}}, \bibinfo {author} {\bibfnamefont {Peter~L.}\ \bibnamefont
  {McMahon}}, \bibinfo {author} {\bibfnamefont {Man-Hong}\ \bibnamefont
  {Yung}}, \bibinfo {author} {\bibfnamefont {Rodney}\ \bibnamefont {Van{
  }Meter}}, \bibinfo {author} {\bibfnamefont {Al\'{a}n}\ \bibnamefont
  {Aspuru-Guzik}}, \ and\ \bibinfo {author} {\bibfnamefont {Yoshihisa}\
  \bibnamefont {Yamamoto}},\ }\href@noop {} {\enquote {\bibinfo {title}
  {Simulating chemistry efficiently on fault-tolerant quantum computers},}\ }
  (\bibinfo {year} {2012}),\ \bibinfo {note} {\emph{Preprint}
  arXiv:1204.0567}\BibitemShut {NoStop}%
\bibitem [{\citenamefont {Bravyi}\ and\ \citenamefont
  {Kitaev}(2005)}]{Bravyi2005}%
  \BibitemOpen
  \bibfield  {author} {\bibinfo {author} {\bibfnamefont {Sergey}\ \bibnamefont
  {Bravyi}}\ and\ \bibinfo {author} {\bibfnamefont {Alexei}\ \bibnamefont
  {Kitaev}},\ }\bibfield  {title} {\enquote {\bibinfo {title} {Universal
  quantum computation with ideal {Clifford} gates and noisy ancillas},}\
  }\href@noop {} {\bibfield  {journal} {\bibinfo  {journal} {Phys. Rev. A}\
  }\textbf {\bibinfo {volume} {71}},\ \bibinfo {pages} {022316} (\bibinfo
  {year} {2005})}\BibitemShut {NoStop}%
\bibitem [{\citenamefont {Kitaev}(1995)}]{Kitaev1995}%
  \BibitemOpen
  \bibfield  {author} {\bibinfo {author} {\bibfnamefont {A.~Yu.}\ \bibnamefont
  {Kitaev}},\ }\href@noop {} {\enquote {\bibinfo {title} {{Quantum measurements
  and the Abelian Stabilizer Problem}},}\ } (\bibinfo {year} {1995}),\ \bibinfo
  {note} {\emph{Preprint} arXiv:quant-ph/9511026v1.}\BibitemShut {Stop}%
\bibitem [{\citenamefont {Cleve}\ \emph {et~al.}(1998)\citenamefont {Cleve},
  \citenamefont {Ekert}, \citenamefont {Macchiavello},\ and\ \citenamefont
  {Mosca}}]{Cleve1998}%
  \BibitemOpen
  \bibfield  {author} {\bibinfo {author} {\bibfnamefont {R.}~\bibnamefont
  {Cleve}}, \bibinfo {author} {\bibfnamefont {A.}~\bibnamefont {Ekert}},
  \bibinfo {author} {\bibfnamefont {C.}~\bibnamefont {Macchiavello}}, \ and\
  \bibinfo {author} {\bibfnamefont {M.}~\bibnamefont {Mosca}},\ }\bibfield
  {title} {\enquote {\bibinfo {title} {Quantum algorithms revisited},}\
  }\href@noop {} {\bibfield  {journal} {\bibinfo  {journal} {P. Roy. Soc.
  A-Math. Phy.}\ }\textbf {\bibinfo {volume} {454}},\ \bibinfo {pages}
  {339--354} (\bibinfo {year} {1998})}\BibitemShut {NoStop}%
\bibitem [{\citenamefont {Vedral}\ \emph {et~al.}(1996)\citenamefont {Vedral},
  \citenamefont {Barenco},\ and\ \citenamefont {Ekert}}]{Vedral1996}%
  \BibitemOpen
  \bibfield  {author} {\bibinfo {author} {\bibfnamefont {Vlatko}\ \bibnamefont
  {Vedral}}, \bibinfo {author} {\bibfnamefont {Adriano}\ \bibnamefont
  {Barenco}}, \ and\ \bibinfo {author} {\bibfnamefont {Artur}\ \bibnamefont
  {Ekert}},\ }\bibfield  {title} {\enquote {\bibinfo {title} {Quantum networks
  for elementary arithmetic operations},}\ }\href@noop {} {\bibfield  {journal}
  {\bibinfo  {journal} {Phys. Rev. A}\ }\textbf {\bibinfo {volume} {54}},\
  \bibinfo {pages} {147--153} (\bibinfo {year} {1996})}\BibitemShut {NoStop}%
\bibitem [{\citenamefont {Cuccaro}\ \emph {et~al.}(2008)\citenamefont
  {Cuccaro}, \citenamefont {Draper}, \citenamefont {Kutin},\ and\ \citenamefont
  {Moulton}}]{Cuccaro2008}%
  \BibitemOpen
  \bibfield  {author} {\bibinfo {author} {\bibfnamefont {Steven~A.}\
  \bibnamefont {Cuccaro}}, \bibinfo {author} {\bibfnamefont {Thomas~G.}\
  \bibnamefont {Draper}}, \bibinfo {author} {\bibfnamefont {Samuel~A.}\
  \bibnamefont {Kutin}}, \ and\ \bibinfo {author} {\bibfnamefont
  {David~Petrie}\ \bibnamefont {Moulton}},\ }\href@noop {} {\enquote {\bibinfo
  {title} {A new quantum ripple-carry addition circuit},}\ } (\bibinfo {year}
  {2008}),\ \bibinfo {note} {\emph{Preprint}
  arXiv:quant-ph/0410184}\BibitemShut {NoStop}%
\bibitem [{\citenamefont {Draper}\ \emph {et~al.}(2006)\citenamefont {Draper},
  \citenamefont {Kutin}, \citenamefont {Rains},\ and\ \citenamefont
  {Svore}}]{Draper2006}%
  \BibitemOpen
  \bibfield  {author} {\bibinfo {author} {\bibfnamefont {Thomas~G.}\
  \bibnamefont {Draper}}, \bibinfo {author} {\bibfnamefont {Samuel~A.}\
  \bibnamefont {Kutin}}, \bibinfo {author} {\bibfnamefont {Eric~M.}\
  \bibnamefont {Rains}}, \ and\ \bibinfo {author} {\bibfnamefont {Krysta~M.}\
  \bibnamefont {Svore}},\ }\bibfield  {title} {\enquote {\bibinfo {title} {A
  logarithmic-depth quantum carry-lookahead adder},}\ }\href@noop {} {\bibfield
   {journal} {\bibinfo  {journal} {Quantum Info. Comput.}\ }\textbf {\bibinfo
  {volume} {6}},\ \bibinfo {pages} {351--369} (\bibinfo {year}
  {2006})}\BibitemShut {NoStop}%
\bibitem [{\citenamefont {Beauregard}(2003)}]{Beauregard2003}%
  \BibitemOpen
  \bibfield  {author} {\bibinfo {author} {\bibfnamefont {St\'{e}phane}\
  \bibnamefont {Beauregard}},\ }\bibfield  {title} {\enquote {\bibinfo {title}
  {Circuit for {S}hor's algorithm using 2n+3 qubits},}\ }\href@noop {}
  {\bibfield  {journal} {\bibinfo  {journal} {Quantum Info. Comput.}\ }\textbf
  {\bibinfo {volume} {3}},\ \bibinfo {pages} {175--185} (\bibinfo {year}
  {2003})}\BibitemShut {NoStop}%
\bibitem [{\citenamefont {Zalka}(2006)}]{Zalka06}%
  \BibitemOpen
  \bibfield  {author} {\bibinfo {author} {\bibfnamefont {Christof}\
  \bibnamefont {Zalka}},\ }\href@noop {} {\enquote {\bibinfo {title} {{S}hor's
  algorithm with fewer pure qubits},}\ } (\bibinfo {year} {2006}),\ \bibinfo
  {note} {\emph{Preprint} arXiv:quant-ph/0601097.}\BibitemShut {Stop}%
\bibitem [{\citenamefont {Takahashi}\ and\ \citenamefont
  {Kunihiro}(2006)}]{Takahashi2006}%
  \BibitemOpen
  \bibfield  {author} {\bibinfo {author} {\bibfnamefont {Yasuhiro}\
  \bibnamefont {Takahashi}}\ and\ \bibinfo {author} {\bibfnamefont {Noboru}\
  \bibnamefont {Kunihiro}},\ }\bibfield  {title} {\enquote {\bibinfo {title} {A
  quantum circuit for {S}hor's factoring algorithm using 2n + 2 qubits},}\
  }\href@noop {} {\bibfield  {journal} {\bibinfo  {journal} {Quantum Info.
  Comput.}\ }\textbf {\bibinfo {volume} {6}},\ \bibinfo {pages} {184--192}
  (\bibinfo {year} {2006})}\BibitemShut {NoStop}%
\bibitem [{\citenamefont {Rivest}\ \emph {et~al.}(1978)\citenamefont {Rivest},
  \citenamefont {Shamir},\ and\ \citenamefont {Adleman}}]{RSA1978}%
  \BibitemOpen
  \bibfield  {author} {\bibinfo {author} {\bibfnamefont {R.~L.}\ \bibnamefont
  {Rivest}}, \bibinfo {author} {\bibfnamefont {A.}~\bibnamefont {Shamir}}, \
  and\ \bibinfo {author} {\bibfnamefont {L.}~\bibnamefont {Adleman}},\
  }\bibfield  {title} {\enquote {\bibinfo {title} {A method for obtaining
  digital signatures and public-key cryptosystems},}\ }\href@noop {} {\bibfield
   {journal} {\bibinfo  {journal} {Commun. ACM}\ }\textbf {\bibinfo {volume}
  {21}},\ \bibinfo {pages} {120--126} (\bibinfo {year} {1978})}\BibitemShut
  {NoStop}%
\bibitem [{\citenamefont {Feynman}(1982)}]{Feynman1982}%
  \BibitemOpen
  \bibfield  {author} {\bibinfo {author} {\bibfnamefont {Richard}\ \bibnamefont
  {Feynman}},\ }\bibfield  {title} {\enquote {\bibinfo {title} {Simulating
  physics with computers},}\ }\href@noop {} {\bibfield  {journal} {\bibinfo
  {journal} {International Journal of Theoretical Physics}\ }\textbf {\bibinfo
  {volume} {21}},\ \bibinfo {pages} {467--488} (\bibinfo {year}
  {1982})}\BibitemShut {NoStop}%
\bibitem [{\citenamefont {Buluta}\ and\ \citenamefont
  {Nori}(2009)}]{Buluta2009}%
  \BibitemOpen
  \bibfield  {author} {\bibinfo {author} {\bibfnamefont {Iulia}\ \bibnamefont
  {Buluta}}\ and\ \bibinfo {author} {\bibfnamefont {Franco}\ \bibnamefont
  {Nori}},\ }\bibfield  {title} {\enquote {\bibinfo {title} {Quantum
  simulators},}\ }\href@noop {} {\bibfield  {journal} {\bibinfo  {journal}
  {Science}\ }\textbf {\bibinfo {volume} {326}},\ \bibinfo {pages} {108--111}
  (\bibinfo {year} {2009})}\BibitemShut {NoStop}%
\bibitem [{\citenamefont {Barreiro}\ \emph {et~al.}(2011)\citenamefont
  {Barreiro}, \citenamefont {M\"{u}ller}, \citenamefont {Schindler},
  \citenamefont {Nigg}, \citenamefont {Monz}, \citenamefont {Chwalla},
  \citenamefont {Hennrich}, \citenamefont {Roos}, \citenamefont {Zoller},\ and\
  \citenamefont {Blatt}}]{Barreiro2011}%
  \BibitemOpen
  \bibfield  {author} {\bibinfo {author} {\bibfnamefont {Julio~T.}\
  \bibnamefont {Barreiro}}, \bibinfo {author} {\bibfnamefont {Markus}\
  \bibnamefont {M\"{u}ller}}, \bibinfo {author} {\bibfnamefont {Philipp}\
  \bibnamefont {Schindler}}, \bibinfo {author} {\bibfnamefont {Daniel}\
  \bibnamefont {Nigg}}, \bibinfo {author} {\bibfnamefont {Thomas}\ \bibnamefont
  {Monz}}, \bibinfo {author} {\bibfnamefont {Michael}\ \bibnamefont {Chwalla}},
  \bibinfo {author} {\bibfnamefont {Markus}\ \bibnamefont {Hennrich}}, \bibinfo
  {author} {\bibfnamefont {Christian~F.}\ \bibnamefont {Roos}}, \bibinfo
  {author} {\bibfnamefont {Peter}\ \bibnamefont {Zoller}}, \ and\ \bibinfo
  {author} {\bibfnamefont {Rainer}\ \bibnamefont {Blatt}},\ }\bibfield  {title}
  {\enquote {\bibinfo {title} {An open-system quantum simulator with trapped
  ions},}\ }\href@noop {} {\bibfield  {journal} {\bibinfo  {journal} {Nature}\
  }\textbf {\bibinfo {volume} {470}},\ \bibinfo {pages} {486--491} (\bibinfo
  {year} {2011})}\BibitemShut {NoStop}%
\bibitem [{\citenamefont {Biamonte}\ \emph {et~al.}(2011)\citenamefont
  {Biamonte}, \citenamefont {Bergholm}, \citenamefont {Whitfield},
  \citenamefont {Fitzsimons},\ and\ \citenamefont
  {Aspuru-Guzik}}]{Biamonte2010}%
  \BibitemOpen
  \bibfield  {author} {\bibinfo {author} {\bibfnamefont {J.D.}\ \bibnamefont
  {Biamonte}}, \bibinfo {author} {\bibfnamefont {V.}~\bibnamefont {Bergholm}},
  \bibinfo {author} {\bibfnamefont {J.D.}\ \bibnamefont {Whitfield}}, \bibinfo
  {author} {\bibfnamefont {J.}~\bibnamefont {Fitzsimons}}, \ and\ \bibinfo
  {author} {\bibfnamefont {A.}~\bibnamefont {Aspuru-Guzik}},\ }\bibfield
  {title} {\enquote {\bibinfo {title} {Adiabatic quantum simulators},}\
  }\href@noop {} {\bibfield  {journal} {\bibinfo  {journal} {AIP Advances}\
  }\textbf {\bibinfo {volume} {1}},\ \bibinfo {pages} {022126} (\bibinfo {year}
  {2011})}\BibitemShut {NoStop}%
\bibitem [{\citenamefont {Zalka}(1998)}]{Zalka1998b}%
  \BibitemOpen
  \bibfield  {author} {\bibinfo {author} {\bibfnamefont {C.}~\bibnamefont
  {Zalka}},\ }\bibfield  {title} {\enquote {\bibinfo {title} {{Simulating
  quantum systems on a quantum computer}},}\ }\href@noop {} {\bibfield
  {journal} {\bibinfo  {journal} {P. Roy. Soc. A-Math. Phy.}\ }\textbf
  {\bibinfo {volume} {454}},\ \bibinfo {pages} {313--322} (\bibinfo {year}
  {1998})},\ \bibinfo {note} {\emph{Preprint}
  arXiv:quant-ph/9603026v2}\BibitemShut {NoStop}%
\bibitem [{\citenamefont {Kassal}\ \emph {et~al.}(2011)\citenamefont {Kassal},
  \citenamefont {Whitfield}, \citenamefont {Perdomo-Ortiz}, \citenamefont
  {Yung},\ and\ \citenamefont {Aspuru-Guzik}}]{Kassal2011}%
  \BibitemOpen
  \bibfield  {author} {\bibinfo {author} {\bibfnamefont {Ivan}\ \bibnamefont
  {Kassal}}, \bibinfo {author} {\bibfnamefont {James~D.}\ \bibnamefont
  {Whitfield}}, \bibinfo {author} {\bibfnamefont {Alejandro}\ \bibnamefont
  {Perdomo-Ortiz}}, \bibinfo {author} {\bibfnamefont {Man-Hong}\ \bibnamefont
  {Yung}}, \ and\ \bibinfo {author} {\bibfnamefont {Al\'{a}n}\ \bibnamefont
  {Aspuru-Guzik}},\ }\bibfield  {title} {\enquote {\bibinfo {title}
  {{Simulating Chemistry Using Quantum Computers}},}\ }\href@noop {} {\bibfield
   {journal} {\bibinfo  {journal} {{Annu. Rev. Phys. Chem.}}\ }\textbf
  {\bibinfo {volume} {62}},\ \bibinfo {pages} {185--207} (\bibinfo {year}
  {2011})}\BibitemShut {NoStop}%
\bibitem [{\citenamefont {Fowler}\ \emph {et~al.}(2012)\citenamefont {Fowler},
  \citenamefont {Whiteside},\ and\ \citenamefont {Hollenberg}}]{Fowler2012}%
  \BibitemOpen
  \bibfield  {author} {\bibinfo {author} {\bibfnamefont {Austin~G.}\
  \bibnamefont {Fowler}}, \bibinfo {author} {\bibfnamefont {Adam~C.}\
  \bibnamefont {Whiteside}}, \ and\ \bibinfo {author} {\bibfnamefont {Lloyd
  C.~L.}\ \bibnamefont {Hollenberg}},\ }\href@noop {} {\enquote {\bibinfo
  {title} {Towards practical classical processing for the surface code: timing
  analysis},}\ } (\bibinfo {year} {2012}),\ \bibinfo {note} {\emph{Preprint}
  arXiv:1202.5602}\BibitemShut {NoStop}%
\bibitem [{\citenamefont {{Texas Instruments}}(2011)}]{TIDMD}%
  \BibitemOpen
  \bibfield  {author} {\bibinfo {author} {\bibnamefont {{Texas Instruments}}},\
  }\href@noop {} {} (\bibinfo {year} {2011}),\ \bibinfo {note}
  {{http://www.dlp.com}}\BibitemShut {NoStop}%
\bibitem [{\citenamefont {Kim}\ \emph {et~al.}(2003)\citenamefont {Kim} \emph
  {et~al.}}]{Kim03}%
  \BibitemOpen
  \bibfield  {author} {\bibinfo {author} {\bibfnamefont {J.}~\bibnamefont
  {Kim}} \emph {et~al.},\ }\bibfield  {title} {\enquote {\bibinfo {title}
  {1100x1100 port {MEMS}-based optical crossconnect with {4-dB} maximum
  loss},}\ }\href@noop {} {\bibfield  {journal} {\bibinfo  {journal} {IEEE
  Photonics Technology Letters}\ }\textbf {\bibinfo {volume} {15}},\ \bibinfo
  {pages} {1537--1539} (\bibinfo {year} {2003})}\BibitemShut {NoStop}%
\bibitem [{\citenamefont {Kim}\ \emph {et~al.}(2007)\citenamefont {Kim},
  \citenamefont {Knoernschild}, \citenamefont {Liu},\ and\ \citenamefont
  {Kim}}]{Kim07}%
  \BibitemOpen
  \bibfield  {author} {\bibinfo {author} {\bibfnamefont {Changsoon}\
  \bibnamefont {Kim}}, \bibinfo {author} {\bibfnamefont {C.}~\bibnamefont
  {Knoernschild}}, \bibinfo {author} {\bibfnamefont {Bin}\ \bibnamefont {Liu}},
  \ and\ \bibinfo {author} {\bibfnamefont {Jungsang}\ \bibnamefont {Kim}},\
  }\bibfield  {title} {\enquote {\bibinfo {title} {Design and characterization
  of {MEMS} micromirrors for ion-trap quantum computation},}\ }\href@noop {}
  {\bibfield  {journal} {\bibinfo  {journal} {IEEE Journal of Selected Topics
  in Quantum Electronics}\ }\textbf {\bibinfo {volume} {13}},\ \bibinfo {pages}
  {322--329} (\bibinfo {year} {2007})}\BibitemShut {NoStop}%
\bibitem [{\citenamefont {Knoernschild}\ \emph {et~al.}(2008)\citenamefont
  {Knoernschild}, \citenamefont {Kim}, \citenamefont {Liu}, \citenamefont
  {Lu},\ and\ \citenamefont {Kim}}]{Knoer08}%
  \BibitemOpen
  \bibfield  {author} {\bibinfo {author} {\bibfnamefont {Caleb}\ \bibnamefont
  {Knoernschild}}, \bibinfo {author} {\bibfnamefont {Changsoon}\ \bibnamefont
  {Kim}}, \bibinfo {author} {\bibfnamefont {Bin}\ \bibnamefont {Liu}}, \bibinfo
  {author} {\bibfnamefont {Felix~P.}\ \bibnamefont {Lu}}, \ and\ \bibinfo
  {author} {\bibfnamefont {Jungsang}\ \bibnamefont {Kim}},\ }\bibfield  {title}
  {\enquote {\bibinfo {title} {{MEMS}-based optical beam steering system for
  quantum information processing in two-dimensional atomic systems},}\
  }\href@noop {} {\bibfield  {journal} {\bibinfo  {journal} {Optics Letters}\
  }\textbf {\bibinfo {volume} {33}},\ \bibinfo {pages} {273--275} (\bibinfo
  {year} {2008})}\BibitemShut {NoStop}%
\bibitem [{\citenamefont {Liu}\ and\ \citenamefont {Zakhor}(1992)}]{Liu92}%
  \BibitemOpen
  \bibfield  {author} {\bibinfo {author} {\bibfnamefont {Y.}~\bibnamefont
  {Liu}}\ and\ \bibinfo {author} {\bibfnamefont {A.}~\bibnamefont {Zakhor}},\
  }\bibfield  {title} {\enquote {\bibinfo {title} {Binary and phase shifting
  mask design for optical lithography},}\ }\href@noop {} {\bibfield  {journal}
  {\bibinfo  {journal} {IEEE T. Semiconduct. M.}\ }\textbf {\bibinfo {volume}
  {5}},\ \bibinfo {pages} {138--152} (\bibinfo {year} {1992})}\BibitemShut
  {NoStop}%
\bibitem [{\citenamefont {Aizenberg}\ \emph {et~al.}(1998)\citenamefont
  {Aizenberg}, \citenamefont {Rogers}, \citenamefont {Paul},\ and\
  \citenamefont {Whitesides}}]{Aizenberg98}%
  \BibitemOpen
  \bibfield  {author} {\bibinfo {author} {\bibfnamefont {Joanna}\ \bibnamefont
  {Aizenberg}}, \bibinfo {author} {\bibfnamefont {John~A.}\ \bibnamefont
  {Rogers}}, \bibinfo {author} {\bibfnamefont {Kateri~E.}\ \bibnamefont
  {Paul}}, \ and\ \bibinfo {author} {\bibfnamefont {George~M.}\ \bibnamefont
  {Whitesides}},\ }\bibfield  {title} {\enquote {\bibinfo {title} {Imaging
  profiles of light intensity in the near field: Applications to phase-shift
  photolithography},}\ }\href@noop {} {\bibfield  {journal} {\bibinfo
  {journal} {Appl. Opt.}\ }\textbf {\bibinfo {volume} {37}},\ \bibinfo {pages}
  {2145--2152} (\bibinfo {year} {1998})}\BibitemShut {NoStop}%
\bibitem [{\citenamefont {Nielson}\ \emph {et~al.}(2007)\citenamefont
  {Nielson}, \citenamefont {Olsson}, \citenamefont {Resnick},\ and\
  \citenamefont {Spahn}}]{NielsonCLEO2007}%
  \BibitemOpen
  \bibfield  {author} {\bibinfo {author} {\bibfnamefont {G.N.}\ \bibnamefont
  {Nielson}}, \bibinfo {author} {\bibfnamefont {R.H.}\ \bibnamefont {Olsson}},
  \bibinfo {author} {\bibfnamefont {P.R.}\ \bibnamefont {Resnick}}, \ and\
  \bibinfo {author} {\bibfnamefont {O.B.}\ \bibnamefont {Spahn}},\ }\bibfield
  {title} {\enquote {\bibinfo {title} {High-speed {MEMS} micromirror
  switching},}\ }\href@noop {} {\bibfield  {journal} {\bibinfo  {journal}
  {Conference on Lasers and Electro-Optics, 2007 (CLEO 2007)}\ ,\ \bibinfo
  {pages} {1--2}} (\bibinfo {year} {2007})}\BibitemShut {NoStop}%
\end{thebibliography}%

\end{document}